\documentclass[11pt]{article}
\usepackage{latexsym,amssymb,amsmath}
\usepackage[dvips]{graphics,graphicx,epsfig,color}
\usepackage[lofdepth,lotdepth]{subfig}
\begin{document}
%
\thispagestyle{empty}
\title{\bf  Forward and inverse acoustic scattering problems involving the mass density}
\author{\bf Armand Wirgin \thanks {
LMA, CNRS, UMR 7031, Aix-Marseille Univ, Centrale Marseille, F-13453 Marseille Cedex 13, France.}}
\date{\today}
\maketitle
\begin{abstract}
This investigation is concerned with the 2D acoustic scattering problem of a plane wave propagating in a non-lossy fluid host and soliciting a linear, isotropic, macroscopically-homogeneous, lossy, flat-plane layer in which the mass density and  wavespeed are different from those of the host. The focus is on the inverse problem of the retrieval of either the layer mass density or the real part of the layer wavespeed. The data is the transmitted pressure field, obtained by simulation (resolution of the forward problem) in exact, explicit form via  separation of variables.  Another form of this solution,  which is exact and more explicit in terms of the mass-density contrast (between the host and layer), is obtained by a domain-integral method. A perturbation technique enables this solution to be cast as a series of powers  of the mass density contrast, the first three terms of which are employed as the trial models in the treatment of the inverse problem. The aptitude of these models to retrieve the mass density contrast and real part of the layer wavespeed is demonstrated both theoretically and numerically.
\end{abstract}Keywords: 2D acoustics, forward scattering, inverse scattering, domain integral equation, constant mass density assumption, small density contrast, retrieval accuracy
\newline
\newline
Abbreviated title: Role of mass density in forward and inverse acoustic scattering
\newline
\newline
Corresponding author: Armand Wirgin, \\ e-mail: wirgin@lma.cnrs-mrs.fr
 \newpage
\tableofcontents
\newpage
\section{Introduction}\label{intro}
Scattering is produced when a material obstacle is solicited by a wave and manifests itself by the transformation  of the incident wave into  several waves  whose amplitudes, directions, and spectral properties are different from those of the solicitation. For instance, when the solicitation is a plane wave and the obstacle is a fluid-like, macroscopically homogeneous, isotropic flat-faced layer situated within another fluid of the same nature, but of different density and/or compressibility, the scattering manifests itself by the splitting of the incident plane wave into a reflected plane wave and a transmitted plane wave in the two half-spaces above and below the layer, accompanied by one upgoing and one downgoing plane wave within the layer itself.

The task, in forward-scattering problems, is to predict the characteristics of the scattered wavefield for a given solicitation, host medium, and obstacle.  This task has occupied  researchers for the last two centuries \cite{ra18,mf53,ri65,pe80,ut86,bh90,de00,kr02a,kr02b} (and continues to occupy them \cite{mt03,ke16,mp17,ts18,wi19}) because the solicitation, and/or the host medium, and/or the obstacle can (and does, in real-world situations) have complicated characteristics. For instance, in the biomedical field, predicting the acoustic wavefield within a human body \cite{st92,mh99}, due to an ultrasonic transducer being applied on the surface of the body, is complicated because the body is a complicated medium/structure composed of many  closely-spaced heterogeneous organs of different nature and size located within a host medium containing many small-scale heterogeneities, and the ultrasonic pulse-like solicitation wavefield emitted by the transducer is not simple either. A first, crude, but very-useful, approximation is to assume that the target organ is similar to a homogeneous, isotropic fluid-like layer, the  wave soliciting this organ within the body is similar to a plane wave, and that it strikes the layer-like organ at normal incidence. It ensues that the scattering manifests itself by a backward-reflected wave redirected towards the skin and then back into the transducer. A relatively-simple analysis shows how the back-scattered pulse is spectrally (i.e., affecting the shape and height of the pulse) and temporally (affecting the time-delay with respect to the incident pulse) modified, this being the information that is then exploited (a problem that is different from the forward problem we have just described) for diagnostic purposes.

The reasons why this crude paradigm is so effective are that: 1) the characteristic frequency of the incident pulse is such that the associated wavelength of the acoustic wavefield within the body is small relative to the size of the organ-like target, 2) most of the incident energy is concentrated (focused) into a beam directed on purpose towards the target and not elsewhere, and 3) the density and compressibility of the target organ are sufficiently-different from those of the host medium (within the body) for the reflected wave to have sufficient amplitude so as to be distinguishable from the background returns \cite{st92,mh99} due to scattering by smaller-scale heterogeneities in the path of the incident and reflected waves.

A less-crude paradigm for the aforementioned biomedical forward problem \cite{mh99} might consist in assuming that: a) the organ is not layer-like but rather potato-like \cite{st92}, but still composed of an isotropic, medium, b) this medium is  piecewise-homogeneous \cite{mh99} and fluid-like, c) the host medium is a homogeneous, isotropic fluid, and d) the solicitation is still a plane wave, propagating in the host medium. The forward problem of predicting the acoustic wavefield outside and within this obstacle turns out to be much more complex than the one involving the layer. It is so complex that in recent years the tendency has been to not even try to solve it mathematically (such as based on low-frequency, high-frequency, canonical shape approximations) but rather numerically (e.g., by finite difference \cite{bh90,ve12}, finite element \cite{dc96,th06,ts99}, spectral element\cite{mo08}, discrete-source \cite{de00} methods, etc., which is not to say that these methods do not involve a lot of mathematics too), justified by the fact that powerful computing facilities and software are increasingly available to researchers. The results of such computations are aesthetically and scientifically-speaking quite impressive in that they reveal the striking complexity and variability of the scattered field  with respect to factors such as frequency, composition and shape \cite{ut86,dc96,ts99,mt03,po12,ke16,mp17,cm17,mi18,ts18,co19}. Making use of this information for  diagnostic purposes such as in biomedicine is another, even more challenging, problem.

The diagnostic problem is more familiar to the layman than the forward problem because it occupies him permanently. For instance, a sound received within his auditory system is transduced, encoded and conveyed to the brain wherein it is analyzed \cite{dh07} so as to provoke a reaction such as withdrawal (reaction to danger as manifested by the loud and/or nearby sound source), pleasure (reaction to a melodious sound) or other types of reaction. The diagnostic problem is therefore one of exploiting the information conveyed by a wave (e.g., sound) to analyze the source (e.g, its distance from the ear) of this wave and/or the medium/structure (e.g., a wall) which the wave encounters before being detected, the purpose of the analysis being to furnish the means for making a decision (e.g., in the biomedical sphere, this might be a decision as to whether an organ submitted to ultrasound is healthy or diseased \cite{wb12,hs11}).

The diagnostic problem, as it relates to the exploitation of the information contained in scattered waves, is known in the scientific community as the inverse scattering problem (because it is the other way around of the forward scattering problem, i.e., determine the source and/or host medium and/or the obstacle from the scattered wavefield). How a human being copes with such a problem in the aforementioned auditory context is nothing less than astounding and certainly enabled by the incredible power and efficiency of the brain. But, a more detailed look at how this functions would probably reveal that the judgements and reactions of a human being to a sound are often either sub-optimal (or even disastrous)  and/or simply biased \cite{cj15}, mostly due to the ill-posed nature of inverse problems \cite{ta77,eg87}. Such errors and bias, which are a consequence of either the non-existence, and/or the non-uniqueness and/or the instability of the solution (i.e., the fundamental properties of ill-posed problems) are generally not admissible in the biomedical diagnostic context, but the question is: can they  be avoided, or, in other terms: if one assumes so-and-so (e.g., certain aspects of the model concerning how the acoustic wave interacts with the environment), what may the consequences be for his diagnosis \cite{gu00}?

 Our initial intention \cite{wi19} was to seek answers to this question in relation to  the inverse scattering problem of the retrival of one or two constitutive properties of the medium of which an obstacle of arbitrary shape (i.e., potato-like) is composed after having been solicited by a wave. If the related forward problem is difficult to solve (and, as written earlier: now usually carried out numerically), the inverse problem is even harder to deal with for several reasons: 1) it is mathematically non-linear even when the associated forward problem is linear, this being one of the causes of ill-posedness, 2) usually, it cannot be solved in a mathematically-sound (e.g., algebraic manner, as for searching for the roots of a polynomial equation) \cite{wi02b,we13}, which means that it is treated algorithmically by what resembles a trial and error optimization technique requiring many resolutions of the associated forward-scattering problem \cite{ro04,wi99,lm13},  3) it is not clear what aspects, and what amount, of the scattered-wave data are necessary to treat the inverse problem in the best manner, 4) in real-world situations (such as in geophysical applications \cite{ve12}), one may dispose of either a very small amount of data (which may be somewhat inappropriate) or a large amount of disparate, unsynchronized  data  gathered by  measurement methods that are not-easily controlled, 5) many of the parameters of the solicitation, the host medium and even of the obstacle (aside, from those that are searched-for) are either not at all, or poorly, known \cite{lm13,wi16c}, and 6) these parameters could also be searched-for by the retrieval scheme, but the latter becomes more difficult as the number of to-be-retrieved parameters increases \cite{sw14}.

 The aforementioned  second point is crucial in that it forces one to think twice before undertaking a purely-numerical resolution of  the inverse problem. This means that the traditional, more, mathematically-oriented methods might have their say, notably in the way each of the potentially-multiple  forward problems are formulated. The latter are so-called trial problems that differ from each other by the values assigned to the to-be-retrieved (in our case, constitutive) parameters. The way a trial problem is treated is called the trial model. What one usually strives for is a trial model that is as mathematically-simple as possible, while being hopefully-able to account for the most important features of the scattered wavefield. Such simple models are often the outcome of certain assumptions and approximations in what is called the domain integral formulation (treated in depth in {\cite{wi19}) of the forward-scattering problem. These assumptions usually have to do with:\\\\
 1) treating an elastic wave problem (in a solid or porous medium) as an acoustic wave problem \cite{ch86,gv86,lv92,gl01,ll07,cc15} (in a so-called equivalent fluid),\\
 2)  treating a microscopically-inhomogeneous (e.g., porous) medium as a macroscopically-homogeneous (effective) medium \cite{bg04,dr07,dr08,dl08,be10,of11,wi18},\\
 3) treating the bioacoustic, marine acoustic, electromagnetic  and geophysical  problem as one in which the mass density, or another constitutive property, is constant everywhere (i.e., is the same and spatially-constant within the obstacle as well as in the host)\cite{ri65,gr80,ra81,ta84,dl85,dt85,le85,ta86,rg86,gv86,wv86,mo87,mc93,dl00,gl01,zb02,bg04,gd07,ns09,be10,to10,hs11,ke16,wi99,zb02},\\
 4) treating  a 3D problem as a 2.5D, 2D or even 1D problem \cite{ta76,el95,wi19} (and many other references)\\
 5) treating the potato as a sphere, coated sphere, cylinder, coated cylinder, layer, multilayer (\cite{kd91,wi02b,ke16,wi16a,bt18}, or other body of canonical shape,\\
 (6) treating the host and/or obstacle as being linear and isotropic \cite{wi19} (and many other references),\\
 7) invoking ad hoc expressions of the scattered field such as the so-called Born, Kirchoff, physical optics, tangent plane approximations \cite{ls50,hh81,wi82,ch86,rg86,wv86,mo87,to99,wi99,dl00,gl01,zb02,gv07,ve12}, etc.\\

  Sometimes,  questions have been raised and addressed as to whether   the acoustic 'approximation', the 2D 'approximation' etc. are really justified (i.e., lead to accurate parameter retrievals), but when this question relates to the retrieval of the density it is understandable that it has not been considered to be interesting since it is not  possible to retrieve the density from the scattered field with a trial model based on the constant-density assumption. Less comprehensible is the fact that  the question of the influence of the constant-density assumption on the quality of the retrieval of another parameter such as the wavespeed in, or compressibility of, the medium in the  obstacle has not received much  attention \cite{mv15,lo18,wi19} either, except insofar as porous media are concerned \cite{go10}. Nevertheless, it should be underlined that in the area of electromagnetic scattering, the (acoustic) constant-density assumption becomes the oft-employed assumption of constant permeability (\cite{ta76,pe80}, p. 2), which is now routinely dropped, notably to account for the exotic properties of metamaterials \cite{cg04,ss02,sv05,et15,wi16b}. Also, more and more studies of acoustic (and elastic wave) inverse scattering no longer appeal to the constant-density assumption \cite{ra81,cc84,gu00,mc93,sw04,dr08,lo09,mv15,wi16a,wi16c,lo18,pa19,wi19}

 A less-radical (than the constant-density) assumption  is that: a) the density of the host is spatially-constant, b) the density of the obstacle, different from that of the host, is also spatially-constant, and c) the mass-density contrast $\epsilon$ (involving the difference of the two constant densities) is small. If the inverse problem is to retrieve the obstacle mass-density (which appears, for instance, to be of interest in certain biomedical applications \cite{zb02,lo09,wb12,mv15,wi16a}),  an important question would appear to be: how small must $\epsilon$ (which is unknown, like the obstacle mass density, when it is the sought-for-parameter) be for the retrieval of $\epsilon$  (or of another constitutive parameter such as the wavespeed in the obstacle) to be reliable if at the outset the small-$\epsilon$ assumption is incorporated in the trial model? This is the principal question that will shall address in our investigation.
\section{The forward-scattering problem}\label{fsp}
%
\subsection{Preliminaries}\label{fsp-prelims}
The material in this section concerns the determination of the scattered pressure field due to a radiated (by a source) pressure field, initially propagating in a lossless fluid, which encounters a fluid-like obstacle placed somewhere outside of the support of the source. The characteristics of the source (or of the field it radiates), as well as the location, geometrical and physical parameters of the obstacle, are assumed to be known, which means that the only unknowns are the pressure fields outside, and within, the obstacle.

We shall assume that: 1) the source density is invariant with respect to $z$, 2) the obstacle  is of infinite extension in the  $z$ direction, 3) the host material surrounding  the obstace is a lossless, homogeneous, isotropic fluid, 4) the fluid-like material within the obstacle is isotropic and possibly lossy, and 5) the eventual heterogeneity  of the obstacle is such that the mass density and wavespeed therein may depend only on $x$ and $y$ but not on $z$. The principal consequence of these assumptions is that the incident (i.e., radiated by the source) and scattered (i.e., due to the presence of the obstacle) pressure fields do not depend on $z$, i.e., the problem is 2D and can henceforth be examined in the $x-y$ (i.e., sagittal) plane.

The time domain governing partial differential equation (PDE) for the total pressure field $p$ (in the entire sagittal plane, i.e., above, within and below the layer obstacle) is the 2D Bergman equation
\begin{equation}\label{intro-10}
\nabla_{\mathbf{x}}\cdot\nabla_{\mathbf{x}} p(\mathbf{x},t)-\frac{1}{c(\mathbf{x})^{2}}\frac{\partial^{2}p(\mathbf{x},t)}{\partial t^{2}}-\frac{\nabla_{\mathbf{x}}\rho(\mathbf{x})}{\rho(\mathbf{x})}\cdot \nabla_{\mathbf{x}} p(\mathbf{x},t)=\rho(\mathbf{x})
\nabla_{\mathbf{x}}\cdot \mathbf{f}(\mathbf{x},t)=0~;~\forall \mathbf{x}\in\mathbb{R}^{2}
 ~,
\end{equation}
wherein  $\mathbf{x}=(x,y)$, $t$ is the time variable, $\rho$ the mass density, $\mathbf{f}$  the applied force associated with the source, and
\begin{equation}\label{intro-20}
\frac{1}{c(\mathbf{x})^{2}}:=\frac{\rho(\mathbf{x})}{K(\mathbf{x})}=\rho(\mathbf{x})\kappa(\mathbf{x})
,
\end{equation}
with $c(\mathbf{x})$, $K(\mathbf{x})$ and $\kappa(\mathbf{x})$ the phase velocity (also termed wavespeed), isentropic bulk modulus and isentropic compressibility within/of the fluid medium.

By expanding the time domain pressure and force in the Fourier integrals
\begin{equation}\label{intro-30}
p(\mathbf{x},t)=\int_{-\infty}^{\infty}p(\mathbf{x},\omega)\exp(-i\omega t)d\omega
 ~,
\end{equation}
\begin{equation}\label{intro-35}
\mathbf{f}(\mathbf{x},t)=\int_{-\infty}^{\infty}\mathbf{f}(\mathbf{x},\omega)\exp(-i\omega t)d\omega
 ~,
\end{equation}
wherein $\omega=2\pi f$ is the angular frequency and $f$ the frequency, one obtains the frequency domain expression of the Bergman wave equation
\begin{equation}\label{intro-40}
\nabla_{\mathbf{x}}\cdot\nabla_{\mathbf{x}}  p(\mathbf{x},\omega)+k^{2}(\mathbf{x},\omega)p(\mathbf{x},\omega)-
\frac{\nabla_{\mathbf{x}}\rho(\mathbf{x},\omega)}{\rho(\mathbf{x},\omega)}\cdot \nabla_{\mathbf{x}} p(\mathbf{x},\omega)=
\rho(\mathbf{x},\omega)
\nabla_{\mathbf{x}}\cdot \mathbf{f}(\mathbf{x},\omega~;~\forall \mathbf{x}\in\mathbb{R}^{2}
~,
\end{equation}
wherein
\begin{equation}\label{intro-50}
k(\mathbf{x},\omega):=\frac{\omega}{c(\mathbf{x},\omega)}
\end{equation}
is the wavenumber, which is a generally-complex quantity due to the fact that $c$ is assumed to be generally-complex (to account for losses in the obstacle portion of the scattering configuration). Eq. (\ref{intro-40}) is the frequency domain version of the 2D Bergman PDE (BPDE).

The BPDE implies certain properties of the pressure field:\\\\
(i) $p(\mathbf{x},\omega)$ is continuous and bounded everywhere in $\mathbb{R}^{2}$ except within the support of the source,\\
(ii) $\frac{\nabla_{\mathbf{x}}\rho(\mathbf{x},\omega)}{\rho(\mathbf{x},\omega)}$ is continuous and bounded everywhere in $\mathbb{R}^{2}$ except within the support of the source.
\\\\
Since the BPDE applies to all of $\mathbb{R}^{2}$, the outer reaches of this infinite domain constitute a sort of boundary on which it turns out to be necessary to specify the behavior of the pressure field. Seen by an observer situated on this 'boundary', the field is considered (on grounds of physical plausibility) to behave as a wave emerging from (rather than converging to)  the sources (both applied due to $\mathbf{f}$, and induced within the obstacle due to its solicitation by the radiated wave), this being what is called the radiation (at infinity) or outgoing wave condition.

The BPDE is the differential form of the frequency domain governing equation of the scattering problem. As is well-known, it is also possible to cast this problem in integral form. The result is the so-called Bergman domain integral equation (BDIE)
\begin{multline}\label{intro-50}
p(\mathbf{x},\omega)=p^{r}(\mathbf{x},\omega)+\\
\int_{\mathbb{R}^{2}}G^{[0]}(\mathbf{x};\mathbf{x'};\omega)
\left(\left[ \big(k(\mathbf{x'},\omega)\big)^{2}-
 \big( k^{[0]}\big ) ^{2}\right] p(\mathbf{x'},\omega)-
 \frac{\nabla_{\mathbf{x'}}\rho(\mathbf{x'},\omega)}{\rho(\mathbf{x'},\omega)}\cdot \nabla_{\mathbf{x'}} p(\mathbf{x'},\omega)\right)d\Omega(\mathbf{x'})~~;~~\forall~\mathbf{x}\in
\mathbb{R}^{2}~,
\end{multline}
wherein $d\Omega$ is the differential area in the sagittal plane and $G^{[0]}(\mathbf{x};\mathbf{x'};\omega)$  the free-space Green's function satisfying
\begin{equation}\label{intro-60}
\nabla_{\mathbf{x}}\cdot\nabla_{\mathbf{x}}G^{[0]}(\mathbf{x};\mathbf{x'};\omega)+
\left(k^{[0]}\right)^{2}G^{[0]}(\mathbf{x};\mathbf{x'};\omega)=-\delta(\mathbf{x}-\mathbf{x'})
~~;~~\forall~\mathbf{x},~\mathbf{x'}\in\mathbb{R}^{2}~,
\end{equation}
and subject (together with $p$) to the radiation condition.

Henceforth, we shall be concerned with the resolution of the BPDE and the BDIE for a specific (i.e., layer-like) obstacle and solicitation (i.e., plane wave), keeping in mind properties (i), (ii) and the radiation condition.
\subsection{Description of the problem of scattering of a plane wave by a layer}\label{fsp-descrip}
The obstacle is a fluid-like, flat-faced layer of infinite extent in the $x$ and $z$ directions. The upper and lower faces of the layer are the planes $y=0$ and $y=-h$ respectively, with $h$ the thickness of the layer. The sagittal ($x-y$) plane view of the scattering configuration is given in fig. \ref{fig1}, wherein $\mathbf{k^{i}}$ is the wavevector of the incident (this word is employed to distinguish it from 'radiated', but the incident and radiated waves play the same role which is to solicit the obstacle) plane wave $p^{i}(\mathbf{x,\omega})=a^{i}(\omega)\exp(i\mathbf{k}^{i}(\omega)\cdot\mathbf{x})$ whose amplitude is $a^{i}(\omega)$.

$\mathbb{R}^{2}$ is initially (i.e., before the introduction of the layer) filled with a {\it homogeneous} isotropic fluid medium
$M^{[0]}(\rho^{[0]},c^{[0]})$, in which $\rho^{[0]}$ and $c^{[0]}$
are position- and frequency-independent.
\begin{figure}[ht]
\begin{center}
\includegraphics[height=2.5in, width=5in ] {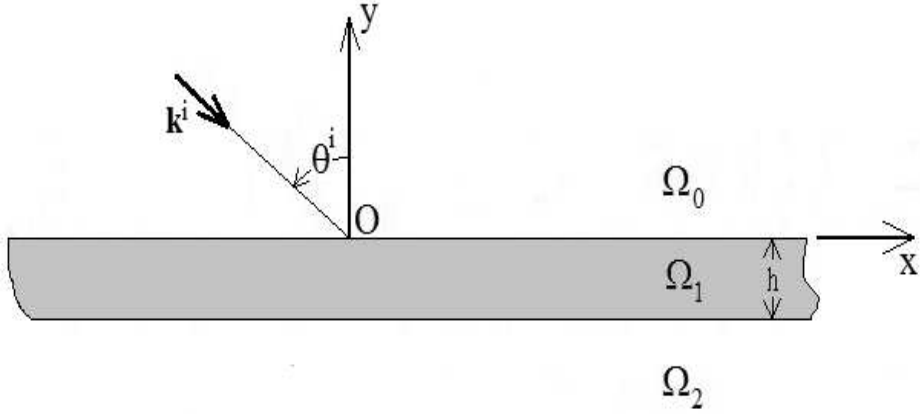}
\caption{Sagittal plane view of a plane acoustic wave striking a fluid-like layer immersed in another fluid.} \label{fig1}
\end{center}
\end{figure}
$M^{[0]}(\rho^{[0]},c^{[0]})$ is termed the host (i.e., to the obstacle) medium. After the introduction of the layer-like obstacle, $M^{[0]}(\rho^{[0]},c^{[0]})$ continues to occupy the spaces above ($\Omega_{0}$) and below ($\Omega_{2}$) the layer, whereas the space ($\Omega_{1}$) within the layer is occupied by the fluid-like medium $M^{[1]}(\rho^{[1]},c^{[1]})$, the latter being
 isotropic,  possibly-heterogeneous (but only with respect to $y$) and generally-lossy, this meaning that $c^{[1]}$ (but also possibly $\rho^{[1]}$ such as in fluid-saturated porous materials) is(are) complex. Thus, $\rho(\mathbf{x},\omega)$ and $c(\mathbf{x},\omega)$ in the BPDE take on the values of $M^{[0]}$, $M^{[1]}$, and $M^{[0]}$ as one progresses from $y=-\infty$ to $y=\infty$ within $\mathbb{R}^{2}$. It may (although this is not mandatory) be convenient to assume that the {\it total} pressure field $p(\mathbf{x},\omega)$ takes on the values $p^{[2]}(\mathbf{x},\omega)$ (within $\Omega_{2}$), $p^{[1]}(\mathbf{x},\omega)$ (within $\Omega_{1}$), and $p^{[0]}\mathbf{x},\omega)$ (within $\Omega_{0}$) as one progresses from $y=-\infty$ to $y=\infty$ within $\mathbb{R}^{2}$. Finally, it is convenient, but again not mandatory, to assume that the scattered pressure field is generally equal to the total pressure field minus the incident pressure field throughout $\mathbb{R}^{2}$. Actually, we will change this rule in the so-called DD-SOV method of solution outlined hereafter.
\subsection{The domain decomposition, separation-of-variables (DD-SOV) solution via the BPDE}\label{fsp-ddsov}
At the heart of the DD-SOV method is the very strong assumption
\begin{equation}\label{ddsov-10}
\nabla_{\mathbf{x}}\rho(\mathbf{x},\omega)=\mathbf{0}~~,~~\nabla_{\mathbf{x}}c(\mathbf{x},\omega)=\mathbf{0}~;~\forall \mathbf{x}\in\Omega_{1}
~,
\end{equation}
which means that $\rho^{[1]}$ and $c^{[1]}$ (and therefore $k^{[1]}=\omega/c^{[1]}$) are constants with respect to $\mathbf{x}$.
The DD-SOV method  accounts explicitly for the fact that $\mathbb{R}^{2}$ is {\it decomposed} into three subdomains, $\Omega_{0},~\Omega_{1},~\Omega_{2}$, each in which  both constitutive functions $\rho(\mathbf{x},\omega)$ and $c(\mathbf{x},\omega)$ (which constitute the coefficients of the BPDE) are constants (with respect to position) so that we are faced with the problem of solving three Helmholtz equations
\begin{equation}\label{ddsov-15}
\left(\frac{\partial^{2}}{\partial x^{2}}+\frac{\partial^{2}}{\partial y^{2}}+\left(k^{[j]}(\omega)\right)^{2}\right)p^{[j]}(\mathbf{x},\omega)=0~;~\forall \mathbf{x}\in\Omega_{j}~,~j=0,1,2
~
\end{equation}
wherein $k^{[2]}=k^{[0]}$ and $k^{[j]}=\omega/c^{[j]}~;~j=0,1$ are constants with respect to $\mathbf{x}$.

For plane-wave incidence, $\mathbf{f}(\mathbf{x},t)=0$, and $p(\mathbf{x},\omega)=p^{i}(\mathbf{x},\omega)+p^{d}(\mathbf{x},\omega)$, wherein $p^{i}(\mathbf{x},\omega)$ is the incident plane-wave pressure field
\begin{equation}\label{ddsov-020}
p^{i}(\mathbf{x},\omega)=a^{i}(\omega)\exp(i\mathbf{k^{i}\cdot}\mathbf{x})=a^{i}(\omega)\exp[i(k_{x}^{i[0]}x-k_{y}^{i[0]}y)]~,
\end{equation}
wherein $\mathbf{k}^{i}=(k_{x}^{i[0]},-k_{y}^{i[0]})=(k^{[0]}\sin\theta^{i},-k^{[0]}\cos\theta^{i})$, $\mathbf{x}=(x,y)$ and $\theta^{i}$ is the angle of incidence. Moreover, $p^{d}(\mathbf{x},\omega)$ is the diffracted (or scattered) pressure field which obeys the radiation condition
\begin{equation}\label{ddsov-030}
p^{d}(\mathbf{x},\omega)\sim \text{outgoing waves}~;~\|\mathbf{x}\|\rightarrow\infty~.
\end{equation}

$p^{[j]}(\mathbf{x},\omega)~;~j=0,1,2$ are the total pressure fields in $\Omega_{j}~;~j=0,1,2$ which (due to the assumption that $\rho^{[2]}=\rho^{[0]}$) are connected via the transmission (boundary) conditions arising from properties (i) and (ii) of sect. \ref{fsp-prelims}:
\begin{equation}\label{ddsov-040}
p^{[0]}(x,0,\omega)-p^{[1]}(x,0,\omega)=0~;~\forall\mathbf{x}\in\mathbb{R}~,
\end{equation}
\begin{equation}\label{ddsov-050}
\frac{1}{\rho^{[0]}}p_{,y}^{[0]}(x,0,\omega)-\frac{1}{\rho^{[1]}}p_{,y}^{[1]}(x,0,\omega)=0~;~\forall\mathbf{x}\in\mathbb{R}~,
\end{equation}
\begin{equation}\label{ddsov-060}
p^{[1]}(x,-h,\omega)-p^{[2]}(x,-h,\omega)=0~;~\forall\mathbf{x}\in\mathbb{R}~,
\end{equation}
\begin{equation}\label{ddsov-070}
\frac{1}{\rho^{[1]}}p_{,y}^{[1]}(x,0,\omega)-\frac{1}{\rho^{[0]}}p_{,y}^{[2]}(x,0,\omega)=0~;~\forall\mathbf{x}\in\mathbb{R}~.
\end{equation}
Applying SOV to the Helmholtz equations  gives rise to the field representations
\begin{equation}\label{ddsov-080}
p^{d[j]}(\mathbf{x},\omega)=a^{[j]}(\omega)\exp[i(k_{x}^{i[0]}x-k_{y}^{i[j]}y)]+b^{[j]}(\omega)\exp[i(k_{x}^{i[0]}x+k_{y}^{i[j]}y)]~,~j=0,1,2~;
\end{equation}
in which $a^{[0]}(\omega)=a^{i}(\omega)$,
\begin{equation}\label{ddsov-080a}
k_{y}^{i[j]}=\sqrt{(k^{[j]})^{2}-(k_{x}^{i[0]})^{2}}~;~\Re(k_{y}^{i[j]}\ge 0~,~\Im(k_{y}^{i[j]}\ge 0~;~\omega\ge 0~,
\end{equation}
whereas the invocation of the radiation (i.e., outgoing-wave) condition (\ref{ddsov-030}) entails
\begin{equation}\label{ddsov-090}
b^{[2]}=0~,
\end{equation}
so that the problem reduces to the determination of the four unknown amplitudes $b^{[0]},a^{[1]},b^{[1]},a^{[2]}$ by means of the four transmission conditions. This gives rise to the four relations:
\begin{equation}\label{ddsov-100}
\begin{array}{cc}
\left(a^{[0]}+b^{[0]}\right)-\left(a^{[l]}-b^{[l]}\right) & =0\\
\frac{ik_{y}^{i[0]}}{\rho^{[0]}}\left(-a^{[0]}+b^{[0]}\right)-\frac{ik_{y}^{i[1]}}{\rho^{[1]}}\left(-a^{[1]}+b^{[1]}\right) & =0\\
\left(a^{[1]}\exp(ik_{y}^{i[1]}h)+b^{[1]}\right)\exp(-ik_{y}^{i[1]}h)-a^{[2]}\exp(ik_{y}^{i[2]}h) & =0\\
\frac{ik_{y}^{i[1]}}{\rho^{[1]}}\left(-a^{[1]}\exp(ik_{y}^{i[1]}h)+b^{[1]}\exp(-ik_{y}^{i[1]}h)\right)-
\frac{ik_{y}^{i[2]}}{\rho^{[2]}}\left(-a^{[2]}\exp(ik_{y}^{i[2]}h)\right) & =0
\end{array}
~,
\end{equation}
from which it is straightforward to extract the solutions:
\begin{equation}\label{ddsov-110}
b^{[0]}=a^{[0]}\left[
\frac
{\left((\gamma^{[0]})^{2}-(\gamma^{[1]})^{2}\right)i\sin(k_{y}^{[1]}h)}
{2\gamma^{[0]}\gamma^{[1]}\cos(k_{y}^{[1]}h)-\left((\gamma^{[0]})^{2}+(\gamma^{[1]})^{2}\right)i\sin(k_{y}^{[1]}h)}
\right]
~,
\end{equation}
\begin{equation}\label{ddsov-120}
a^{[1]}=a^{[0]}\left[
\frac
{\gamma^{[0]}\left(\gamma^{[1]}+\gamma^{[0]}\right)\exp(-ik_{y}^{[1]}h)}
{2\gamma^{[0]}\gamma^{[1]}\cos(k_{y}^{[1]}h)-\left((\gamma^{[0]})^{2}+(\gamma^{[1]})^{2}\right)i\sin(k_{y}^{[1]}h)}
\right]
~,
\end{equation}
\begin{equation}\label{ddsov-130}
b^{[1]}=a^{[0]}\left[
\frac
{\gamma^{[0]}\left(\gamma^{[1]}-\gamma^{[0]}\right)\exp(ik_{y}^{[1]}h)}
{2\gamma^{[0]}\gamma^{[1]}\cos(k_{y}^{[1]}h)-\left((\gamma^{[0]})^{2}+(\gamma^{[1]})^{2}\right)i\sin(k_{y}^{[1]}h)}
\right]
~,
\end{equation}
\begin{equation}\label{ddsov-140}
a^{[2]}=a^{[0]}\left[
\frac
{2\gamma^{[0]}\gamma^{[1]}\exp(-ik_{y}^{[0]}h)}
{2\gamma^{[0]}\gamma^{[1]}\cos(k_{y}^{[1]}h)-\left((\gamma^{[0]})^{2}+(\gamma^{[1]})^{2}\right)i\sin(k_{y}^{[1]}h)}
\right]
~,
\end{equation}
wherein $\gamma^{[j]}=k_{y}^{i[j]}/\rho^{[j]}$.

It is not evident  how the mass density contrast $\epsilon=\left(\rho^{[1]}-\rho^{[0]}\right)/\rho^{[1]}$ intervenes in these expressions. For this reason, we shall search for the solution of the scattering problem by another method.
\subsection{The domain integral (DI) solution}\label{di}
The point of departure is now the Bergman integral equation (\ref{intro-50}) wherein $G^{[0]}$ is the 2D Helmholtz operator free-space Green's function for medium $M^{[0]}$ given by \cite{mf53}
\begin{equation}\label{di-30}
G^{[0]}(\mathbf{x};\mathbf{x'})=\frac{i}{4}H_{0}^{(1)}(k^{[0]}(\mathbf{x};\mathbf{x'}))~,
\end{equation}
wherein $H_{0}^{(1)}$ is the zeroth-order Hankel function of the first kind \cite{as68}. Note that we have dropped the $\omega$-dependence in this relation and shall do so from now on for all functions that depend on this variable. This Green's function admits the following representation in cartesian coordinates ((\cite{mf53}, p. 823), with $\mathbf{x}=(x,y)$ and $\mathbf{x'}=(x',y')$:
\begin{equation}\label{di-40}
G^{[0]}(\mathbf{x};\mathbf{x'})=
\frac{i}{4\pi}\int_{-\infty}^{\infty}\exp[i(k_{x}(x-x')+k_{y}^{[0]}|y-y'|)]\frac{dk_{x}}{k_{y}^{[0]}}
~,
\end{equation}
in which
\begin{equation}\label{di-50}
k_{y}^{[j]}=\sqrt{(k^{[j})^{2}-(k_{x})^{2}}~;~\Re k_{y}^{[j]}\ge 0~,~\Im k_{y}^{[j]}\ge~;~\omega\ge 0
~.
\end{equation}

Note that when the mass density is everywhere the same,  the Bergman integral equation reduces to what is usually termed the Lippmann-Schwinger (LS) equation
\begin{equation}\label{di-52}
p(\mathbf{x})=p^{i}(\mathbf{x})+
\int_{\mathbb{R}^{2}}G^{[0]}(\mathbf{x};\mathbf{x'})
\left[ \big(k(\mathbf{x'})\big)^{2}-
 \big( k^{0}\big ) ^{2}\right] p(\mathbf{x'})d\Omega(\mathbf{x'})~~;~~\forall~\mathbf{x}\in
\mathbb{R}^{2}~,
\end{equation}
which, of course, is easier to deal with than the Bergman integral equation. We shall return to the LS integral equation further on, and continue, for the moment, to concentrate our attention on the Bergman integral equation.
\subsubsection{The assumption of invariance with respect to $x$}
Recall that $M^{[0]}$ was assumed to be homogeneous and isotropic. Although the BPDE (\ref{intro-50}) and (\ref{di-30})-(\ref{di-50}) hold for an arbitrary isotropic, {\it}inhomogeneous, although invariant with respect to $x_{3}=z$, medium  occupying the 'obstacle' (i.e., the layer), we shall henceforth assume $M^{[1]}$ to (also) be invariant with respect to $x$, i.e.,
\begin{equation}\label{di-55}
k(\mathbf{x})=k(y)~~,~~\rho(\mathbf{x})=\rho(y)
~.
\end{equation}
The consequence of this in (\ref{intro-50}) is, on account of the plane wave nature of the incident pressure field:
\begin{equation}\label{di-60}
p(\mathbf{x})=p(y)\exp(ik_{x}^{i[0]}x)
~
\end{equation}
which, after projection of (\ref{intro-50}) and employment of the identity
\begin{equation}\label{di-70}
\int_{-\infty}^{\infty}\exp[i(k_{x}^{i[0]}-k_{x})x']dx'=2\pi\delta(k_{x}^{i[0]}-k_{x})
~,
\end{equation}
(with $\delta(~)$ the Dirac delta distribution), leads to the 1D integral equation
\begin{multline}\label{di-80}
p(y)=p^{i}(y)+\\
\frac{i}{2k_{y}^{i[0]}}\int_{-\infty}^{\infty}dy'\left(\left[\left(k(y')\right)^{2}-\left(k^{[0]}\right)^{2}\right] p(y')-
\frac{\rho_{y'}(y')}{\rho(y')} p_{,y'}(y')\right)\exp(ik_{y}^{i[0]}|y-y'|)~;~\forall y\in\mathbb{R}~.
\end{multline}
\subsubsection{The assumption of piecewise-constant  density with respect to y}
We now make the even more radical (but less radical than in the DD-SOV method) assumption that the mass density (but not the wavespeed) is piecewise-constant with respect to $y$. This translates to the relation
\begin{equation}\label{di-90}
\rho(y')=\rho^{[0]}+\left(\rho^{[1]}-\rho^{[0}\right)[H(y'+h)-H(y')]~,
\end{equation}
in which $\rho^{[0]}$ and $\rho^{[1]}$ are constants with respect to the spatial coordinates and $H(~)$ is the Heaviside distribution. It follows that
\begin{equation}\label{di-100}
\rho_{,y'}(y')=\left(\rho^{[1]}-\rho^{[0}\right)[\delta(y'+h)-\delta(y')]~,
\end{equation}
so that (\ref{di-80}) becomes
\begin{multline}\label{di-110}
p(y)=p^{i}(y)+\frac{i}{2k_{y}^{i[0]}}\int_{-\infty}^{\infty}dy'\Big(\left[\left(k(y')\right)^{2}-\left(k^{[0]}\right)^{2}\right] p(y')-\\
\frac{\left(\rho^{[1]}-\rho^{[0}\right)[\delta(y'+h)-\delta(y')]}{\rho(y')} p_{,y'}(y')\Big)\exp(ik_{y}^{i[0]}|y-y'|)=p^{i}(y)+I(y)-K(y)~;~\forall y\in\mathbb{R}~,
\end{multline}
wherein
\begin{equation}\label{di-120}
I(y)=\frac{i}{2k_{y}^{i[0]}}\int_{-\infty}^{\infty}dy'\left[\left(k(y')\right)^{2}-\left(k^{[0]}\right)^{2}\right] p(y')\exp(ik_{y}^{i[0]}|y-y'|)~;~\forall y\in\mathbb{R}~,
\end{equation}
\begin{equation}\label{di-130}
K(y)=\frac{i}{2k_{y}^{i[0]}}\left(\rho^{[1]}-\rho^{[0}\right)\int_{-\infty}^{\infty}dy')[\delta(y'+h)-\delta(y')] \frac{p_{,y'}(y')}{\rho(y')})\exp(ik_{y}^{i[0]}|y-y'|)~;~\forall y\in\mathbb{R}~,
\end{equation}
Using the fact that since we assumed at the outset that $c(\mathbf{x})=c^{[0]}$ in $\Omega_{0}$ and $\Omega_{2}$ it follows that
\begin{equation}\label{di-135}
I(y)=\frac{i}{2k_{y}^{i[0]}}\int_{-h}^{0}dy'\left[\left(k(y')\right)^{2}-\left(k^{[0]}\right)^{2}\right] p(y')\exp(ik_{y}^{i[0]}|y-y'|)~;~\forall y\in\mathbb{R}~.
\end{equation}
Now, let us return to $K(y)$. Using the sifting property of the delta distributions we obtain
\begin{equation}\label{di-140}
K(y)=\frac{i}{2k_{y}^{i[0]}}\left(\rho^{[1]}-\rho^{[0}\right)\left[\frac{p_{,y'}(-h)}{\rho(-h)}\exp(ik_{y}^{i[0]}|y+h|)-
\frac{p_{,y'}(0)}{\rho(0)}\exp(ik_{y}^{i[0]}|y|)\right]~;~\forall y\in\mathbb{R}~,
\end{equation}
We cannot deal with $I$ and $K$ in analytic manner beyond this point unless further assumptions are made.
\subsubsection{The assumption of constant wavespeed within the layer}
We now make the further assumption (in agreement with what is assumed in the DD-SOV approach) that
\begin{equation}\label{di-150}
c(\mathbf{x})=c^{[1]}=\text{const.}~\Rightarrow k(\mathbf{x})=k^{[1]}=\text{const.}~,
\end{equation}
the principal consequence of which is that it seems legitimate to suppose (this is actually a consequence of SOV applied to the Helmholtz PDE in $\Omega_{1}$) that the pressure field within the layer can be represented by the sum of two plane waves:
\begin{equation}\label{di-160}
p(y)=a^{[1]}\exp(-ik_{y}^{i[1]}y)+b^{[1]}\exp(ik_{y}^{i[1]}y)~;~\forall y\in[-h,0]~,
\end{equation}
($a^{[1]}$ and $b^{[1]}$ are constants) from which it follows (in addition to the continuity properties of $\frac{1}{\rho}p_{,y}$) that
\begin{equation}\label{di-170}
\begin{array}{cc}
\frac{p_{,y}(-h)}{\rho(-h)}= & \frac{ik_{y}^{i[1]}}{\rho^{[1]}}\left[-a^{[1]}\exp(ik_{y}^{i[1]}h)+b^{[1]}\exp(-ik_{y}^{i[1]}h)\right]\\
\frac{p_{,y}(0)}{\rho(0)}= & \frac{ik_{y}^{i[1]}}{\rho^{[1]}}\left[-a^{[1]}+b^{[1]}\right]
\end{array}
~.
\end{equation}
This suggests that we focus our attention on the integral equation
\begin{equation}\label{di-180}
p(y)= I(y)-K(y)~;~\forall y\in ]-h,0[
~.
\end{equation}
It is easy to find via (\ref{di-160}):
\begin{multline}\label{di-190}
I(y)=\frac{i}{2k_{y}^{i[0]}}\left[\left(k^{[1]}\right)^{2}-\left(k^{[0]}\right)^{2}\right]\times\\
\left(
\int_{-h}^{y}dy'p(y')\exp[ik_{y}^{i[0]}(y-y')]+
\int_{y}^{0}dy'p(y')\exp[-ik_{y}^{i[0]}(y-y')]
\right)~;~\forall y\in]-h,0[~,
\end{multline}
or, on account of (\ref{di-160}) and the fact that $\left(k_{y}^{i[1]}\right)^{2}-\left(k_{y}^{i[0]}\right)^{2}=\left(k^{[1]}\right)^{2}-\left(k^{[0]}\right)^{2}$:
\begin{multline}\label{di-200}
I(y)=a^{[1]}\exp[-ik_{y}^{i[1]}y]+b^{[1]}\exp[ik_{y}^{i[1]}y]+\\
\left[-a^{[1]}\left(\frac{k_{y}^{i[0]}-k_{y}^{i[1]}}{2k_{y}^{i[0]}}\right)\exp[i(k_{y}^{i[1]}+k_{y}^{i[0]})h]-
b^{[1]}\left(\frac{k_{y}^{i[0]}+k_{y}^{i[1]}}{2k_{y}^{i[0]}}\right)\exp[-i(k_{y}^{i[1]}-k_{y}^{i[0]})h]\right]\exp[ik_{y}^{i[0]}y]+\\
\left[-a^{[1]}\left(\frac{k_{y}^{i[0]}+k_{y}^{i[1]}}{2k_{y}^{i[0]}}\right)-
b^{[1]}\left(\frac{k_{y}^{i[0]}-k_{y}^{i[0]}}{2k_{y}^{i[0]}}\right)\right]\exp[-ik_{y}^{i[0]}y]~;~\forall y\in]-h,0[~.
\end{multline}
It is also readily found via (\ref{di-170}) that:
\begin{multline}\label{di-210}
K(y)=
\frac{k_{y}^{i[1]}}{2k_{y}^{i[0]}}\left(\frac{\rho^{i[1]}-\rho^{i[0]}}{\rho^{[1]}}\right)\left[a^{[1]}
\exp[i(k_{y}^{i[1]}+k_{y}^{i[0]})h]-
b^{[1]}\exp[-i(k_{y}^{i[1]}-k_{y}^{i0})h]\right]\exp[ik_{y}^{i0}y]-\\
\frac{k_{y}^{i[1]}}{2k_{y}^{i[0]}}\left(\frac{\rho^{i[1]}-\rho^{i[0]}}{\rho^{[1]}}\right)\left[a^{[1]}
-b^{[1]}\right]\exp[-ik_{y}^{i[0]}y]~;~\forall y\in]-h,0[~.
\end{multline}
It follows that (\ref{di-180}) takes the form:
\begin{multline}\label{di-220}
a^{[1]}\exp[-ik_{y}^{i[1]}y]+b^{[1]}\exp[ik_{y}^{i[1]}y]=a^{[0]}\exp[-ik_{y}^{i[0]}y]+
a^{[1]}\exp[-ik_{y}^{i[1]}y]+b^{[1]}\exp[ik_{y}^{i[1]}y]+\\
\left[-a^{[1]}\left(\frac{k_{y}^{i[0]}-k_{y}^{i[1]}}{2k_{y}^{i[0]}}\right)\exp[i(k_{y}^{i[1]}+k_{y}^{i[1]})h]-
b^{[1]}\left(\frac{k_{y}^{i[0]}+k_{y}^{i[1]}}{2k_{y}^{i[0]}}\right)\exp[-i(k_{y}^{i[1]}-k_{y}^{i[0]})h]\right]\exp[ik_{y}^{i[0]}y]+\\
\left[-a^{[1]}\left(\frac{k_{y}^{i[0]}+k_{y}^{i[1]}}{2k_{y}^{i[0]}}\right)-
b^{[1]}\left(\frac{k_{y}^{i[0]}-k_{y}^{i[1]}}{2k_{y}^{i[0]}}\right)\right]\exp[-ik_{y}^{i[0]}y]-\\
\frac{k_{y}^{i[1]}}{2k_{y}^{i[0]}}\left(\frac{\rho^{i[1]}-\rho^{i[0]}}{\rho^{[1]}}\right)\left[a^{[1]}
\exp[i(k_{y}^{i[1]}+k_{y}^{i0})h]-
b^{[1]}\exp[-i(k_{y}^{i[1]}-k_{y}^{i0})h]\right]\exp[ik_{y}^{i[0]}y]-\\
\frac{k_{y}^{i[1]}}{2k_{y}^{i[0]}}\left(\frac{\rho^{i[1]}-\rho^{i[0]}}{\rho^{[1]}}\right)\left[a^{[1]}
-b^{[1]}\right]\exp[-ik_{y}^{i[0]}y]~;~\forall y\in]-h,0[~,
\end{multline}
from which, after cancelations of like terms, and the fact that the term multiplying $\exp\big[-ik_{y}^{i[0]}\big]$ must be equal to $a^{[0]}$ and the terms multiplying $\exp\big[ik_{y}^{i[0]}\big]$ must vanish if the relation (\ref{di-220}) is to hold for every $y$ in $]-h,0[$, that we are left with the two results:
\begin{equation}\label{di-230}
a^{[0]}=\left[a^{[1]}\left(\frac{k_{y}^{i[0]}+k_{y}^{i[1]}}{2k_{y}^{i[0]}}\right)+
b^{[1]}\left(\frac{k_{y}^{i[0]}-k_{y}^{i[1]}}{2k_{y}^{i[0]}}\right)\right]-
\left[a^{[1]}-b^{[1]}\right]\frac{k_{y}^{i[1]}}{2k_{y}^{i[0]}}
\left(\frac{\rho^{i[1]}-\rho^{i[0]}}{\rho^{[1]}}\right)
~,
\end{equation}
\begin{multline}\label{di-240}
0=\left[a^{[1]}\left(\frac{k_{y}^{i[0]}-k_{y}^{i[1]}}{2k_{y}^{i[0]}}\right)\exp[i(k_{y}^{i[1]}+k_{y}^{i0})h]+
b^{[1]}\left(\frac{k_{y}^{i[0]}+k_{y}^{i[1]}}{2k_{y}^{i[0]}}\right)\exp[-i(k_{y}^{i[1]}-k_{y}^{i0})h]\right]+\\
\frac{k_{y}^{i[1]}}{2k_{y}^{i[0]}}\left(\frac{\rho^{i[1]}-\rho^{i[0]}}{\rho^{[1]}}\right)
\left[a^{[1]}\exp[i(k_{y}^{i[1]}+k_{y}^{i0})h]-b^{[1]}\exp[-i(k_{y}^{i[1]}-k_{y}^{i0})h]\right]
~.
\end{multline}
These two equations can be put in the form:
\begin{equation}\label{di-250}
a^{[0]}2\frac{k_{y}^{i[0]}}{\rho^{[0]}}=
a^{[1]}\left[\left(\frac{k_{y}^{i[0]}}{\rho^{[0]}}+\frac{k_{y}^{i[1]}}{\rho^{[0]}}\right)-
\frac{k_{y}^{i[1]}}{\rho^{[0]}}\epsilon\right]+
b^{[1]}\left[\left(\frac{k_{y}^{i[0]}}{\rho^{[0]}}-\frac{k_{y}^{i[1]}}{\rho^{[0]}}\right)+
\frac{k_{y}^{i[1]}}{\rho^{[0]}}\epsilon\right]
~,
\end{equation}
\begin{equation}\label{di-260}
0=a^{[1]}e^{ik_{y}^{i[1]}h}\left[\left(\frac{k_{y}^{i[0]}}{\rho^{[0]}}-\frac{k_{y}^{i[1]}}{\rho^{[0]}}\right)+
\frac{k_{y}^{i[1]}}{\rho^{[0]}}\epsilon\right]+
b^{[1]}e^{-ik_{y}^{i[1]}h}\left[\left(\frac{k_{y}^{i[0]}}{\rho^{[0]}}+\frac{k_{y}^{i[1]}}{\rho^{[0]}}\right)-
\frac{k_{y}^{i[1]}}{\rho^{[0]}}\epsilon\right]
~,
\end{equation}
wherein
\begin{equation}\label{di-270}
\epsilon=\frac{\rho^{[1]}-\rho^{[0]}}{\rho^{[1]}}
~,
\end{equation}
is the {\it mass density contrast}.  The solution of the two linear equations (\ref{di-250})-(\ref{di-260}) in the two unknowns $a^{[1]}$, $b^{[1]}$ is then readily found to be:
\begin{equation}\label{di-280}
a^{[1]}=a^{[0]}
\left(
\frac
{k_{y}^{i[0]}\left[k_{y}^{i[0]}+k_{y}^{i[1]}(1-\epsilon)\right]e^{-ik_{y}^{i[1]}h}}
{2k_{y}^{i[1]}k_{y}^{i[0]}(1-\epsilon)\cos(k_{y}^{i[1]}h)-
\left[\left(k_{y}^{i[0]}\right)^{2}+\left(k_{y}^{i[1]}\right)^{2}(1-\epsilon)^2\right]i\sin(k_{y}^{i[1]}h)}
\right)
~,
\end{equation}
\begin{equation}\label{di-290}
b^{[1]}=a^{[0]}
\left(
\frac
{k_{y}^{i[0]}\left[k_{y}^{i[1]}(1-\epsilon)-k_{y}^{i[0]}\right]e^{ik_{y}^{i[1]}h}}
{2k_{y}^{i[1]}k_{y}^{i[0]}(1-\epsilon)\cos(k_{y}^{i[1]}h)-
\left[\left(k_{y}^{i[0]}\right)^{2}+\left(k_{y}^{i[1]}\right)^{2}(1-\epsilon)^2\right]i\sin(k_{y}^{i[1]}h)}
\right)
~.
\end{equation}
Since $1-\epsilon=\rho^{[0]}/\rho^{[1]}$ it is easy to see that (\ref{di-280})-(\ref{di-290}) are identical to their DD-SOV counterparts (\ref{ddsov-120})-(\ref{ddsov-130}. Moreover, (\ref{di-280})-(\ref{di-290} provide the constant-density (i.e., $\epsilon=0$) solution of the Lippmann-Schwinger integral equation
\begin{equation}\label{di-300}
p(y)=p^{i}(y)+I(y)~;~\forall y\in ]-h,0[
~.
\end{equation}
We now turn to the expressions of the pressure fields in $\Omega_{0}$ and $\Omega_{2}$ which are necessarily an outcome of the solution obtained for the field within the layer, as is expressed by the Bergman (and also LS) integral representations of these fields:
\begin{equation}\label{di-310}
p(y)=p^{i}(y)+I(y)-K(y)~;~\forall y\in ]-\infty,-h[~,~\forall y \in ]0,\infty[
~.
\end{equation}
wherein $I(y)$ and $K(y)$ (=0 for $\epsilon=0$, as for the LS integral equation) are functions of the previously-found $p(y)~;~y \in ]-h,0[$.

Let us first consider the field in $\Omega_{0}$. Proceeding as previously, we find
\begin{multline}\label{di-320}
I(y)=\frac{1}{2}e^{ik_{y}^{i[0]}y}\times\\
\Big[
a^{[1]}\left(\frac{k_{y}^{i[0]}-k_{y}^{i[1]}}{k_{y}^{i[0]}}\right)+
b^{[1]}\left(\frac{k_{y}^{i[0]}+k_{y}^{i[1]}}{k_{y}^{i[0]}}\right)-
a^{[1]}\left(\frac{k_{y}^{i[0]}-k_{y}^{i[1]}}{k_{y}^{i[0]}}\right)e^{i(k_{y}^{i[0]}+k_{y}^{i[1]})h}\\-
b^{[1]}\left(\frac{k_{y}^{i[0]}+k_{y}^{i[1]}}{k_{y}^{i[0]}}\right)e^{i(k_{y}^{i[0]}-k_{y}^{i[1]})h}
\Big]~;~\forall  y \in ]0,\infty[
~,
\end{multline}
\begin{multline}\label{di-330}
K(y)=-\frac{\epsilon}{2}\frac{k_{y}^{i[1]}}{k_{y}^{i[0]}}e^{ik_{y}^{i[0]}y}
\Big[
-a^{[1]}e^{i(k_{y}^{i[0]}+k_{y}^{i[1]})h}+
b^{[1]}e^{i(k_{y}^{i[0]}-k_{y}^{i[1]})h}+a^{[1]}-b^{[1]}
\Big]~;~\forall  y \in ]0,\infty[
~,
\end{multline}
so that
\begin{multline}\label{di-340}
p(y)=p^{i}(y)-\frac{1}{2k_{y}^{i[0]}}e^{ik_{y}^{i[0]}y}\times\\
\Big(
a^{[1]}\left[k_{y}^{i[0]}-k_{y}^{i[1]}(1-\epsilon)\right]+
b^{[1]}\left[k_{y}^{i[0]}+k_{y}^{i[1]}(1-\epsilon)\right]+
a^{[1]}\left[-k_{y}^{i[0]}+k_{y}^{i[1]}(1-\epsilon)\right]e^{i(k_{y}^{i[0]}+k_{y}^{i[1]})h}+\\
b^{[1]}\left[-k_{y}^{i[0]}-k_{y}^{i[1]}(1-\epsilon)\right]e^{i(k_{y}^{i[0]}-k_{y}^{i[1]})h}
\Big)~;~\forall  y \in ]0,\infty[
~,
\end{multline}
which, upon the introduction of (\ref{di-280})-(\ref{di-290}), yields
\begin{equation}\label{di-345}
p(y)=b^{[0]}\exp(ik_{y}^{i[0]}y)~;~\forall  y \in ]0,\infty[
~,
\end{equation}
wherein
\begin{equation}\label{di-350}
b^{[0]}=a^{[0]}
\left(
\frac
{\left[\left(k_{y}^{i[1]}\right)^{2}(1-\epsilon)^{2}-\left(k_{y}^{i[0]}\right)^{2}\right]i\sin(k_{y}^{i[1]}h)}
{2k_{y}^{i[1]}k_{y}^{i[0]}(1-\epsilon)\cos(k_{y}^{i[1]}h)-
\left[\left(k_{y}^{i[0]}\right)^{2}+\left(k_{y}^{i[1]}\right)^{2}(1-\epsilon)^2\right]i\sin(k_{y}^{i[1]}h)}
\right)
~.
\end{equation}
Since $1-\epsilon=\rho^{[0]}/\rho^{[1]}$ it is easy to see that (\ref{di-350}) is identical to its DD-SOV counterpart (\ref{ddsov-110}).

Let us next consider the field in $\Omega_{2}$. Proceeding as previously, we find
\begin{multline}\label{di-360}
I(y)=-\frac{1}{2}e^{-ik_{y}^{i[0]}y}\times\\
\left[
a^{[1]}\left(\frac{k_{y}^{i[0]}+k_{y}^{i[1]}}{k_{y}^{i[0]}}\right)\left[1-e^{-i(k_{y}^{i[0]}-k_{y}^{i[1]})h}\right]+
b^{[1]}\left(\frac{k_{y}^{i[0]}-k_{y}^{i[1]}}{k_{y}^{i[0]}}\right)\left[1-e^{-i(k_{y}^{i[0]}+k_{y}^{i[1]})h}\right]
\right]\\
~;~\forall  y \in ]-\infty,-h[
~,
\end{multline}
\begin{multline}\label{di-370}
K(y)=-\frac{\epsilon}{2}\frac{k_{y}^{i[1]}}{k_{y}^{i[0]}}e^{-ik_{y}^{i[0]}y}
\Big[
-a^{[1]}e^{-i(k_{y}^{i[0]}-k_{y}^{i[1]})h}+
b^{[1]}e^{-i(k_{y}^{i[0]}+k_{y}^{i[1]})h}+a^{[1]}-b^{[1]}
\Big]~;~\forall  y \in ]-\infty,-h[
~,
\end{multline}
so that
\begin{multline}\label{di-380}
p(y)=p^{i}(y)-\frac{1}{2k_{y}^{i[0]}}e^{-ik_{y}^{i[0]}y}\times\\
\Big(
a^{[1]}\left[k_{y}^{i[0]}+k_{y}^{i[1]}(1-\epsilon)\right]+
b^{[1]}\left[k_{y}^{i[0]}-k_{y}^{i[1]}(1-\epsilon)\right]+
a^{[1]}\left[-k_{y}^{i[0]}-k_{y}^{i[1]}(1-\epsilon)\right]e^{-i(k_{y}^{i[0]}-k_{y}^{i[1]})h}+\\
b^{[1]}\left[-k_{y}^{i[0]}+k_{y}^{i[1]}(1-\epsilon)\right]e^{-i(k_{y}^{i[0]}+k_{y}^{i[1]})h}
\Big)~;~\forall  y \in ]-\infty,-h[
~,
\end{multline}
which, upon the introduction of (\ref{di-280})-(\ref{di-290}), yields
\begin{equation}\label{di-385}
p(y)=a^{[2]}\exp(-ik_{y}^{i[0]}y)~;~\forall  y \in ]-\infty,-h[
~,
\end{equation}
wherein
\begin{equation}\label{di-390}
a^{[2]}=a^{[0]}
\left(
\frac
{2k_{y}^{i[1]}k_{y}^{i[0]}(1-\epsilon)\exp(-ik_{y}^{i[0]}h)}
{2k_{y}^{i[1]}k_{y}^{i[0]}(1-\epsilon)\cos(k_{y}^{i[1]}h)-
\left[\left(k_{y}^{i[0]}\right)^{2}+\left(k_{y}^{i[1]}\right)^{2}(1-\epsilon)^2\right]i\sin(k_{y}^{i[1]}h)}
\right)
~.
\end{equation}
Since $1-\epsilon=\rho^{[0]}/\rho^{[1]}$ it is easy to see that (\ref{di-390}) is identical to its DD-SOV counterpart (\ref{ddsov-140}). Moreover, (\ref{di-345})-(\ref{di-350} on the one hand, and (\ref{di-385})-(\ref{di-390}) on the other hand, provide the constant-density (i.e., $\epsilon=0$) solution of the Lippmann-Schwinger integral equations
\begin{equation}\label{di-300}
p(y)=p^{i}(y)+I(y)~;~\forall y\in ]0,\infty[~~,~~p(y)=p^{i}(y)+I(y)~;~\forall y\in ]-\infty,-h[
~,
\end{equation}
respectively.

The important point to note here is that the DI expressions for $b^{[0]},a^{[1]},b^{[1]},a^{[2]}$ enable, contrary to their DD-SOV counterparts, to account, in very explicit manner, for the mass density contrast (expressed by $\epsilon$) of the layer obstacle configuration. We shall see further on that this has important consequences, especially in the inverse scattering context.
\subsubsection{The expansion of the transmission coefficient in a series of powers of $\epsilon$}
What will be offered here for the transmission coefficient $a^{[2]}$ is applicable, as well to the other coefficients $b^{[0]},a^{[1]},b^{[1]}$.

Eq. (\ref{di-390}) is of the form
\begin{equation}\label{spe-020}
a^{[2]}=\frac{a+b\epsilon}{c+d\epsilon+f\epsilon^2}
~,
\end{equation}
in which:
\begin{equation}\label{spe-030}
\begin{array}{cc}
a= & a^{[0]}2k_{y}^{i[1]}k_{y}^{i[0]}\exp(-ik_{y}^{i[0]}h)\\\\
b= & -a\\\\
c= & 2k_{y}^{i[1]}k_{y}^{i[0]}\cos(k_{y}^{i[1]}h)-\left[\left(k_{y}^{i[0]}\right)^{2}+\left(k_{y}^{i[1]}\right)^{2}\right]i\sin(k_{y}^{i[1]}h)\\\\
d= & -2k_{y}^{i[1]}k_{y}^{i[0]}\cos(k_{y}^{i[1]}h)+2\left(k_{y}^{i[1]}\right)^{2}i\sin(k_{y}^{i[1]}h)\\\\
f= & -\left(k_{y}^{i[1]}\right)^{2}i\sin(k_{y}^{i[1]}h)
\end{array}
~.
\end{equation}
so that
\begin{equation}\label{spe-040}
a^{[2]}=\left(
\frac{a}{c}+\frac{b}{c}\epsilon
\right)
\sum_{l=0}^{\infty}\left[
-\epsilon\left(
\frac{d}{c}+\frac{f}{c}\epsilon\right)
\right]^{l}
~.
\end{equation}
This relation suggests that the transmission coefficient can be expressed in iterative manner as follows:

\begin{equation}\label{spe-050}
\begin{array}{cc}
a^{[2](0)}= & \frac{a}{c}\\\\
a^{[2](1)}= & a^{[2](0)}+\left(\frac{b}{c}-\frac{ad}{c^{2}}\right)\epsilon\\\\
a^{[2](2)}= & a^{[2](1)}+\left(-\frac{af}{c^{2}}-\frac{bd}{c^{2}}+\frac{ad^{2}}{c^{3}}\right)\epsilon^{2}\\\\
. ~ . ~ . ~ .     & . ~ . ~ . ~ .
\end{array}
~.
\end{equation}
It is readily-verified that the zeroth-order approximation $a^{[2](0)}$ of $a^{[2]}$ is  the constant-density (i.e., the $\epsilon=0$) solution for this coefficient. Similarly, the first-order approximation $a^{[2](1)}$ of $a^{[2]}$ constitutes the first correction, presumably-valid for very small $\epsilon$, of the constant-density solution. This approximation is further corrected via the second-order approximation $a^{[2](2)}$, presumably-valid for small $\epsilon$, and so on.

The questions, to which numerical answers will be given in the next section, are (a) how reliable is the zeroth-order (constant-density) approximation (which does not depend on $\epsilon$)?, and (b) to what order of approximation must we go to get a decent approximation of $a^{[2]}$ for a given $\epsilon$ which is not necessarily small?
\subsection{Numerical results for the comparison of the exact $a^{[2]}$ with the zeroth, first and second-order (in terms of $\epsilon$) approximations thereof}
The aim of the computations was to compare numerically the exact  acoustic transmitted amplitude $a^{[2]}$ to: (a) the zeroth-order $\epsilon$ prediction $a^{[2](0)}$ (equal to the constant-density) prediction of $a^{[2]}$, (b)  the first-order $\epsilon$ prediction $a^{[2](1)}$ prediction of $a^{[2]}$, and (c)  the second-order $\epsilon$ prediction $a^{[2](2)}$  prediction of $a^{[2]}$, as a function of the various parameters $\epsilon$, $f$, $\theta^{i}$, $\Re\left(c^{[1]}\right)$, $\Im\left(c^{[1]}\right)$, the other parameters being fixed at the following values: $\rho^{[0]}=1000~Kgm^{-3}$, $c^{[0]}=1500~ms^{-1}$, $h=0.2~m$, $a^{[0]}=1$.
\subsubsection{Zeroth, first, and second-order $\epsilon$ approximations of $a^{[2]}$ compared to the exact expression thereof as a function of the density contrast $\epsilon$}
\begin{figure}[ht]
\begin{center}
\includegraphics[width=0.75\textwidth]{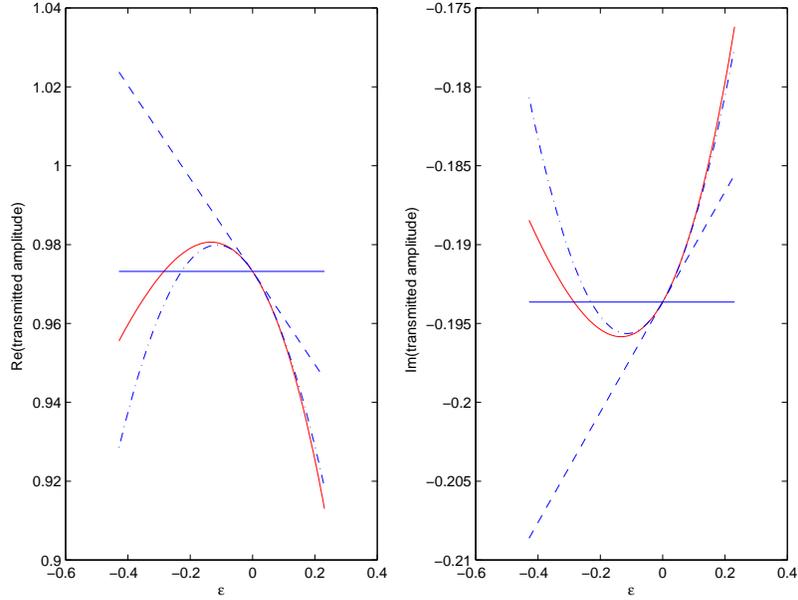}
\caption{Transmitted amplitude as a function of density contrast $\epsilon$.  The  left(right)-hand panels depict the real(imaginary) parts of  $a^{[2]}$ (red-----), $a^{[2](0)}$ (blue ------), $a^{[2](1)}$ (blue - - - -), $a^{[2](2)}$ (blue -.-.-.-).
Case $f=2000~Hz$, $\theta^{i}=0^{\circ}$, $c^{[1]}=1700-0i~ms^{-1}$.}
\label{fig01-01}
\end{center}
\end{figure}
\begin{figure}[ptb]
\begin{center}
\includegraphics[width=0.75\textwidth]{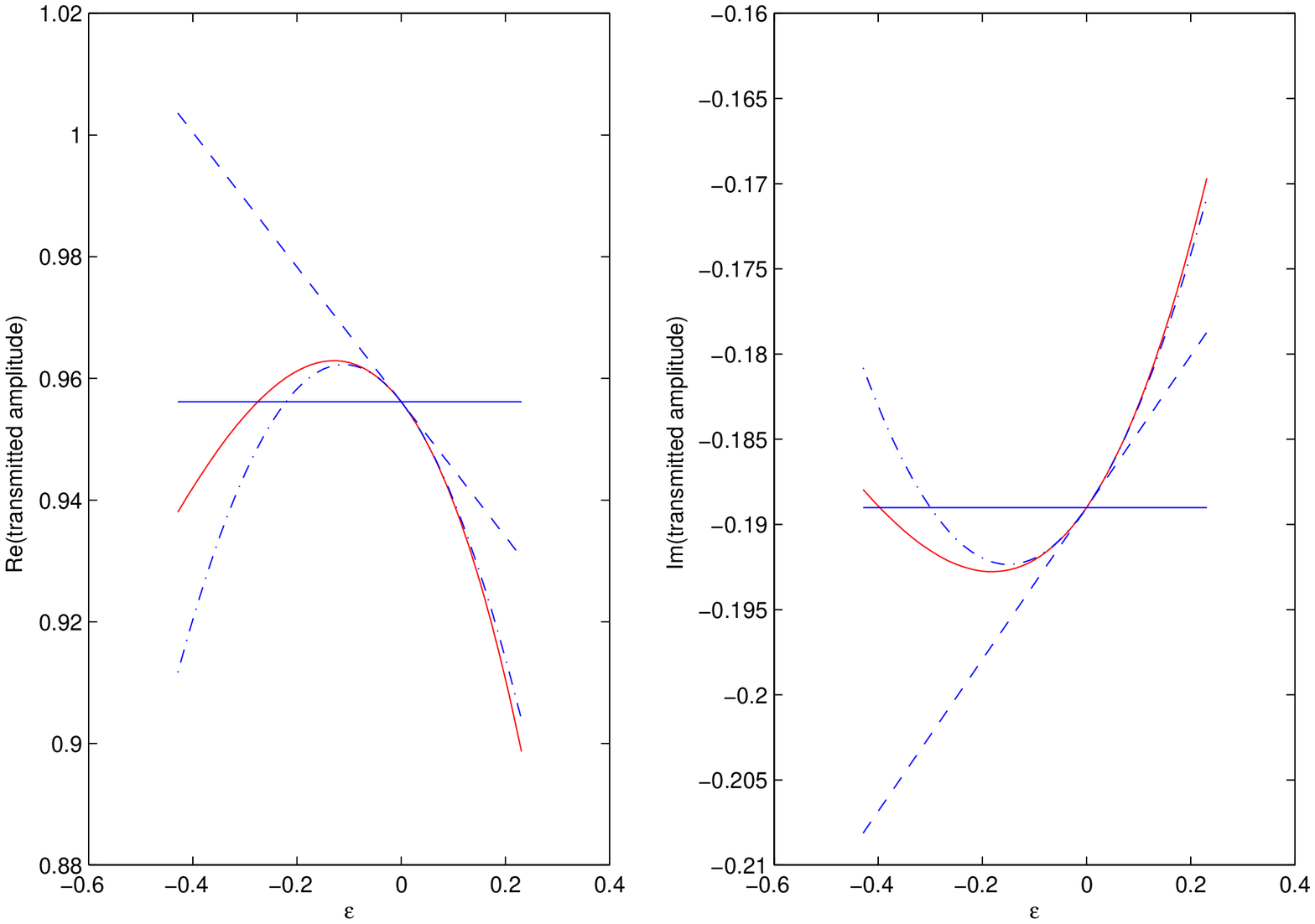}
\caption{Same as fig. \ref{fig01-01} except that $c^{[1]}=1700-21i~ms^{-1}$.}
\label{fig01-02}
\end{center}
\end{figure}
\begin{figure}[ptb]
\begin{center}
\includegraphics[width=0.75\textwidth]{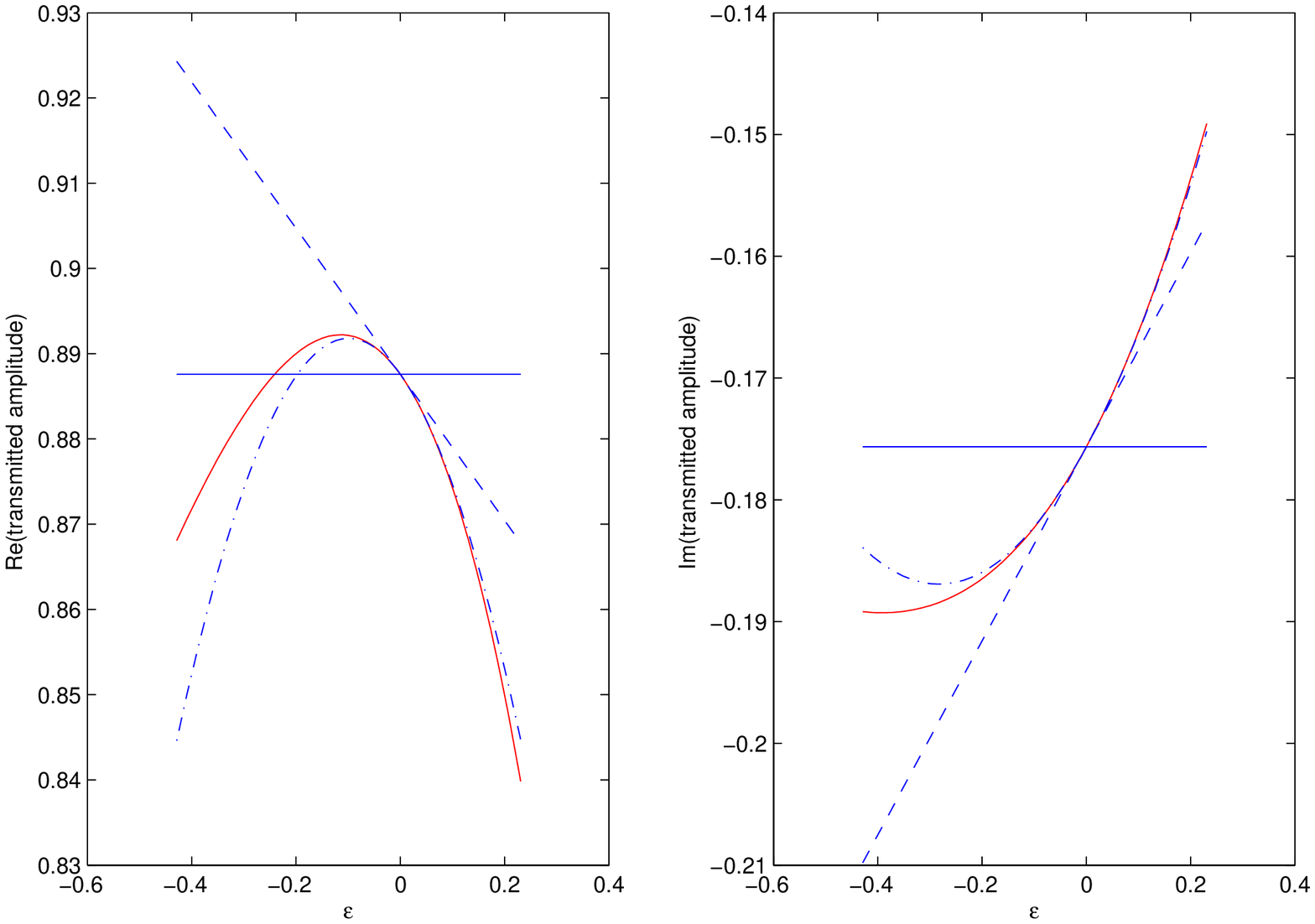}
\caption{Same as fig. \ref{fig01-01} except that $c^{[1]}=1700-110i~ms^{-1}$.}
\label{fig01-03}
\end{center}
\end{figure}
\begin{figure}[ptb]
\begin{center}
\includegraphics[width=0.75\textwidth]{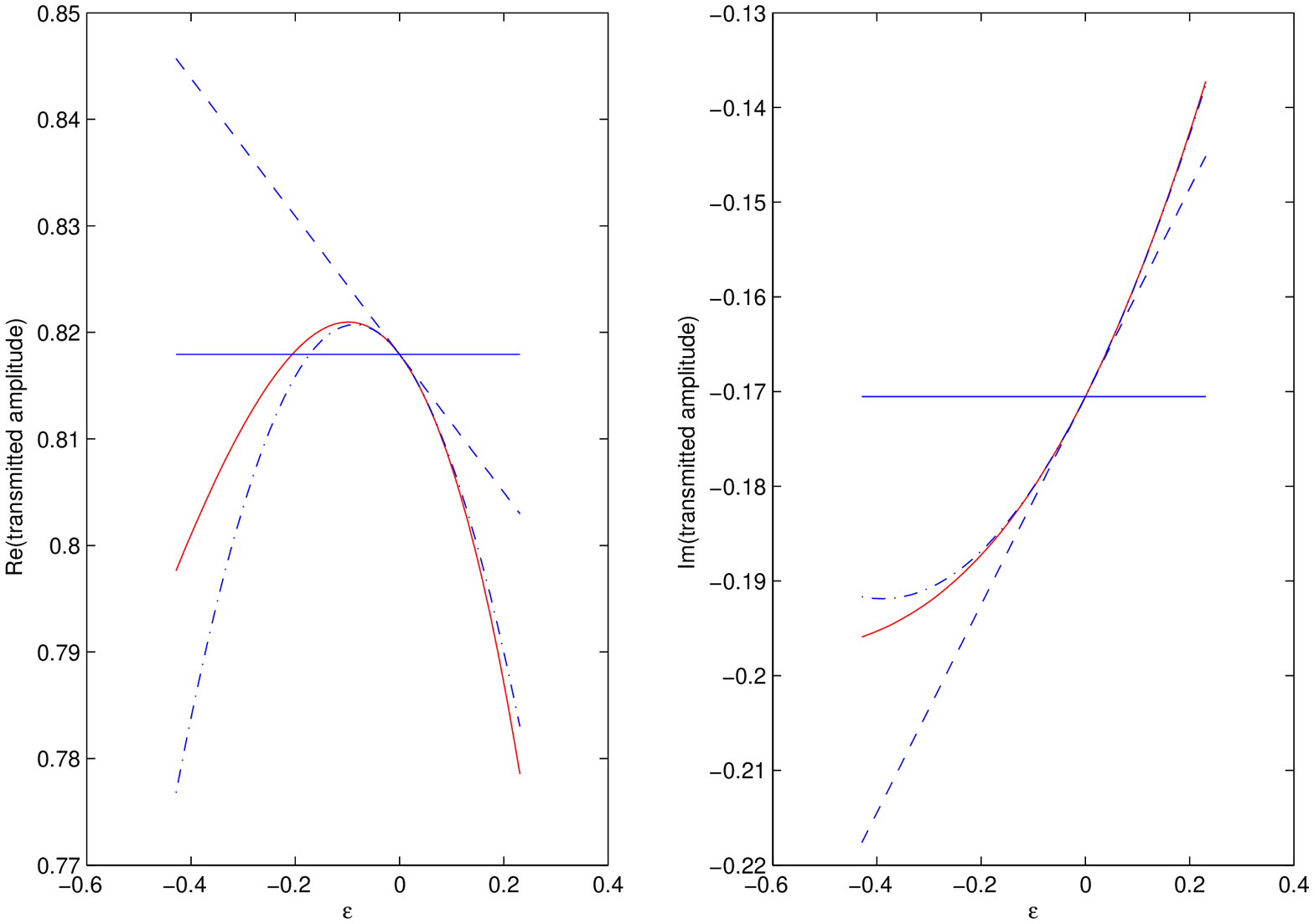}
\caption{Same as fig. \ref{fig01-01} except that $c^{[1]}=1700-210i~ms^{-1}$.}
\label{fig01-04}
\end{center}
\end{figure}
\begin{figure}[ptb]
\begin{center}
\includegraphics[width=0.75\textwidth]{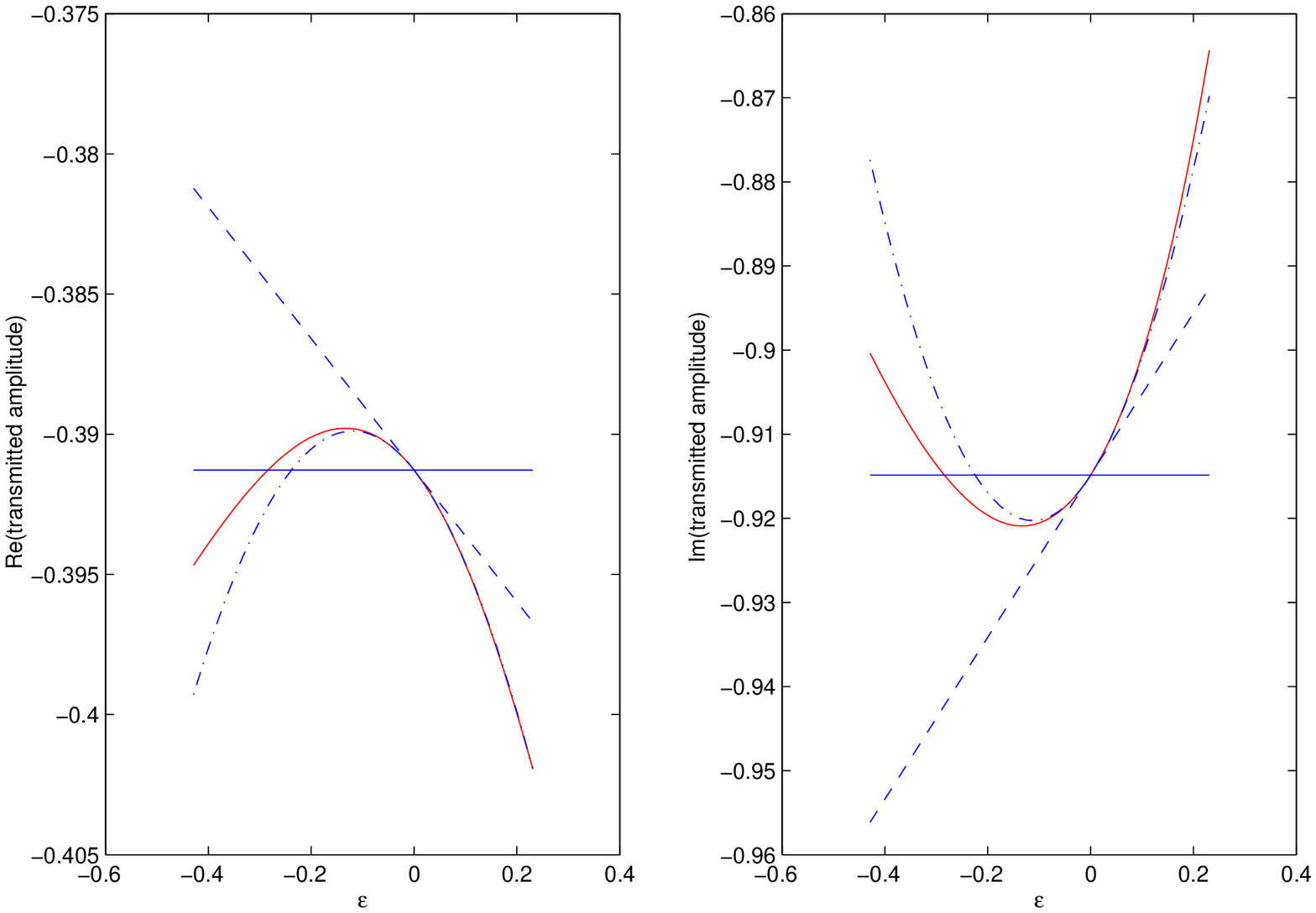}
\caption{Transmitted amplitude as a function of density contrast $\epsilon$.  The  left(right)-hand panels depict the real(imaginary) parts of  $a^{[2]}$ (red-----), $a^{[2](0)}$ (blue ------), $a^{[2](1)}$ (blue - - - -), $a^{[2](2)}$ (blue -.-.-.-).
Case $f=20000~Hz$, $\theta^{i}=0^{\circ}$, $c^{[1]}=1700-0i~ms^{-1}$.}
\label{fig02-01}
\end{center}
\end{figure}
\clearpage
\begin{figure}[ptb]
\begin{center}
\includegraphics[width=0.75\textwidth]{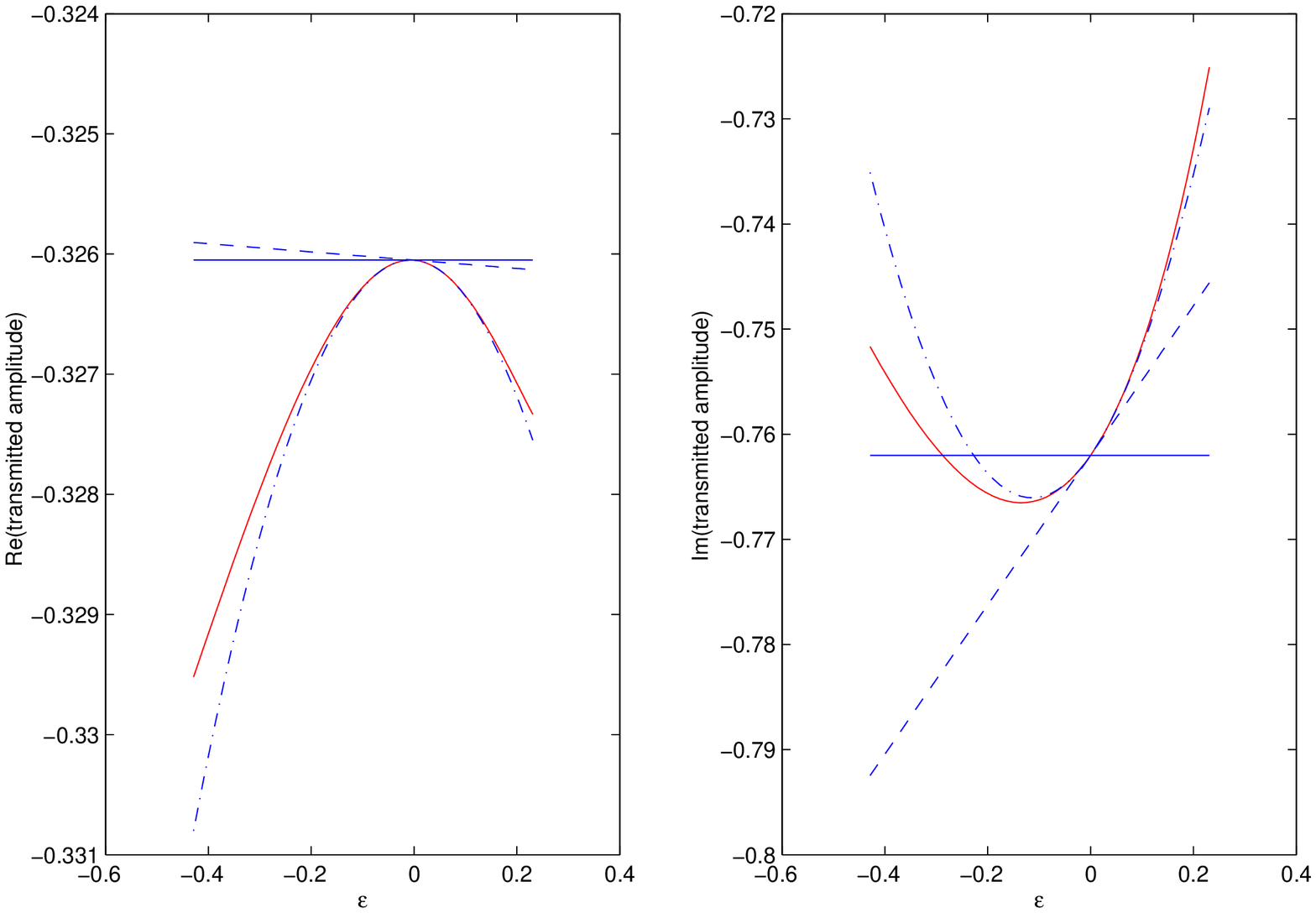}
\caption{Same as fig. \ref{fig02-01} except that $c^{[1]}=1700-21i~ms^{-1}$.}
\label{fig02-02}
\end{center}
\end{figure}
\begin{figure}[ptb]
\begin{center}
\includegraphics[width=0.75\textwidth]{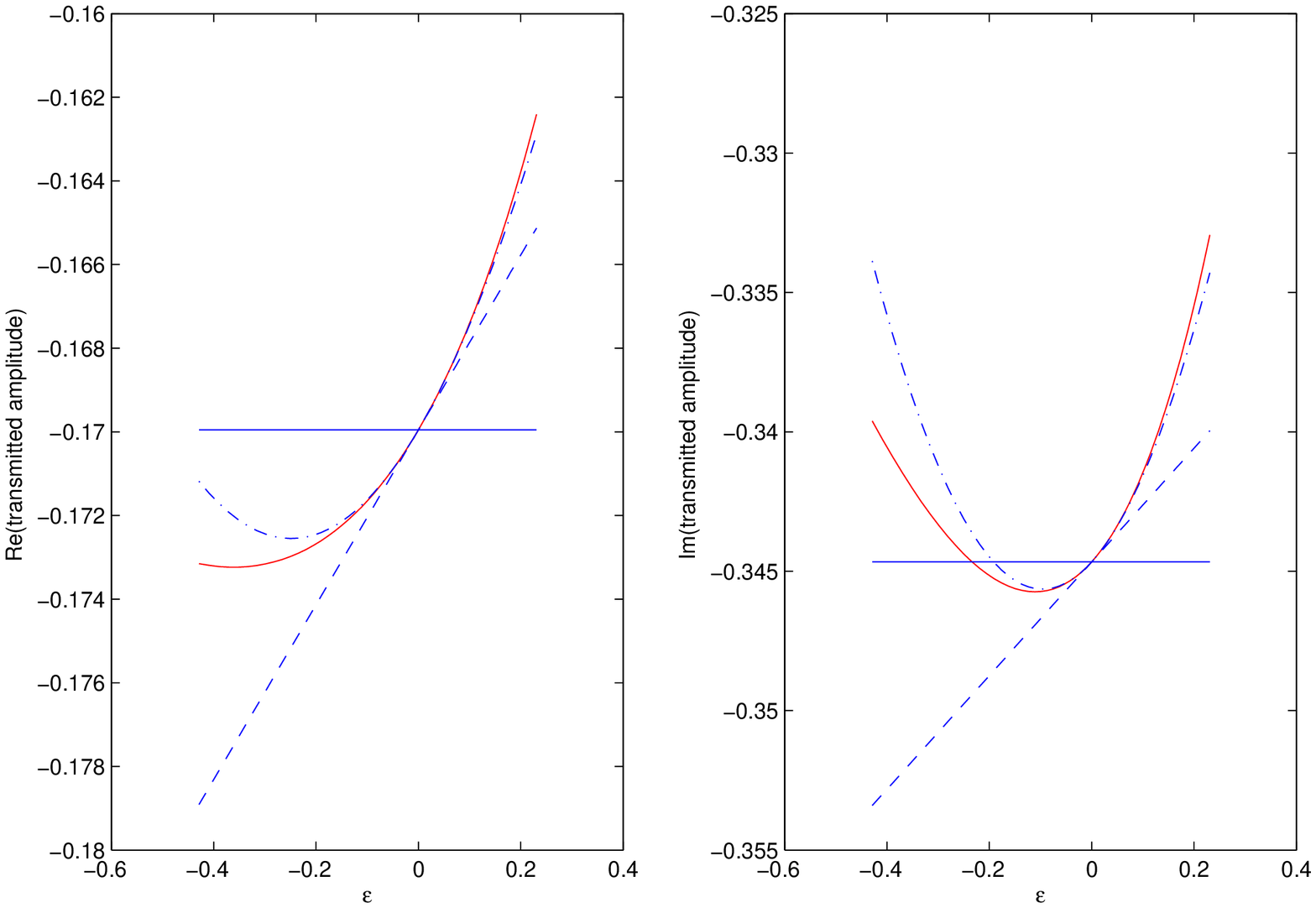}
\caption{Same as fig. \ref{fig02-01} except that $c^{[1]}=1700-110i~ms^{-1}$.}
\label{fig02-03}
\end{center}
\end{figure}
\begin{figure}[ptb]
\begin{center}
\includegraphics[width=0.75\textwidth]{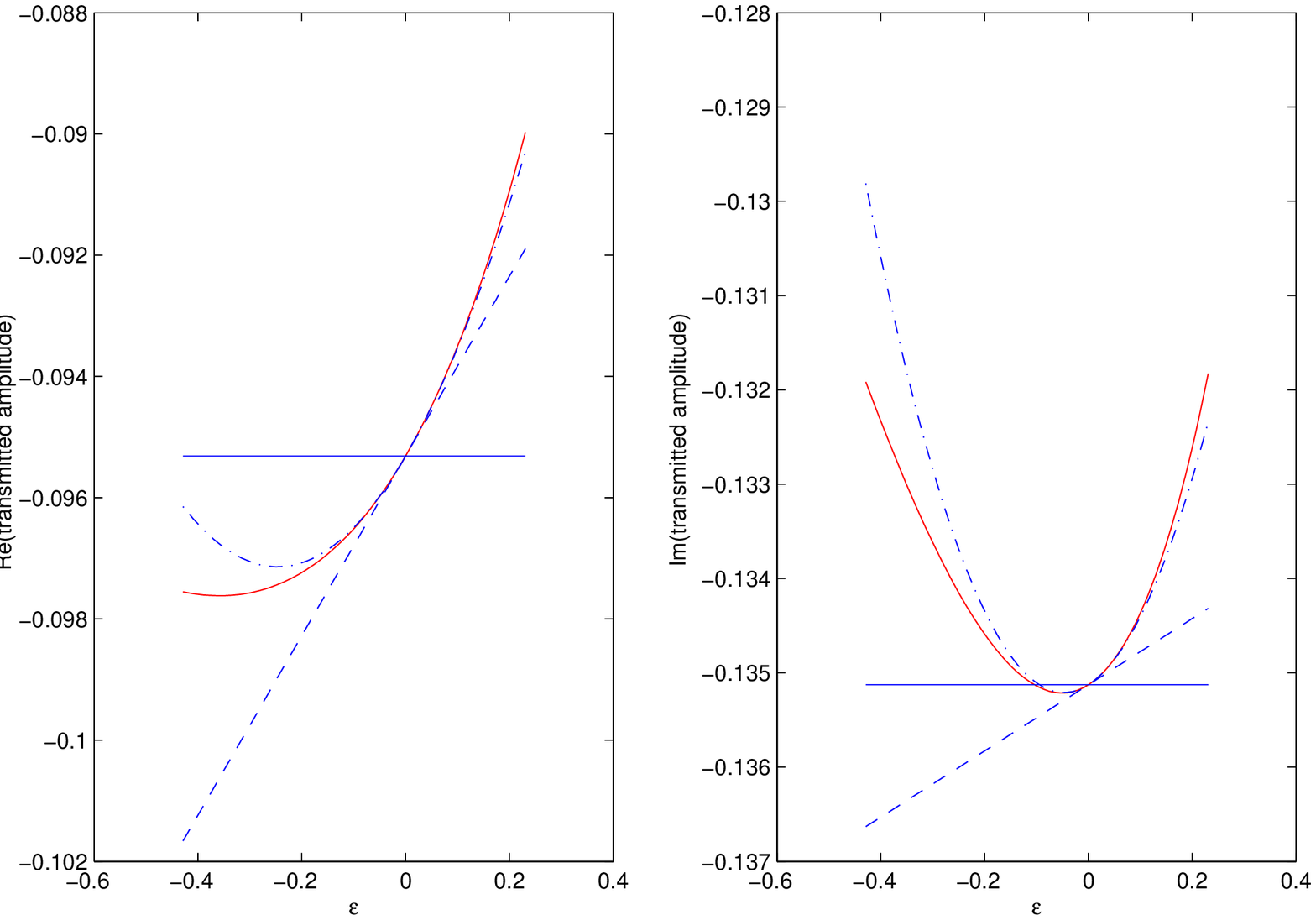}
\caption{Same as fig. \ref{fig02-01} except that $c^{[1]}=1700-210i~ms^{-1}$.}
\label{fig02-04}
\end{center}
\end{figure}
\begin{figure}[ptb]
\begin{center}
\includegraphics[width=0.75\textwidth]{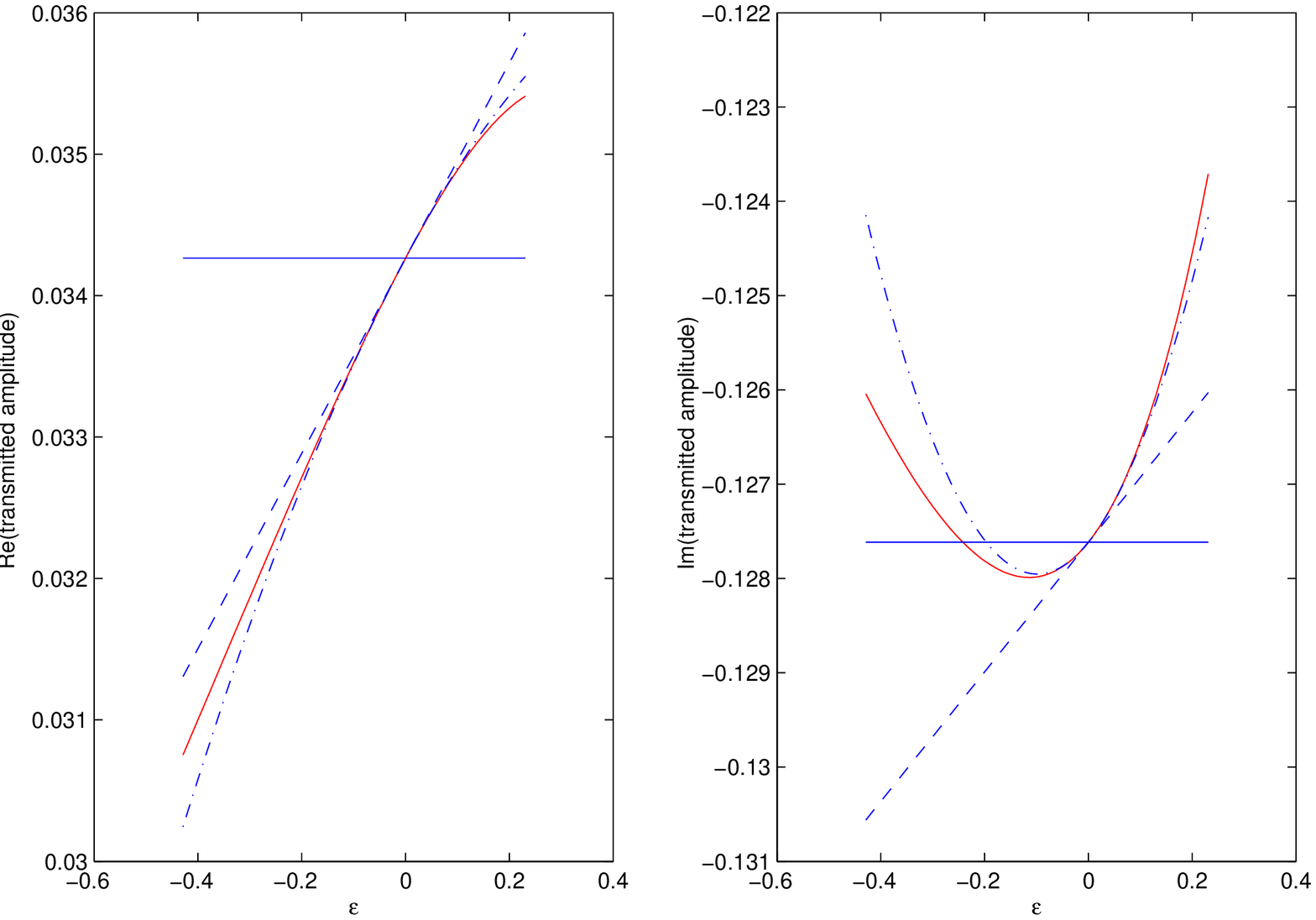}
\caption{Transmitted amplitude as a function of density contrast $\epsilon$.  The  left(right)-hand panels depict the real(imaginary) parts of  $a^{[2]}$ (red-----), $a^{[2](0)}$ (blue ------), $a^{[2](1)}$ (blue - - - -), $a^{[2](2)}$ (blue -.-.-.-).
Case $f=20000~Hz$, $\theta^{i}=0^{\circ}$, $c^{[1]}=1600-210i~ms^{-1}$.}
\label{fig03-01}
\end{center}
\end{figure}
\begin{figure}[ptb]
\begin{center}
\includegraphics[width=0.75\textwidth]{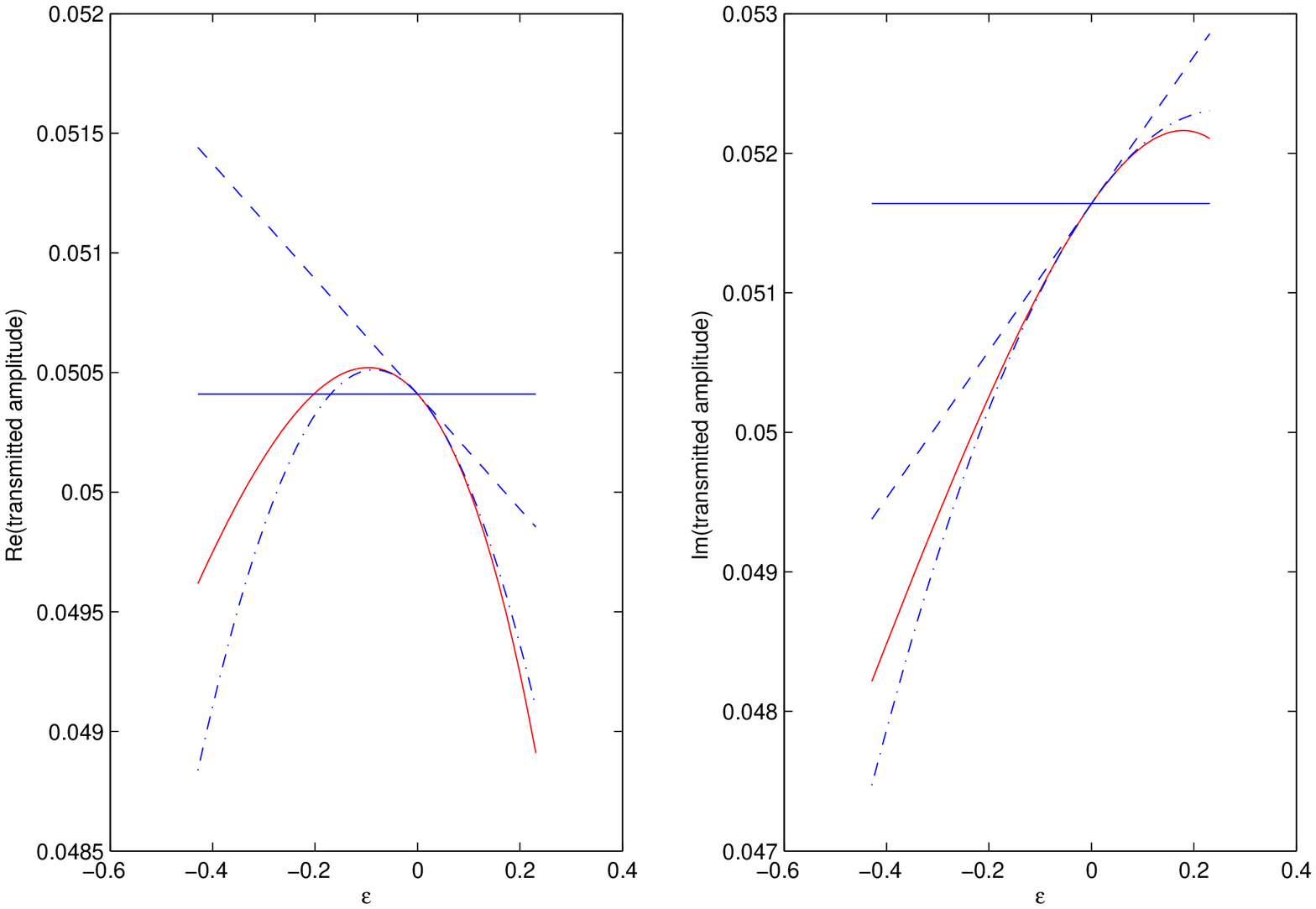}
\caption{Same as fig. \ref{fig03-01} except that $c^{[1]}=1400-210i~ms^{-1}$.}
\label{fig03-02}
\end{center}
\end{figure}
\begin{figure}[ptb]
\begin{center}
\includegraphics[width=0.75\textwidth]{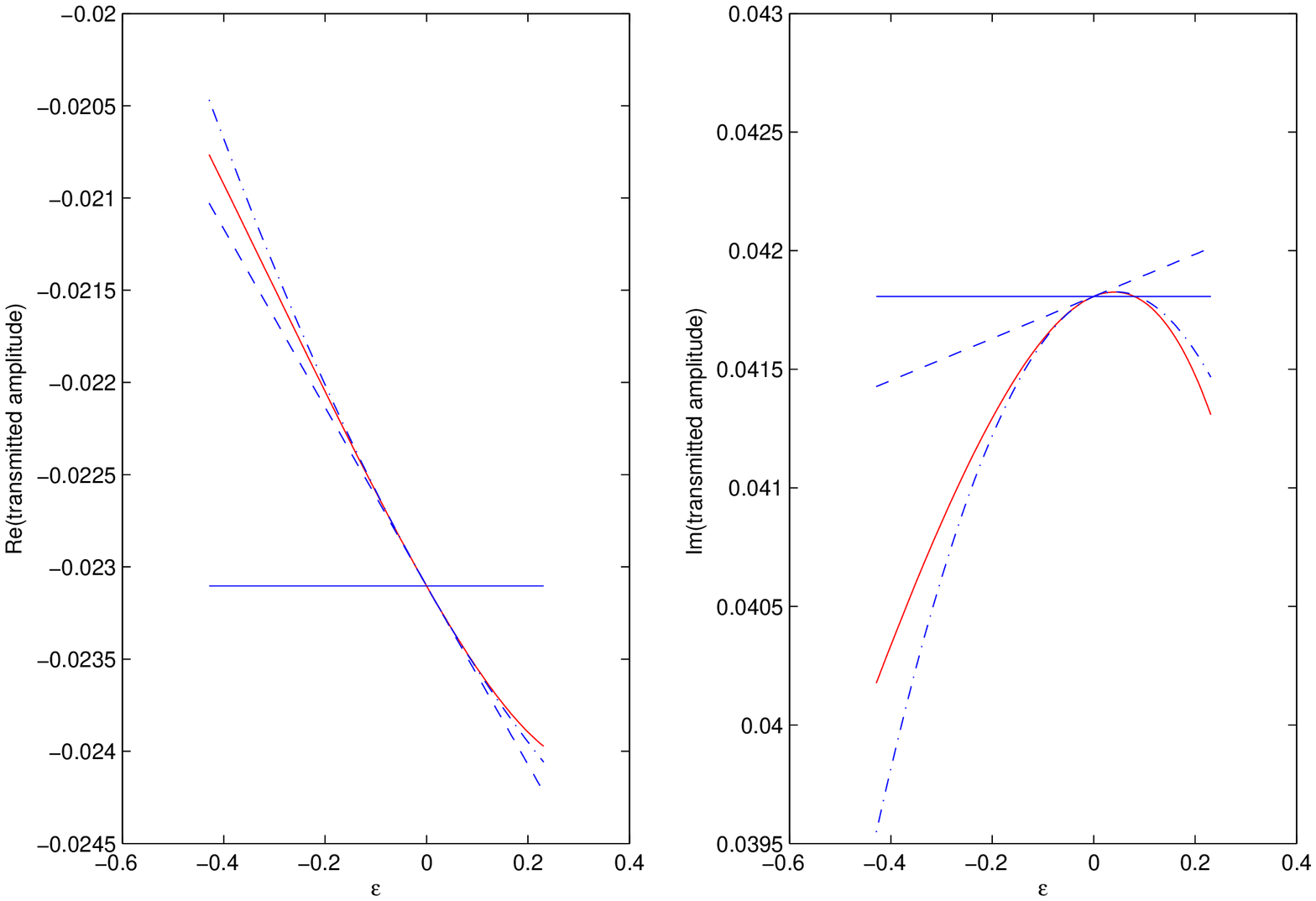}
\caption{Same as fig. \ref{fig03-01} except that $c^{[1]}=1300-210i~ms^{-1}$.}
\label{fig03-03}
\end{center}
\end{figure}
\clearpage
\begin{figure}[ptb]
\begin{center}
\includegraphics[width=0.75\textwidth]{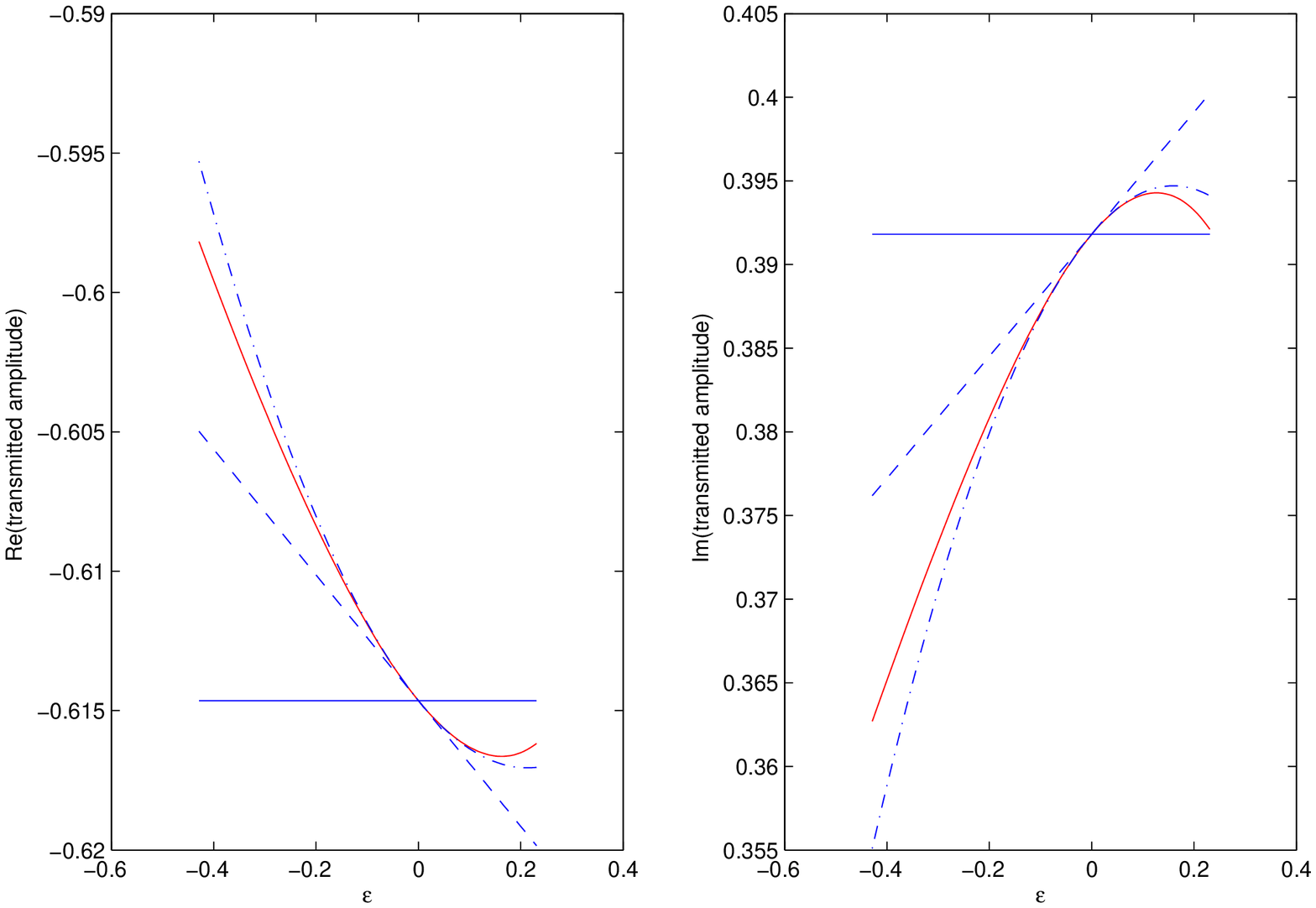}
\caption{Transmitted amplitude as a function of density contrast $\epsilon$.  The  left(right)-hand panels depict the real(imaginary) parts of  $a^{[2]}$ (red-----), $a^{[2](0)}$ (blue ------), $a^{[2](1)}$ (blue - - - -), $a^{[2](2)}$ (blue -.-.-.-).
Case $f=20000~Hz$, $\theta^{i}=0^{\circ}$, $c^{[1]}=1300-21i~ms^{-1}$..}
\label{fig04-01}
\end{center}
\end{figure}
\begin{figure}[ptb]
\begin{center}
\includegraphics[width=0.75\textwidth]{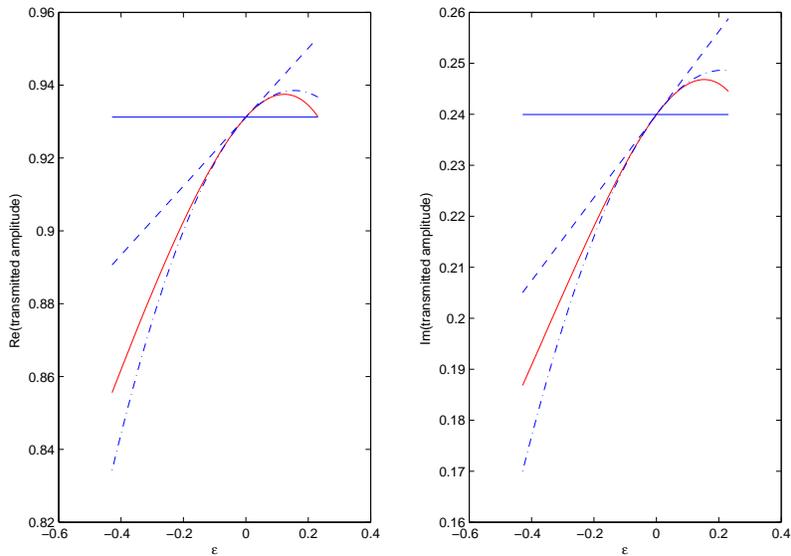}
\caption{Same as fig. \ref{fig04-01} except that $f=2000~Hz$.}
\label{fig04-02}
\end{center}
\end{figure}
\begin{figure}[ptb]
\begin{center}
\includegraphics[width=0.75\textwidth]{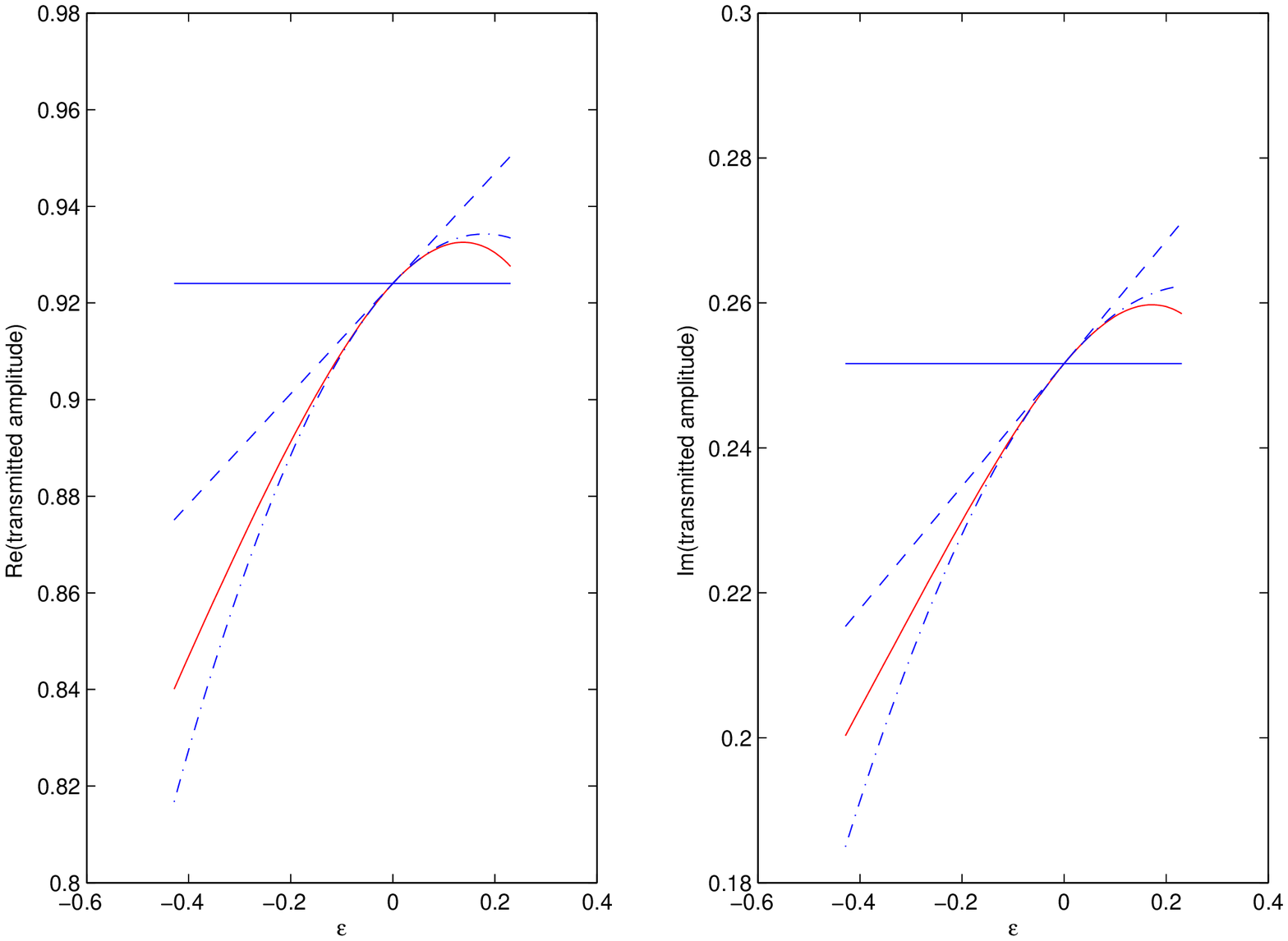}
\caption{Transmitted amplitude as a function of density contrast $\epsilon$.  The  left(right)-hand panels depict the real(imaginary) parts of  $a^{[2]}$ (red-----), $a^{[2](0)}$ (blue ------), $a^{[2](1)}$ (blue - - - -), $a^{[2](2)}$ (blue -.-.-.-).
Case $f=2000~Hz$, $\theta^{i}=20^{\circ}$, $c^{[1]}=1300-21i~ms^{-1}$.}
\label{fig05-01}
\end{center}
\end{figure}
\begin{figure}[ptb]
\begin{center}
\includegraphics[width=0.75\textwidth]{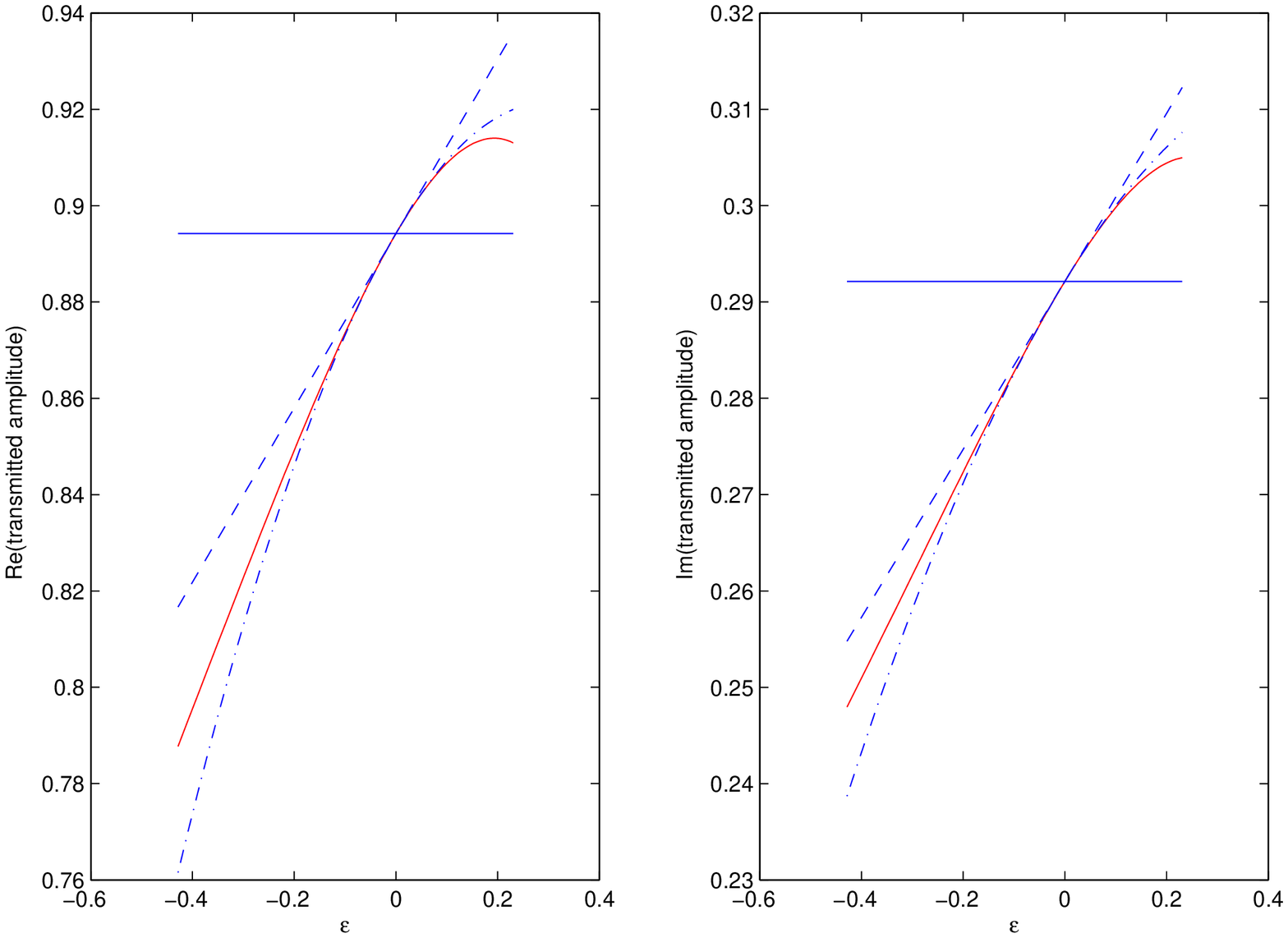}
\caption{Same as fig. \ref{fig05-01} except that $\theta^{i}=40^{\circ}$.}
\label{fig05-02}
\end{center}
\end{figure}
\begin{figure}[ptb]
\begin{center}
\includegraphics[width=0.75\textwidth]{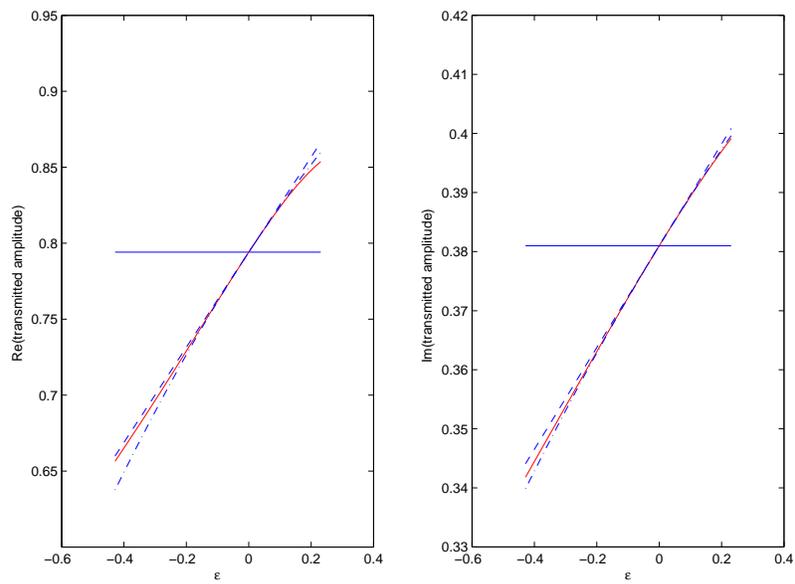}
\caption{Same as fig. \ref{fig05-01} except that $\theta^{i}=60^{\circ}$.}
\label{fig05-03}
\end{center}
\end{figure}
\clearpage
\newpage
What changes between the four figures figs. \ref{fig01-01}-\ref{fig01-04} (for the low frequency $f=2~KHz$) and the four figures \ref{fig02-01}-\ref{fig02-04} (for the higher frequency $f=20~KHz$) is the imaginary part of the layer wavespeed and it is observed in these figures that these changes have very little influence on the relative qualities of the three approximations of the transmitted amplitude. In fact, the constant mass density approximation of $a^{[2]}$ is manifestly both functionally and numerically incorrect, the linear approximation is numerically correct only within a very small interval around $\epsilon=0$, but the quadratic approximation is both functionally and numerically correct for $\epsilon$ in the interval $\epsilon\approx[-0.2,0.2]$ to which corresponds the rather large range $\rho^{[1]}\approx[1250~ms^{-1},1875~ms^{-1}]$ (when, as is here the case, $\rho^{[0]}=1500~ms^{-1})$ that includes  the obstacle wavespeeds encountered in most bioacoustic applications, except those regarding teeth and bones (which, anyway, are more solid-like than fluid-like).

What changes between the three figures figs. \ref{fig03-01}-\ref{fig03-03} (for the frequency $f=20~KHz$) is the real part of the layer wavespeed and it is seen in these figures that the said changes have very little influence on the relative qualities of the three approximations of the transmitted amplitude, so that the same comment as previously applies to these three figures, as to fig. \ref{fig04-01} and \ref{fig04-02} relative to a frequency that changes from $f=20~KHz$ to $f=2~KHz$.

Finally, what changes between the four figures figs. \ref{fig04-02}-\ref{fig03-03} is the incident angle, which too is seen to have little influence on the relative qualities of the three approximations of the transmitted amplitude, so that the same comment as previously regarding the acceptable range of $\epsilon$ applies to these four figures as well.
\subsubsection{Zeroth, first, and second-order $\epsilon$ approximations of $a^{[2]}$ compared to the exact expression thereof as a function of the incident angle $\theta^{i}$}
\begin{figure}[ht]
\begin{center}
\includegraphics[width=0.75\textwidth]{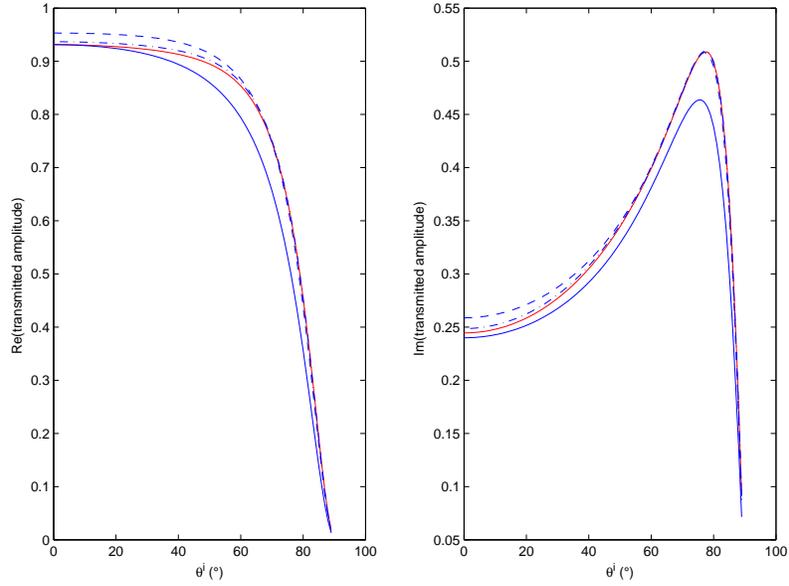}
\caption{Transmitted amplitude as a function of incident angle $\theta^{i}$.  The  left(right)-hand panels depict the real(imaginary) parts of  $a^{[2]}$ (red-----), $a^{[2](0)}$ (blue ------), $a^{[2](1)}$ (blue - - - -), $a^{[2](2)}$ (blue -.-.-.-).
Case $f=2000~Hz$, $\rho^{[1]}=1300~Kg/m^{3}$, $c^{[1]}=1300-21i~ms^{-1}$.}
\label{fig06-01}
\end{center}
\end{figure}
\begin{figure}[ptb]
\begin{center}
\includegraphics[width=0.75\textwidth]{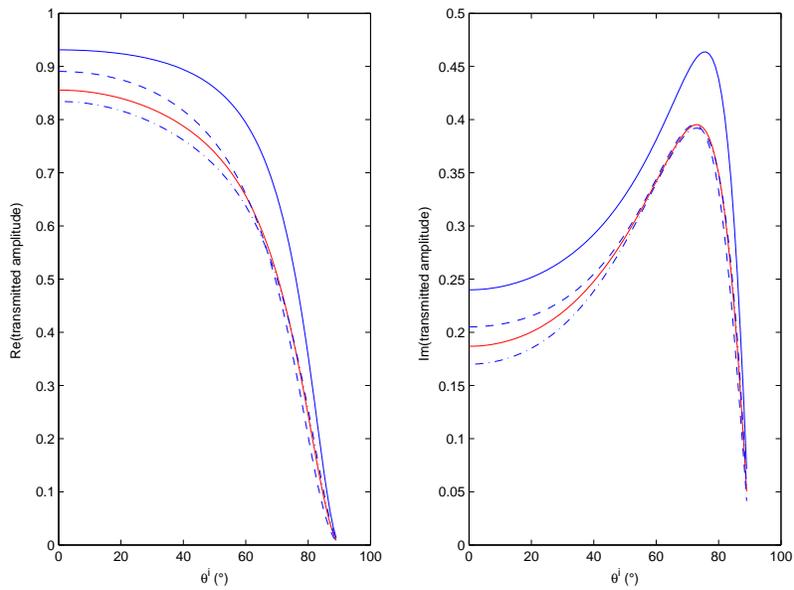}
\caption{Same as fig. \ref{fig06-01} except that $\rho^{[1]}=700~Kg/m^{3}$.}
\label{fig06-02}
\end{center}
\end{figure}
\clearpage
\newpage
What changes between the two figures figs. \ref{fig06-01}-\ref{fig06-02} (both of which concern the effect of the variation of the incident angle) is the layer mass density. It is observed in these figures that, reardless of the layer mass density,   the constant mass density approximation of $a^{[2]}$ is  functionally correct but numerically rather incorrect, whereas the linear approximation, and more so the quadratic approximation, are both functionally and numerically correct for most incident angles, this of course, being true for the chosen priors.
\subsubsection{Zeroth, first, and second-order $\epsilon$ approximations of $a^{[2]}$ compared to the exact expression thereof as a function of the frequency $f$}
\begin{figure}[ht]
\begin{center}
\includegraphics[width=0.75\textwidth]{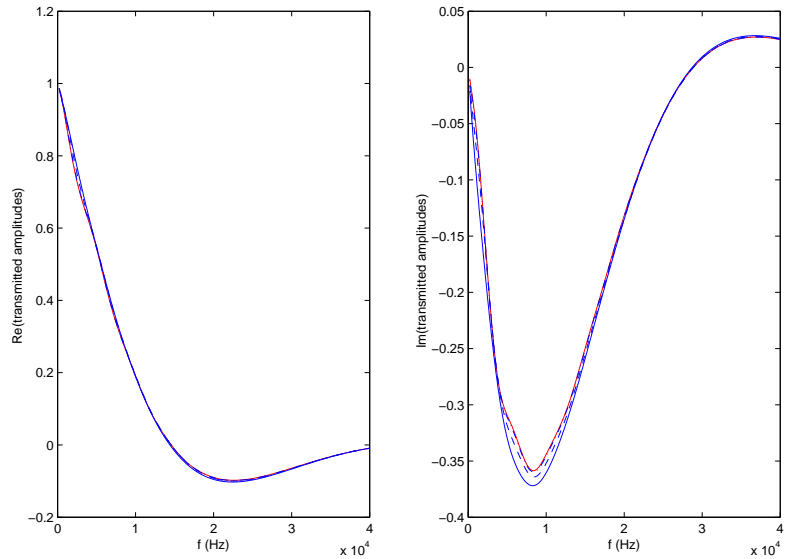}
\caption{Transmitted amplitude as a function of frequency $f$.  The  left(right)-hand panels depict the real(imaginary) parts of  $a^{[2]}$ (red-----), $a^{[2](0)}$ (blue ------), $a^{[2](1)}$ (blue - - - -), $a^{[2](2)}$ (blue -.-.-.-).
Case $\theta^{i}=0^{\circ}$, $\rho^{[1]}=1300~Kg/m^{3}$, $c^{[1]}=1700-210i~ms^{-1}$.}
\label{fig07-01}
\end{center}
\end{figure}
\begin{figure}[ptb]
\begin{center}
\includegraphics[width=0.75\textwidth]{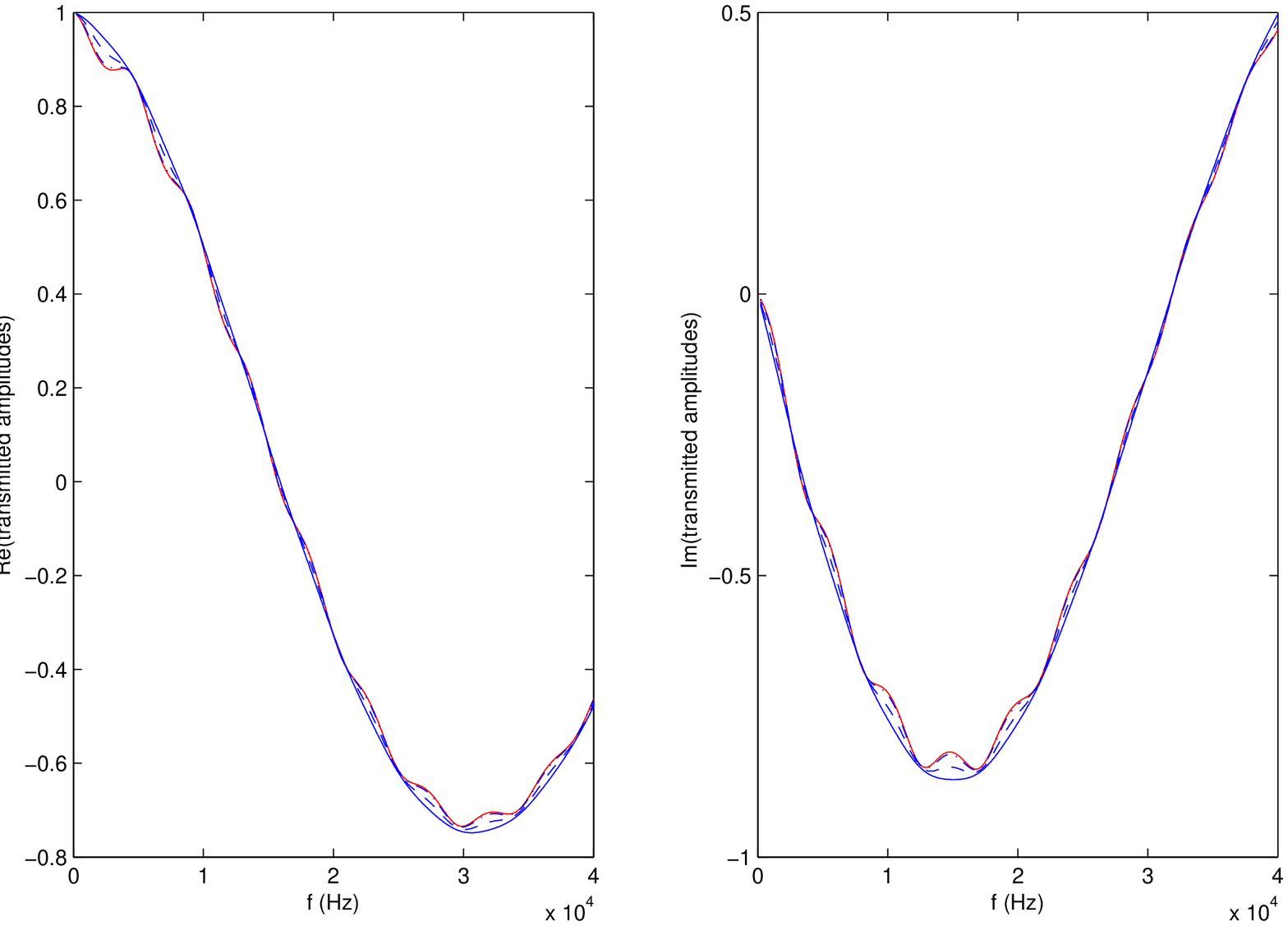}
\caption{Same as fig. \ref{fig07-01} except that $c^{[1]}=1700-21i~ms^{-1}$..}
\label{fig07-02}
\end{center}
\end{figure}
\begin{figure}[ptb]
\begin{center}
\includegraphics[width=0.75\textwidth]{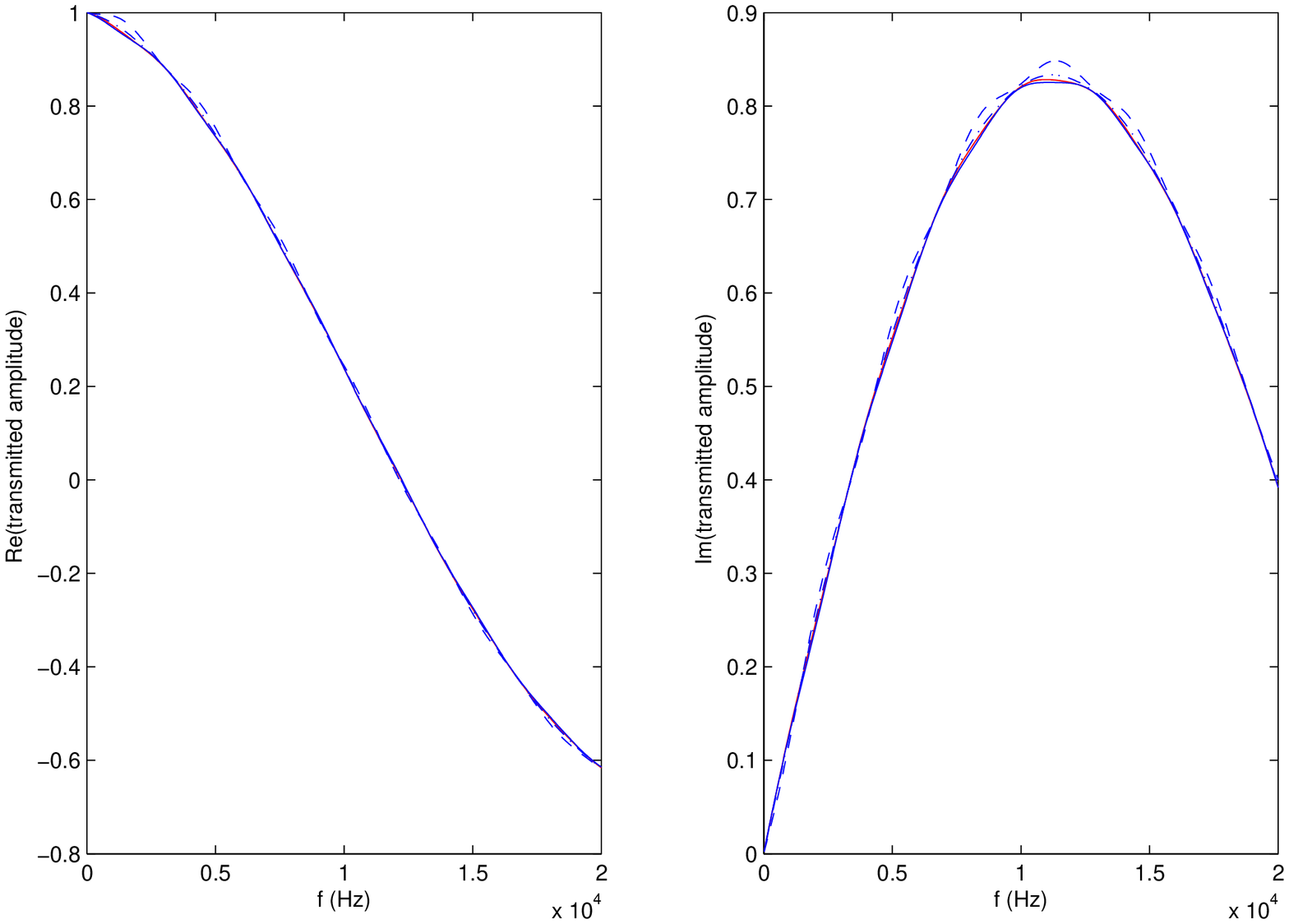}
\caption{Same as fig. \ref{fig07-01} except that $c^{[1]}=1300-21i~ms^{-1}$.}
\label{fig07-03}
\end{center}
\end{figure}
\begin{figure}[ptb]
\begin{center}
\includegraphics[width=0.75\textwidth]{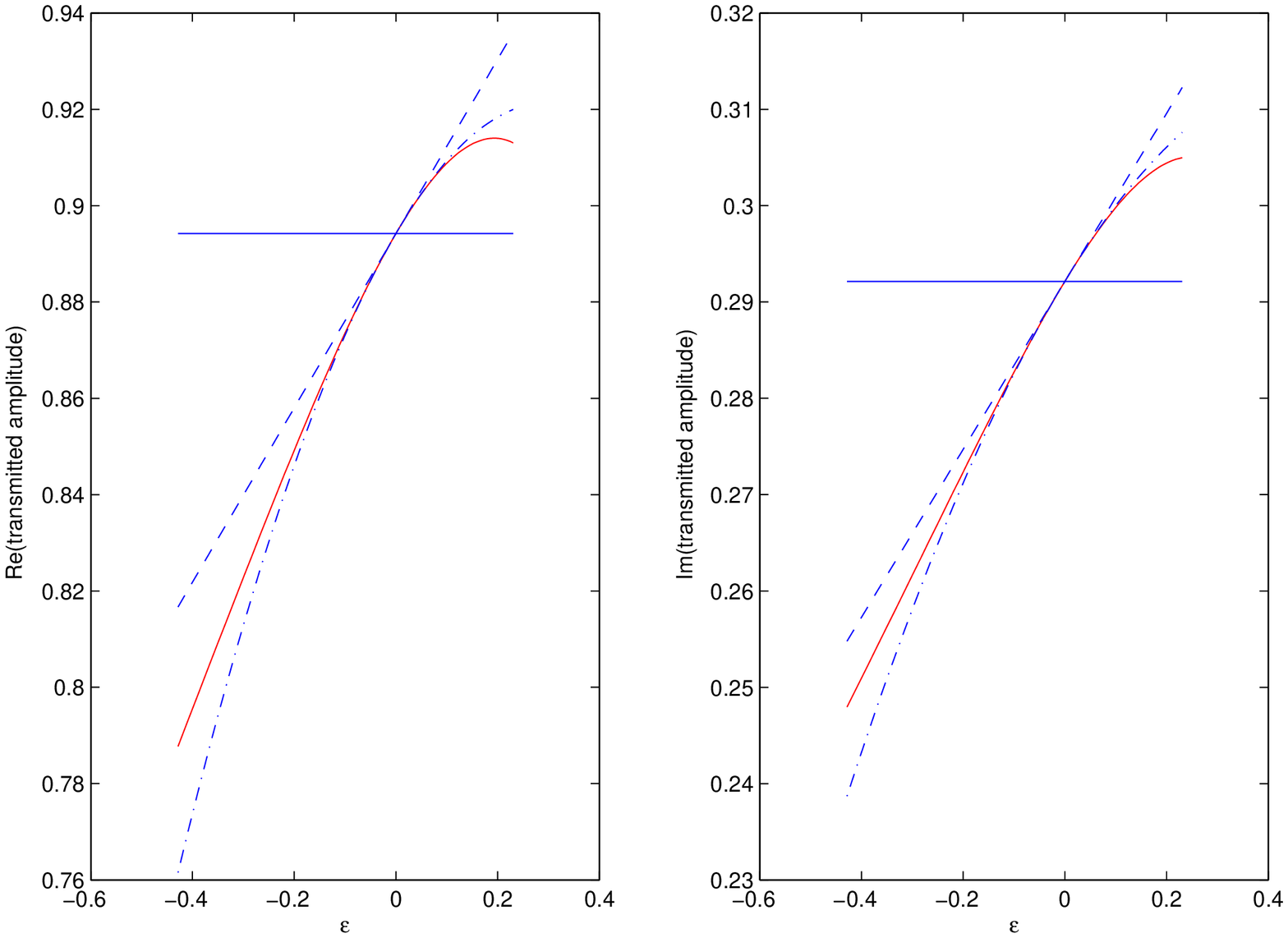}
\caption{Transmitted amplitude as a function of frequency $f$.  The  left(right)-hand panels depict the real(imaginary) parts of  $a^{[2]}$ (red-----), $a^{[2](0)}$ (blue ------), $a^{[2](1)}$ (blue - - - -), $a^{[2](2)}$ (blue -.-.-.-).
Case $\theta^{i}=0^{\circ}$, $\rho^{[1]}=700~Kg/m^{3}$, $c^{[1]}=1700-21i~ms^{-1}$.}
\label{fig08-01}
\end{center}
\end{figure}
\begin{figure}[ptb]
\begin{center}
\includegraphics[width=0.75\textwidth]{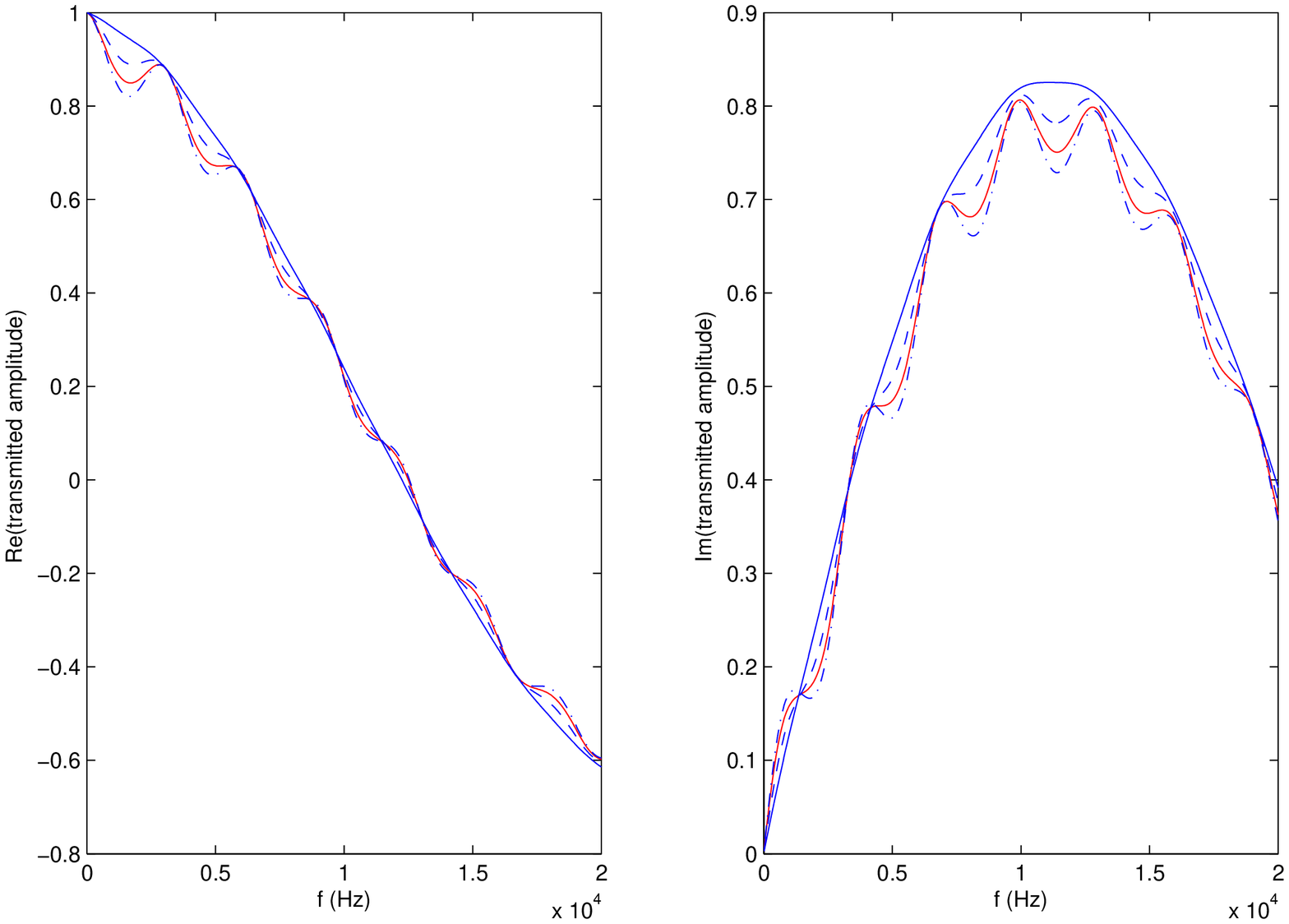}
\caption{Same as fig. \ref{fig08-02} except that $c^{[1]}=1300-21i~ms^{-1}$.}
\label{fig08-02}
\end{center}
\end{figure}
\begin{figure}[ptb]
\begin{center}
\includegraphics[width=0.75\textwidth]{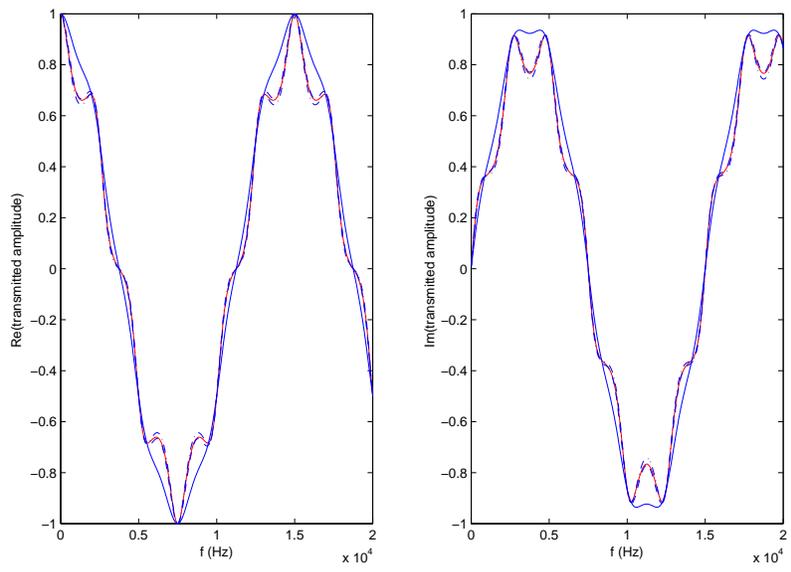}
\caption{Same as fig. \ref{fig08-03} except that $c^{[1]}=1000-0i~ms^{-1}$.}
\label{fig08-03}
\end{center}
\end{figure}
\clearpage
\begin{figure}[ptb]
\begin{center}
\includegraphics[width=0.75\textwidth]{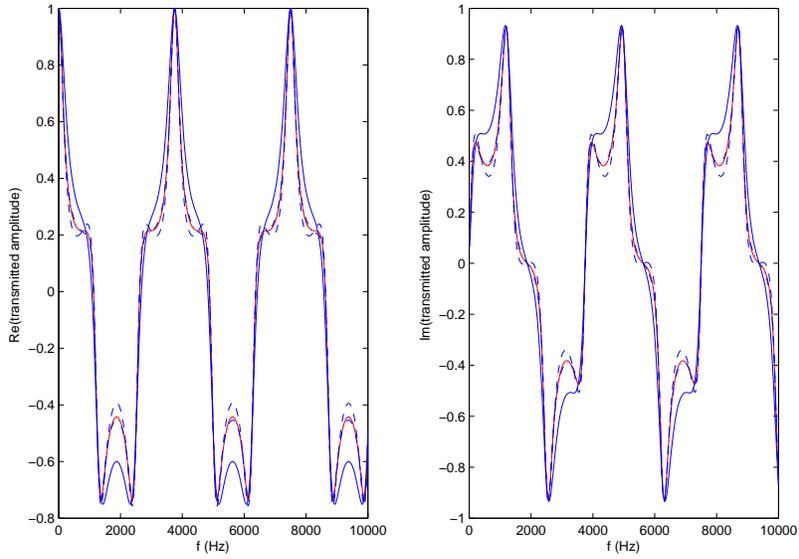}
\caption{Same as fig. \ref{fig08-04} except that $c^{[1]}=500-0i~ms^{-1}$.}
\label{fig08-04}
\end{center}
\end{figure}
\begin{figure}[ptb]
\begin{center}
\includegraphics[width=0.75\textwidth]{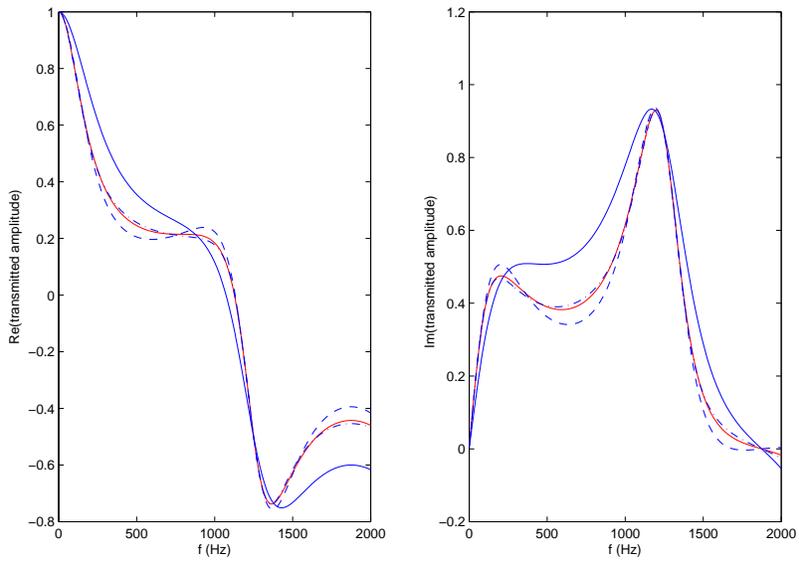}
\caption{Same as fig. \ref{fig08-05} except that range of $f$ is smaller.}
\label{fig08-05}
\end{center}
\end{figure}
\clearpage
\newpage
What changes between the two figures figs. \ref{fig07-01}-\ref{fig07-02} is the imaginary part of the layer wavespeed which is seen to obliterate the small-scale dispersive effects for the larger $|\Im c^{[1]}|$. The zeroth-order approximation of the transmission amplitude is seen to be functionally-correct for the large-scale dispersive effects only, the first-order approximation to be both functionally and numerically rather correct for both the large and small-scale dispersive effects, and the second-order approximation to be be even more correct for both of the dispersive effects.

What changes between the two figures figs. \ref{fig07-02}-\ref{fig07-03} (both for $\rho^{[1}=1300~Kgm^{-3}$) is the real part of the layer wavespeed which is seen to not affect the aforementioned relative qualities of the three approximations of the transmitted amplitude.

What changes between the two figures figs. \ref{fig08-01}-\ref{fig08-02} (both for $\rho^{[1}=700~Kgm^{-3}$) is again the real part of the layer wavespeed which is seen to  affect the aforementioned relative qualities of the three approximations of the transmitted amplitude to the extent that now even the second-order approximation has trouble of accurately-accounting for the small-scale dispersive effects, especially in the first of these two figures.

Figs. \ref{fig08-03}-\ref{fig08-04} (wherein the changes concern the real part of the layer wavespeed) apply to a lossless layer so that the small-scale dispersive effects are amplified. Now, the real part of the layer wavespeed is smaller than in the previous two figures, which fact improves the quality of the three approximations, especially the second-order one, even to the extent of accounting quite well for the small-scale dispersive effects. Fig. \ref{fig08-05} is a zoom of fig. \ref{fig08-04} whose purpose is to show that the small-scale dispersive effects  of the transmitted amplitude are of pseudo-resonant (not resonant because the response is not infinite for a lossless configuration) nature.
\subsubsection{Zeroth, first, and second-order $\epsilon$ approximations of $a^{[2]}$ compared to the exact expression thereof as a function of the real part of the wavespeed $\Re\left(c^{[1]}\right)$ in the layer}
\begin{figure}[ht]
\begin{center}
\includegraphics[width=0.75\textwidth]{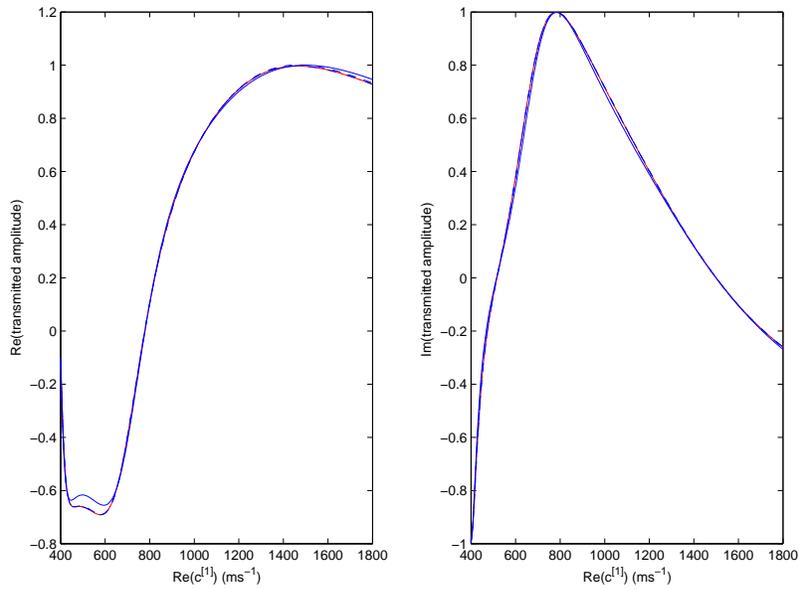}
\caption{Transmitted amplitude as a function of the real part of the wavespeed $\Re\left(c^{[1]}\right)$ in the layer .  The  left(right)-hand panels depict the real(imaginary) parts of  $a^{[2]}$ (red-----), $a^{[2](0)}$ (blue ------), $a^{[2](1)}$ (blue - - - -), $a^{[2](2)}$ (blue -.-.-.-).
Case $f=2000~Hz$, $\theta^{i}=0^{\circ}$, $\rho^{[1]}=1100~Kg/m^{3}$, $\Im\left(c^{[1]}\right)=0~ms^{-1}$.}
\label{fig09-01}
\end{center}
\end{figure}
\begin{figure}[ptb]
\begin{center}
\includegraphics[width=0.75\textwidth]{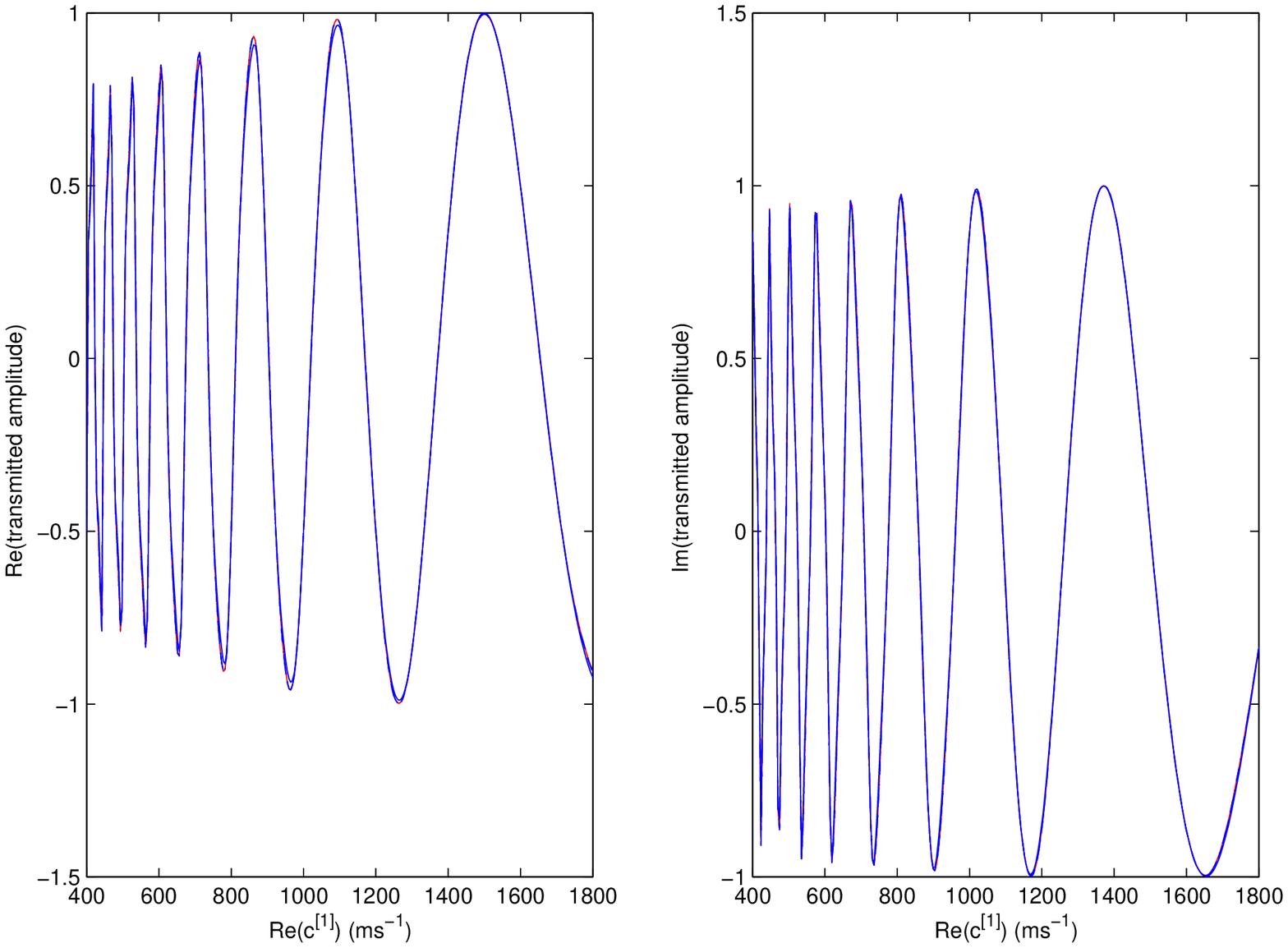}
\caption{Same as fig. \ref{fig09-01} except that $f=20000~Hz$.}
\label{fig09-02}
\end{center}
\end{figure}
\begin{figure}[ptb]
\begin{center}
\includegraphics[width=0.75\textwidth]{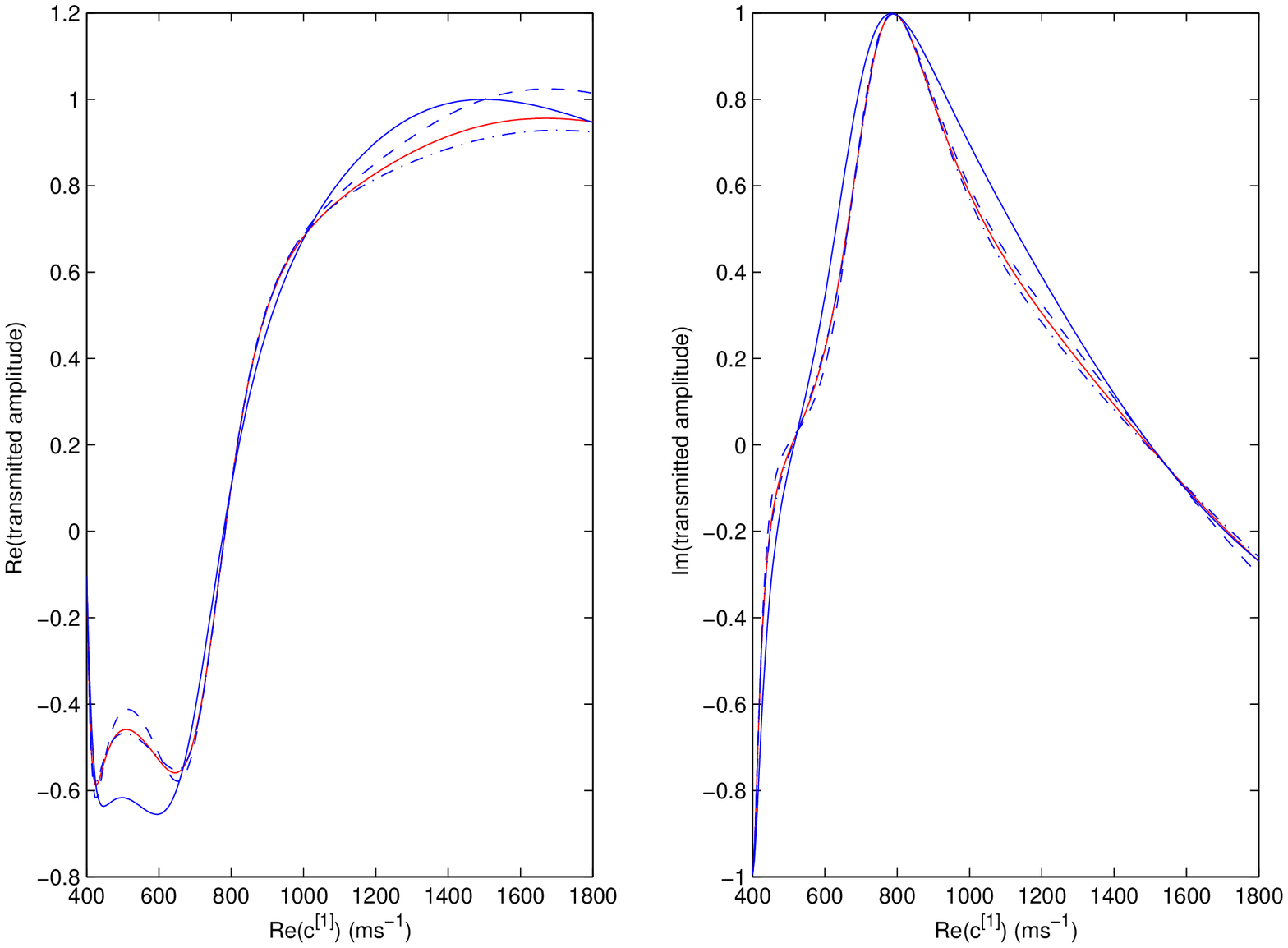}
\caption{Same as fig. \ref{fig09-01} except that $\rho^{[1]}=700~Kgm^{-3}$.}
\label{fig09-03}
\end{center}
\end{figure}
\begin{figure}[ptb]
\begin{center}
\includegraphics[width=0.75\textwidth]{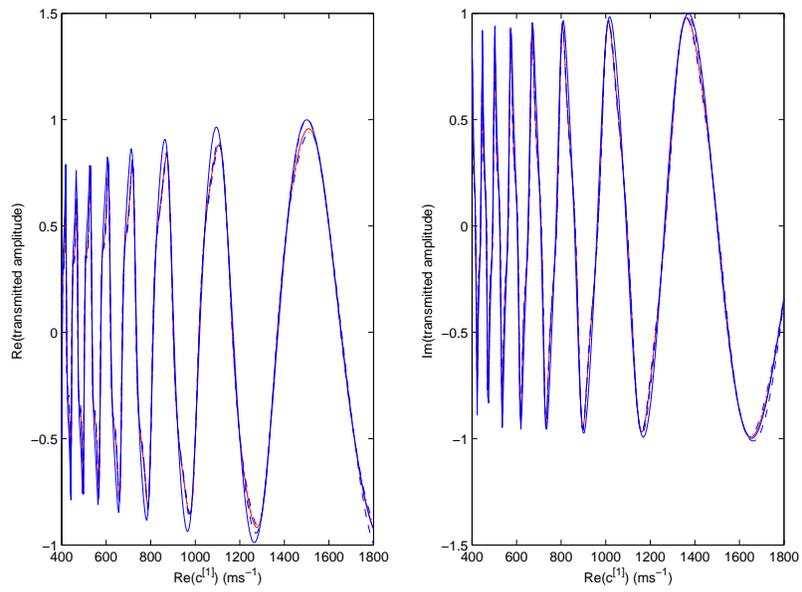}
\caption{Same as fig. \ref{fig09-03} except that $f=20000~Hz$.}
\label{fig09-04}
\end{center}
\end{figure}
\clearpage
\newpage
What changes between figs. \ref{fig09-01} and \ref{fig09-02} (for $\rho^{[1]}=1100~Kgm^{-3}$) is the frequency; likewise between figs. \ref{fig09-03} and \ref{fig09-04} (for $\rho^{[1]}=700~Kgm^{-3}$). In all four of these figures it is apparent that the zeroth-order appoximation of the transmitted amplitude is functionally quite correct, the first-order even more so and numerically quite correct, and the second-order approximation nearly-coincident with the exact transmitted amplitude over the considered range of $\Re c^{[1]}$.
\subsubsection{Zeroth, first, and second-order $\epsilon$ approximations of $a^{[2]}$ compared to the exact expression thereof as a function of the imaginary part of the wavespeed $\Im\left(c^{[1]}\right)$ in the layer}
\begin{figure}[ht]
\begin{center}
\includegraphics[width=0.75\textwidth]{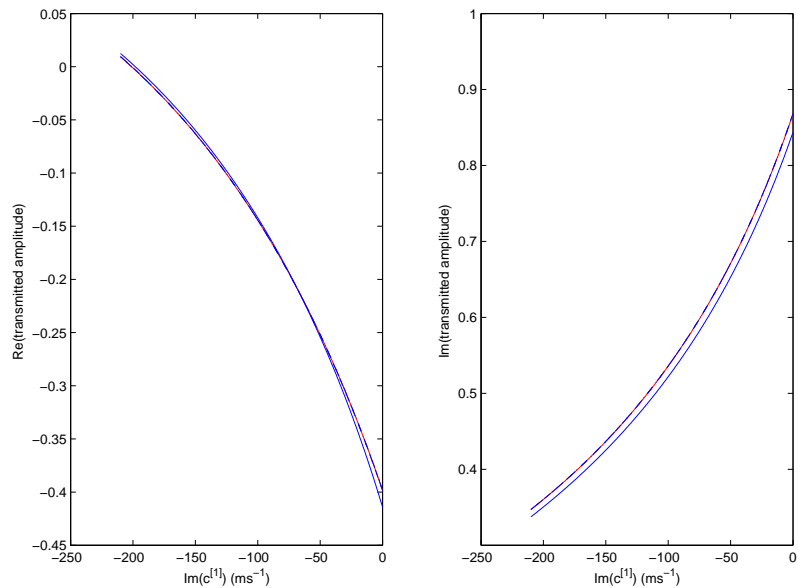}
\caption{Transmitted amplitude as a function of the imaginary part of the wavespeed $\Im\left(c^{[1]}\right)$ in the layer.  The  left(right)-hand panels depict the real(imaginary) parts of  $a^{[2]}$ (red-----), $a^{[2](0)}$ (blue ------), $a^{[2](1)}$ (blue - - - -), $a^{[2](2)}$ (blue -.-.-.-).
Case $f=2000~Hz$. $\theta^{i}=0^{\circ}$, $\rho^{[1]}=1100~Kg/m^{3}$, $\Re\left(c^{[1]}\right)=700~ms^{-1}$.}
\label{fig10-01}
\end{center}
\end{figure}
\begin{figure}[ptb]
\begin{center}
\includegraphics[width=0.75\textwidth]{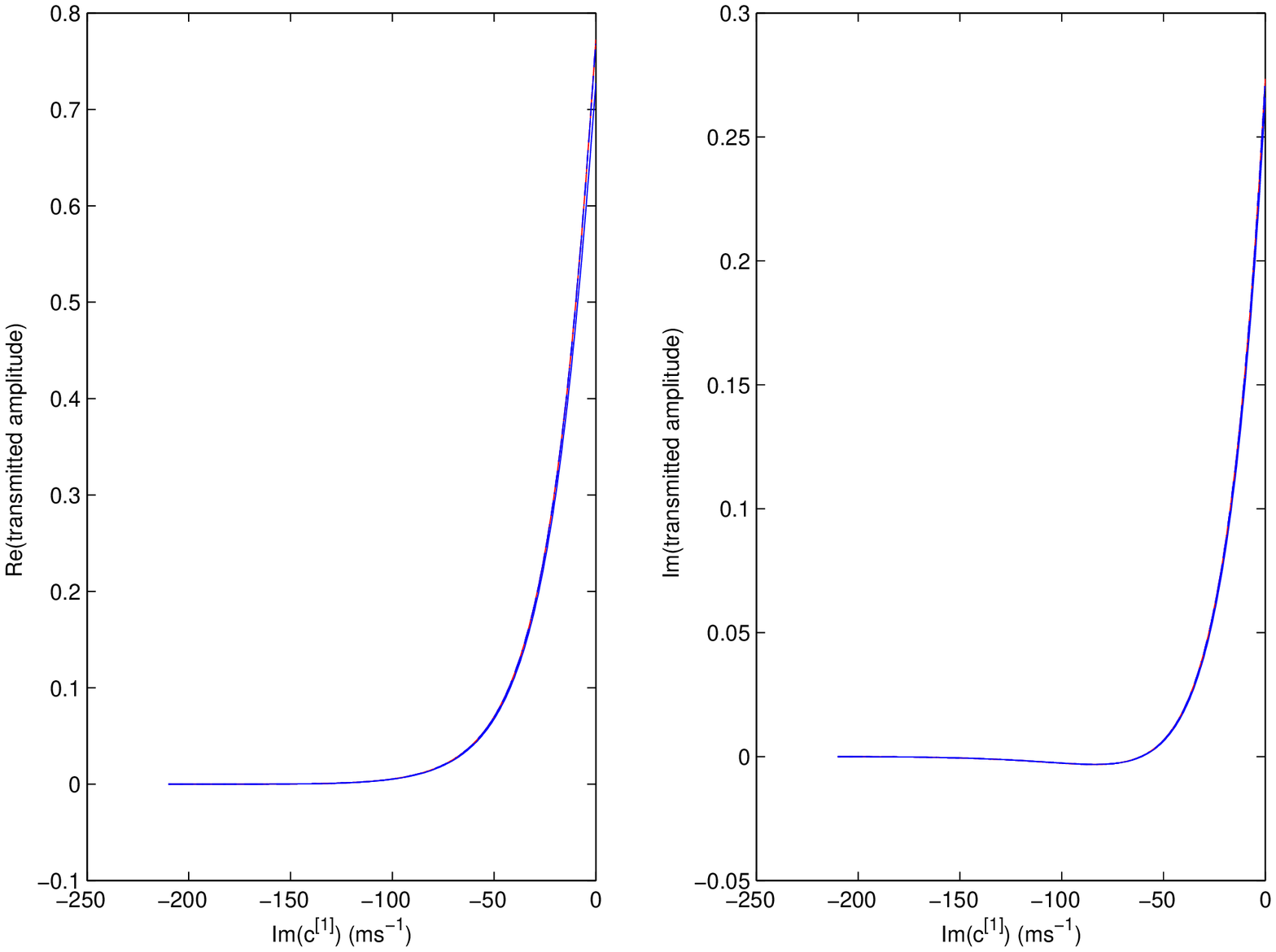}
\caption{Same as fig. \ref{fig10-01} except that $f=20000~Hz$.}
\label{fig10-02}
\end{center}
\end{figure}
\begin{figure}[ptb]
\begin{center}
\includegraphics[width=0.75\textwidth]{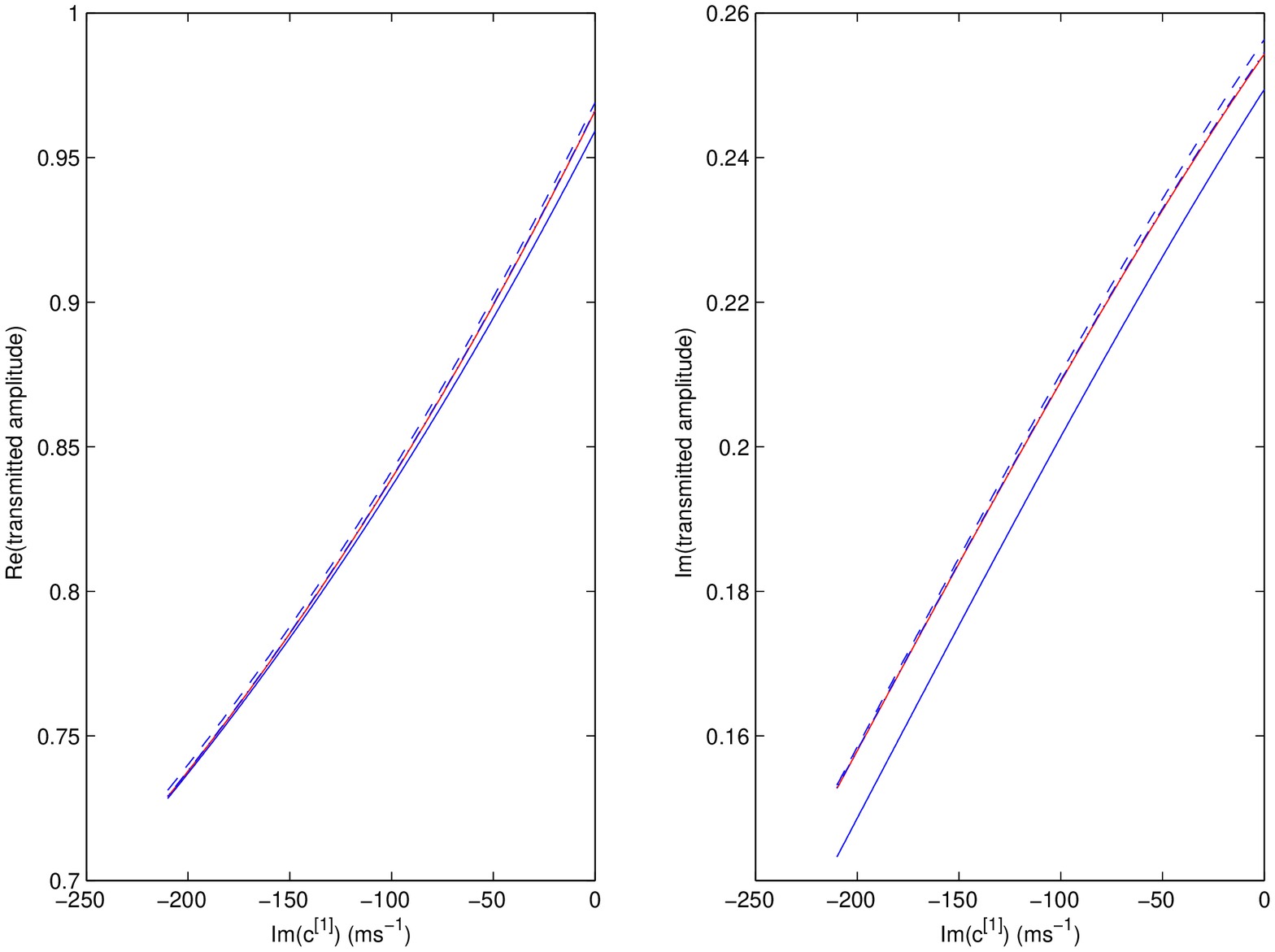}
\caption{Same as fig. \ref{fig10-01} except that $\Re\left(c^{[1]}\right)=1300~ms^{-1}$.}
\label{fig10-03}
\end{center}
\end{figure}
\begin{figure}[ptb]
\begin{center}
\includegraphics[width=0.75\textwidth]{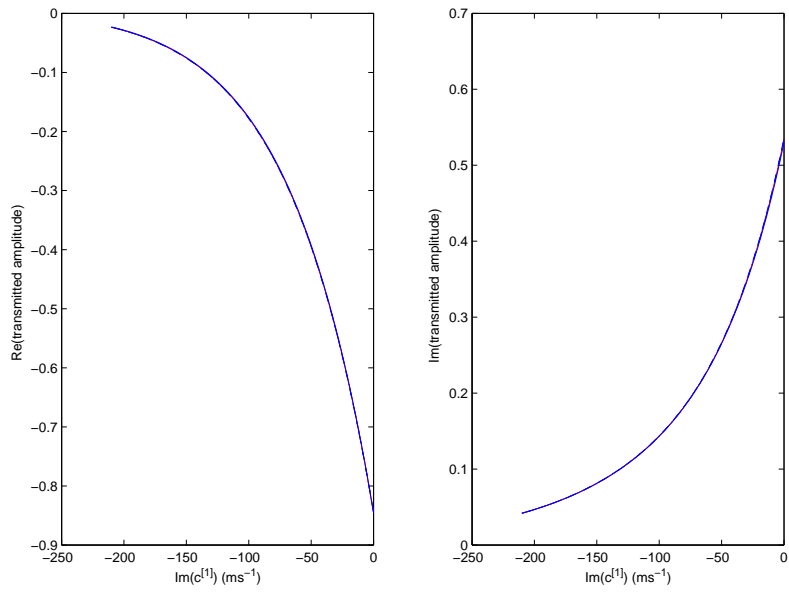}
\caption{Same as fig. \ref{fig10-03} except that $f=20000~Hz$.}
\label{fig10-04}
\end{center}
\end{figure}
\clearpage
\newpage
What changes between figs. \ref{fig10-01} and \ref{fig10-02} (for $\Re c^{[1]}=700~ms^{-1}$) is the frequency; likewise between figs. \ref{fig10-03} and \ref{fig10-04} (for $\Re c^{[1]}=1300~ms^{-1}$). In all four of these figures it is apparent that the zeroth-order appoximation of the transmitted amplitude is functionally quite correct, the first-order even more so and numerically quite correct, and the second-order approximation coincident with the exact transmitted amplitude over the considered range of $\Im c^{[1]}$.
\section{The inverse-scattering problem}
\subsection{Preliminaries}
The following material concerns the retrieval of either $\epsilon$ or $\Re\left(c^{[1]}\right)$ from data relating to the measured  field in $\Omega_{2}$.

Actually, we shall not measure this field, but rather obtain it by simulation, the latter appealing to the exact solution of the forward problem.

Furthermore, we shall assume that the configurations of the forward and inverse problems are exactly the same, this meaning that (a) the obstacle is the same layer (i.e., same: thickness, macroscopic homogeneity invariant with respect to the three space coordinates, location in space, but differing a priori only by the mass density contrast and/or the wavespeed), (b) the obstacle is submitted to the same radiation (i.e., same plane wave), and (c) the measurement location and device are the same (actually a device capable of detecting with perfect accuracy the amplitude and phase of the pressure on the lower face of the layer).

Inversion of this sort normally is the outcome of the comparison of two pressure fields (at the above-mentioned location): 1) the one predicted by a trial model of the layer involving a trial value of the to-be-retrieved parameter (which may initially be far-removed from the actual value of this parameter as it intervenes in the data) and 2) the simulated pressure field. If the difference (the meaning of this word will be detailed further on) of the trial field from the measured (simulated) data field is larger than some prescribed value, then the trial parameter in the trial model is changed in some hopefully-rational manner and the  new trial model is again compared to the data. This comparison procedure is repeated as many times as necessary to obtain a difference equal to, or smaller than, a prescribed value. We shall employ this procedure to retrieve  $\Re\left(c^{[1]}\right)$. However, inversion is simpler when the difference between the predictions of (the pressure fields) trial model and the simulated data are imposed at the outset to vanish and the equation translating this fact can be solved in explicit, mathematical manner. This will be the procedure \cite{wi02b} for retrieving the density contrast $\epsilon$.

In all the following material, upper-case letters will designate trial parameters (and the corresponding trial fields) and lower-case letters will designate the actual parameters (and data fields) involved in the simulation model employed to generate the data. When a parameter in the trial model is not varied (i.e., not the object of the retrieval), it is termed 'prior', and due to the aforementioned assumptions, all the priors are considered to be equal to the corresponding parameters in the simulation model. This means, in particular, that when trying to retrieve the layer wavespeed, we assume the mass density contrast prior to be equal to its value in the simulation model, and when trying to retrieve the mass density contrast, we assume the layer wavespeed to be equal to its value in the simulation model.

The simulation model employed to generate the data appeals to the exact DD-SOV solutions. We shall employ three trial models to retrieve the density contrast: these are based on the zeroth, first and second-order $\epsilon$ approximations of the pressure field, and we shall show that these three inversions can be carried out in explicit, mathematical manner. We shall employ the same three trial models to retrieve the real part of the layer wavespeed, but these three inversions cannot be carried out in explicit mathematical manner. In fact, they appeal to the numerical, algorithmic comparison procedure described previously.

This will enable us to show that the employment of approximate trial models, such as the one based on constant-density assumption, lead to impossible or inaccurate retrievals, much as do the employment of exact trial models with one or several trial priors being different from the corresponding parameters employed in the simulation model.
\subsection{Retrieval of the mass density contrast}
The simulated (via the DD-SOV model) field in the transmission half-space is
\begin{equation}\label{inv-010}
p(x,z)=a^{[2]}\exp[i(k_{y}^{i[0]}x-k_{y}^{i[0]}y)]
~.
\end{equation}
We assume the data is collected at $x=0,~y=-h$, so that the simulated pressure data is
\begin{equation}\label{inv-020}
p(0,-h)=a^{[2]}(\epsilon)\exp[ik_{y}^{i[0]}h)]
~.
\end{equation}
The trial fields, at the same location, for the three trial models, are of the general form
\begin{equation}\label{inv-030}
P(0,-h)=A^{[2](l)}(E)\exp[iK_{y}^{i[0]}H]~;~l=0,1,2
~,
\end{equation}
wherein $E$ is the trial mass density contrast, and due to the assumption of the identity of the priors, $K_{y}^{i[0]}=k_{y}^{i[0]}$, $H=h$, so that
\begin{equation}\label{inv-040}
P^{(l)}(0,-h)=A^{[2](l)}\exp[ik_{y}^{i[0]}h]~;~l=0,1,2
~,
\end{equation}
wherein, it must be recalled:
\begin{equation}\label{inv-045}
A^{[2](l)}=\sum_{j=0}^{l}A_{j}^{[2]}E^{j}~;~l=0,1,2
~,
\end{equation}
with the coefficients $A_{j}^{[2]}$ not depending on $E$.

The so-called {\it cost function} is a function of the difference of $P^{(l)}(0,-h)$ from $p(0,-h)$
\begin{equation}\label{inv-050}
\kappa^{(l)}(E^{(l)})=\mathcal{F}\left(P^{(l)}(0,-h)-p(0,-h)\right)~;~l=0,1,2
~,
\end{equation}
the object being to find the trial density contrast $E^{(l)}$ that minimizes $\kappa^{(l)}$ for each $l$.
For reasons that will become obvious, we shall choose $\mathcal{F}\left((P^{(l)}(0,-h)-p(0,-h)\right)=P^{(l)}(0,-h)-p(0,-h))$ and the minimum to be zero, so that the inverse problems reduce to finding $E^{(l)}$ from
\begin{equation}\label{inv-060}
\kappa^{(l)}(E^{(l)})=P^{(l)}(0,-h)-p(0,-h)=0
~,
\end{equation}
or, more explicitly,
\begin{equation}\label{inv-070}
\left(A^{[2](l)}-a^{[2]}\right)\exp[ik_{y}^{i[0]}h]=0
\end{equation}
which implies
\begin{equation}\label{inv-080}
A^{[2](l)}-a^{[2]}=0~,
\end{equation}

First consider the case $l=0$. On account of (\ref{inv-045}) the previous equation reduces to
\begin{equation}\label{inv-090}
A_{0}^{[2]}-a^{[2]}=0~,
\end{equation}
which possesses {\it no solution} for the mass density contrast since $A_{0}^{[2]}$ does not depend on this parameter. Thus, the somewhat-trivial conclusion is that it is not possible to retrieve the density contrast when the constant-density assumption is made.

Next consider the case $l=1$. On account of (\ref{inv-045}) the comparison equation (\ref{inv-080}) reduces to the linear equation (in terms of the mass  density contrast)
\begin{equation}\label{inv-100}
A_{0}^{[2]}+A_{1}^{[2]}E^{(1)}-a^{[2]}=0~,
\end{equation}
whose {\it single solution} is simply
\begin{equation}\label{inv-110}
E^{(1)}=-\left(\frac{A_{0}^{[2]}-a^{[2]}}{A_{1}^{[2]}}\right)~.
\end{equation}
This shows that the employment of the linearized approximation $A^{[2](1)}$ of $A^{[2]}$ as the trial model enables: (1) not only to solve the inverse problem in explicit, mathematical form, (2) but also to obtain a {\it unique} solution of this problem. These are the reasons why linearization is often employed in dealing with inverse problems. A last property of this solution: it may be complex even when $\epsilon$ is real, this being due to the fact that the different terms in (\ref{inv-110}) are generally-complex.

Finally, consider the $l=2$ case. On account of (\ref{inv-045}) the comparison equation (\ref{inv-080}) reduces to the quadratic equation (in terms of the mass density contrast)
\begin{equation}\label{inv-120}
A_{0}^{[2]}+A_{1}^{[2]}E^{(2)}+A_{2}^{[2]}\left(E^{(2)}\right)^{2}-a^{[2]}=0~,
\end{equation}
whose {\it two solutions} are simply
\begin{equation}\label{inv-130}
E^{(2)\pm}=-\frac{A_{1}^{[2]}}{2A_{2}^{[2]}}\pm
\frac
{\sqrt{\left(A_{1}^{[2]}\right)^{2}-4A_{2}^{[2]}\left(A_{0}^{[2]}-a^{[2]}\right)}}
{2A_{2}^{[2]}}~.
\end{equation}
This shows that the employment of the second-order  approximation $A^{[2](2)}$ of $A^{[2]}$ as the trial model enables: (1) not only to solve the inverse problem in explicit, mathematical form, (2) but also to obtain a {\it non-unique} solution of this problem. In fact, we find two solutions to this problem, and there does not appear to exist any easily-discernible relation between these two solutions on the one hand and the sole solution of the the $l=1$ problem on the other hand. A last property of the second-order solution: it may be complex even when $\epsilon$ is real, this being due to the fact that the different terms in (\ref{inv-130}) are generally-complex.

It has been suggested that to alleviate the non-uniqueness problem, and/or improve the accuracy of the retrievals, one should take advantage of the fact that a retrieved constitutive parameter should not depend on the characteristics of the solicitation because of our assumption of isotropy. We translate this remark here to: $E^{[2]}$ should not depend on the angle of incidence $\theta^{i}$, so that to possibly alleviate the non-uniqueness problem and/or increase the accuracy of the retrieval, we generalize the comparison equation (\ref{inv-090}) to $N$ realizations which differ from each other only by the choice of incident angle.

The chosen incident angles form the set
\begin{equation}\label{inv-150}
\boldsymbol{\Theta}^{i}=\{\theta^{i}_{1},\theta^{i}_{2},...,\theta^{i}_{N}\}~,
\end{equation}
and for the $n$-th realization, the simulated transmission coefficient is $a^{[2]}(\theta^{i}_{n})$ while its  trial counterpart is $A^{[2]}(\theta^{i}_{n})$ so that the comparison equation now takes the form
\begin{equation}\label{inv-155}
\sum_{n=1}^{N}\left(A^{[2]}(\theta^{i}_{n})-a^{[2]}(\theta^{i}_{n})\right)=0~.
\end{equation}
which is the same as (\ref{inv-090}) when $N=1$.

Proceeding as previously, we make the expansion (for each incident angle realization)
\begin{equation}\label{inv-160}
A^{[2](l)}(\theta^{i}_{n})=\sum_{m=1}^{l}A_{m}^{[2]}(\theta^{i}_{n})\left(E^{(l)}\right)^{m}~,
\end{equation}
whence (\ref{inv-155}) becomes
\begin{equation}\label{inv-165}
\sum_{n=1}^{N}\left(\sum_{m=1}^{l}A_{m}^{[2]}(\theta^{i}_{n})\left(E^{(l)}\right)^{m}-a^{[2]}(\theta^{i}_{n})\right)=0~.
\end{equation}
For $l=1$ this equation yields the unique solution
\begin{equation}\label{inv-170}
E^{(1)}=-\left(\frac{\mathcal{A}_{0}^{[2]}-\alpha^{[2]}}{\mathcal{A}_{1}^{[2]}}\right)~,
\end{equation}
whereas for $l=2$ it gives rise to the two solutions
\begin{equation}\label{inv-175}
E^{(2)\pm}=-\frac{\mathcal{A}_{1}^{[2]}}{2\mathcal{A}_{2}^{[2]}}\pm
\frac
{\sqrt{\left(\mathcal{A}_{1}^{[2]}\right)^{2}-4\mathcal{A}_{2}^{[2]}\left(\mathcal{A}_{0}^{[2]}-\alpha^{[2]}\right)}}
{2\mathcal{A}_{2}^{[2]}}~.
\end{equation}
wherein:
\begin{equation}\label{inv-180}
\mathcal{A}_{m}^{[2]}=\sum_{n=1}^{N}A_{m}^{[2]}(\theta_{n}^{i})~~,~~\alpha^{[2]}=\sum_{n=1}^{N}a^{[2]}(\theta_{n}^{i})~.
\end{equation}
As previously, the $l=1$ and $l=2$ retrievals for the mass density contrast may turn out to be complex even though $\epsilon$ is real, this being due to the fact that the different terms in (\ref{inv-170}) and (\ref{inv-175}) are generally-complex.
\subsection{Numerical results for the retrieval of the mass density contrast}
The aim of the computations was to compare numerically the actual density contrast $\epsilon$ to the retrievals thereof: (a) $E^{(1)})$ obtained via the first-order $\epsilon$ trial model, and (b) $E^{(2)}$ obtained via the second-order $\epsilon$ trial model, as a function of the various parameters $\Re\left(c^{[1]}\right)$, $\Im\left(c^{[1]}\right)$, $\epsilon$, $f$, $h$ and $\theta^{i}$,   the other parameters being fixed at the following values: $\rho^{[0]}=1000~Kgm^{-3}$, $c^{[0]}=1500~ms^{-1}$, $a^{[0]}=1$. As concerns the set of incident angle(s), we shall designate it in condensed form by $\boldsymbol{\Theta}^{i}=(\theta^{i}_{b},\theta^{i}_{e},N)$, which means that $N$ equally-spaced values of $\theta^{i}$ are chosen for the retrieval ranging from $\theta^{i}_{b}$ to $\theta^{i}_{e}$.
\subsubsection{First and second-order retrievals as a function of the frequency $f$}
\begin{figure}[ht]
\begin{center}
\includegraphics[width=0.65\textwidth]{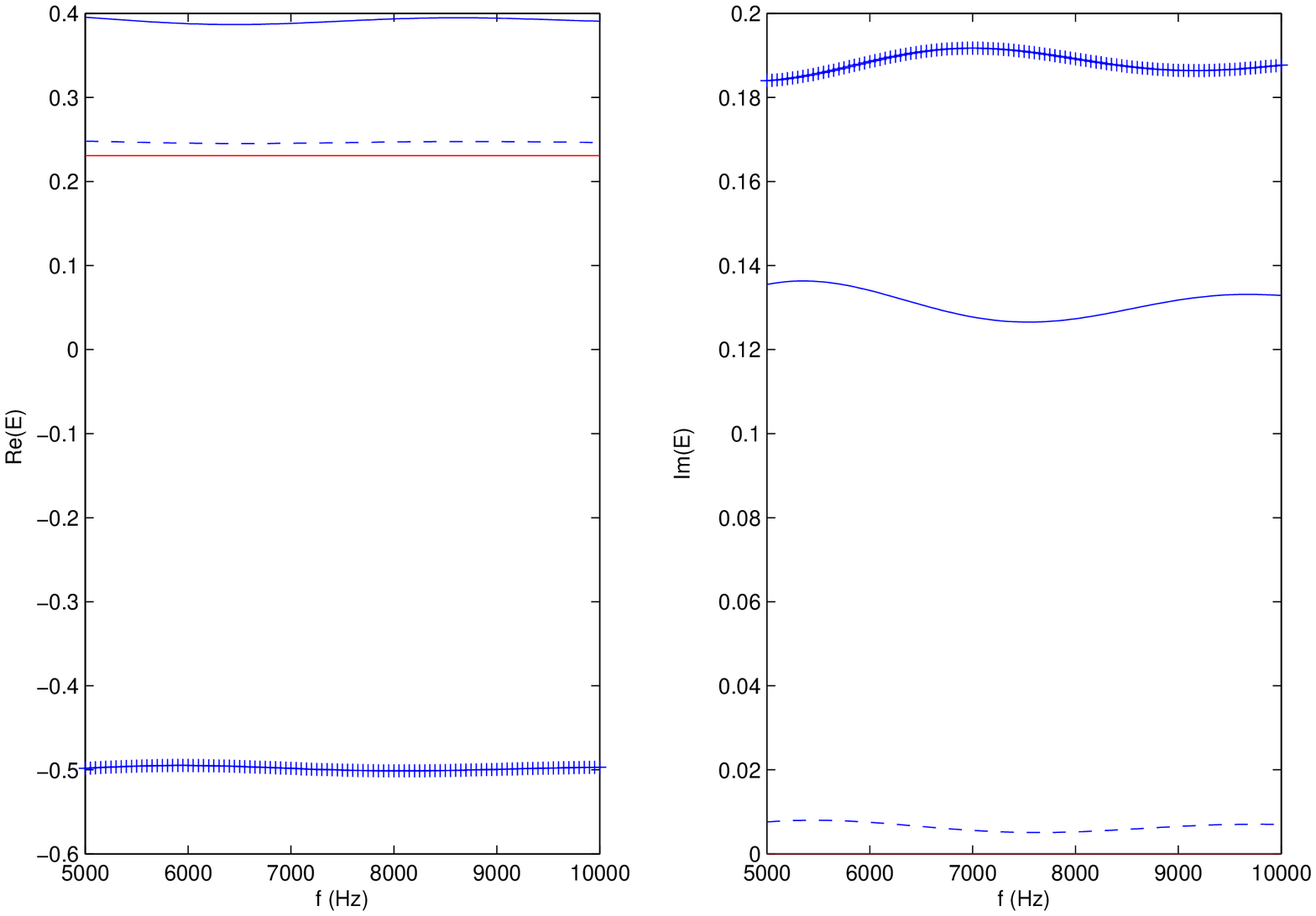}
\caption{Retrieved ($E$) mass density contrast compared to the actual ($\epsilon$) mass density contrast as a function of the frequency $f$.  The  left(right)-hand panels depict the real(imaginary) parts of  $\epsilon$ (red-----), $E^{(1)}$ (blue ------), $E^{(2)-}$ (blue - - - -), $E^{(2)+}$ (blue + + + +). Case $\rho^{[1]}=1300~Kgm^{-3}$, $c^{[1]}=1700-210i~ms^{-1}$, $h=0.2~m$, $\boldsymbol{\Theta}^{i}=(0^{\circ},0^{\circ},1)$.}
\label{fig20-01}
\end{center}
\end{figure}
\begin{figure}[ptb]
\begin{center}
\includegraphics[width=0.65\textwidth]{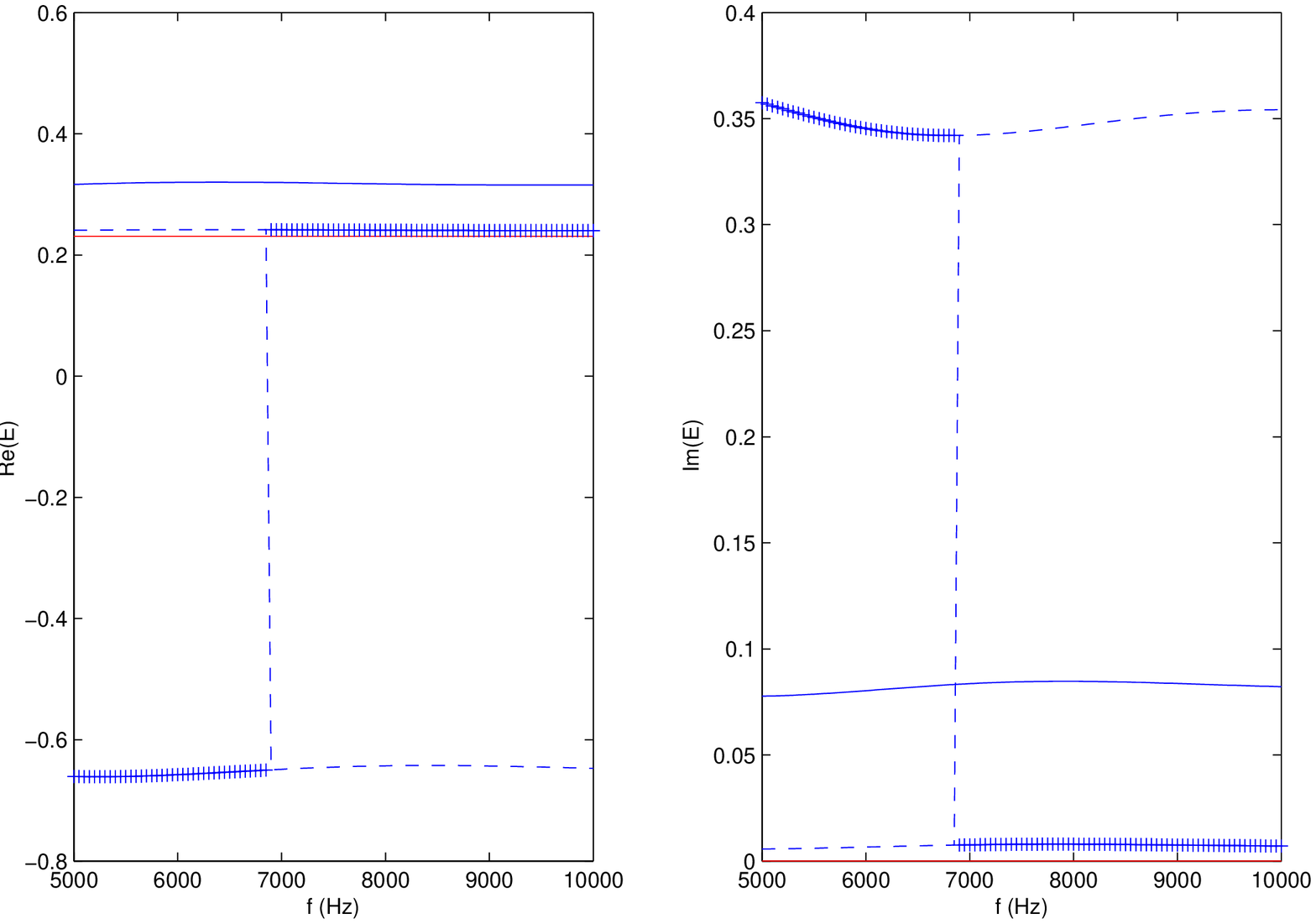}
\caption{Same as fig. \ref{fig20-01} except that $\boldsymbol{\Theta}^{i}=(40^{\circ},40^{\circ},1)$.}
\label{fig20-02}
\end{center}
\end{figure}
\begin{figure}[ptb]
\begin{center}
\includegraphics[width=0.65\textwidth]{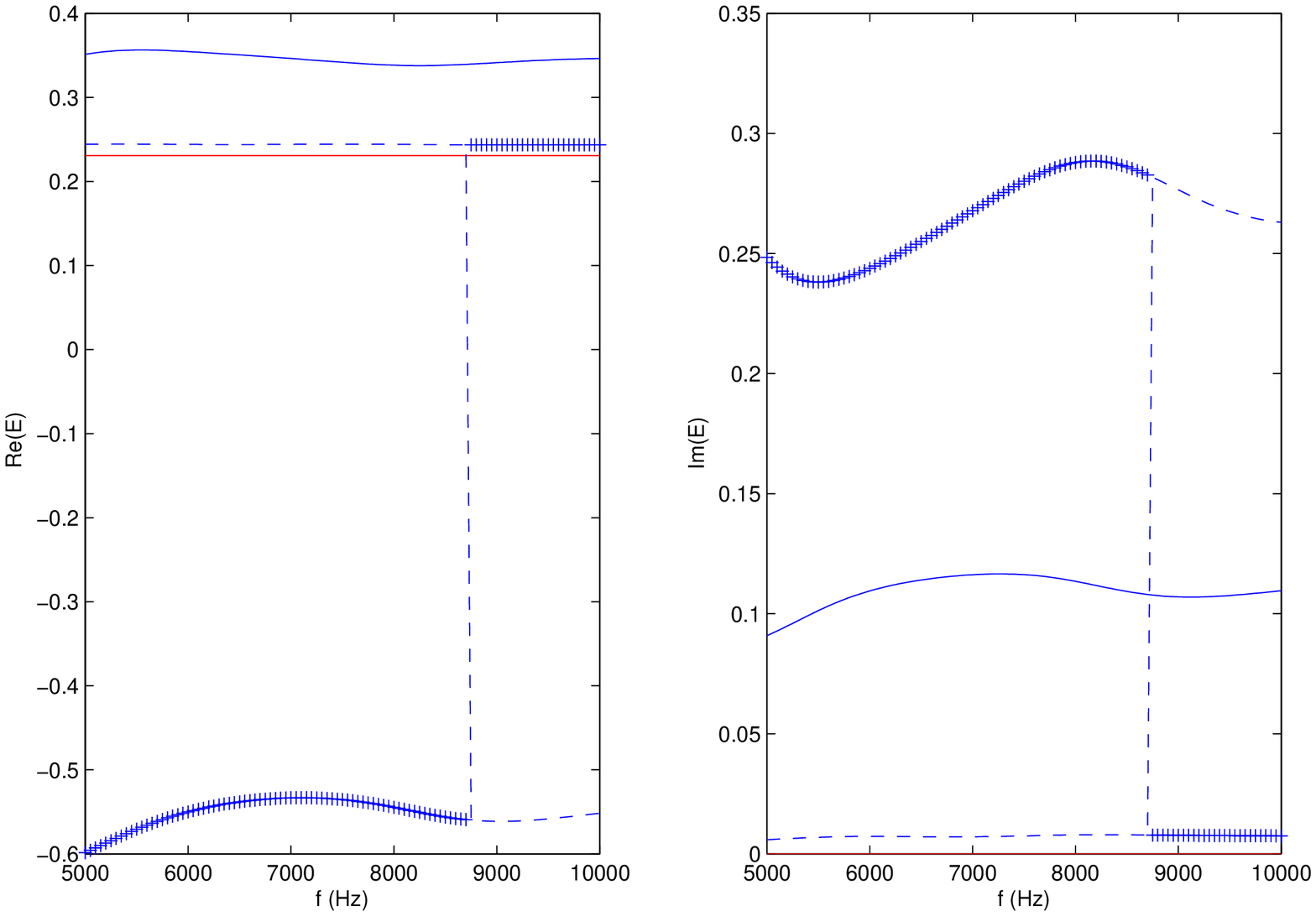}
\caption{Same as fig. \ref{fig20-01} except that $\boldsymbol{\Theta}^{i}=(0^{\circ}, 40^{\circ},2)$.}
\label{fig20-03}
\end{center}
\end{figure}
\begin{figure}[ptb]
\begin{center}
\includegraphics[width=0.65\textwidth]{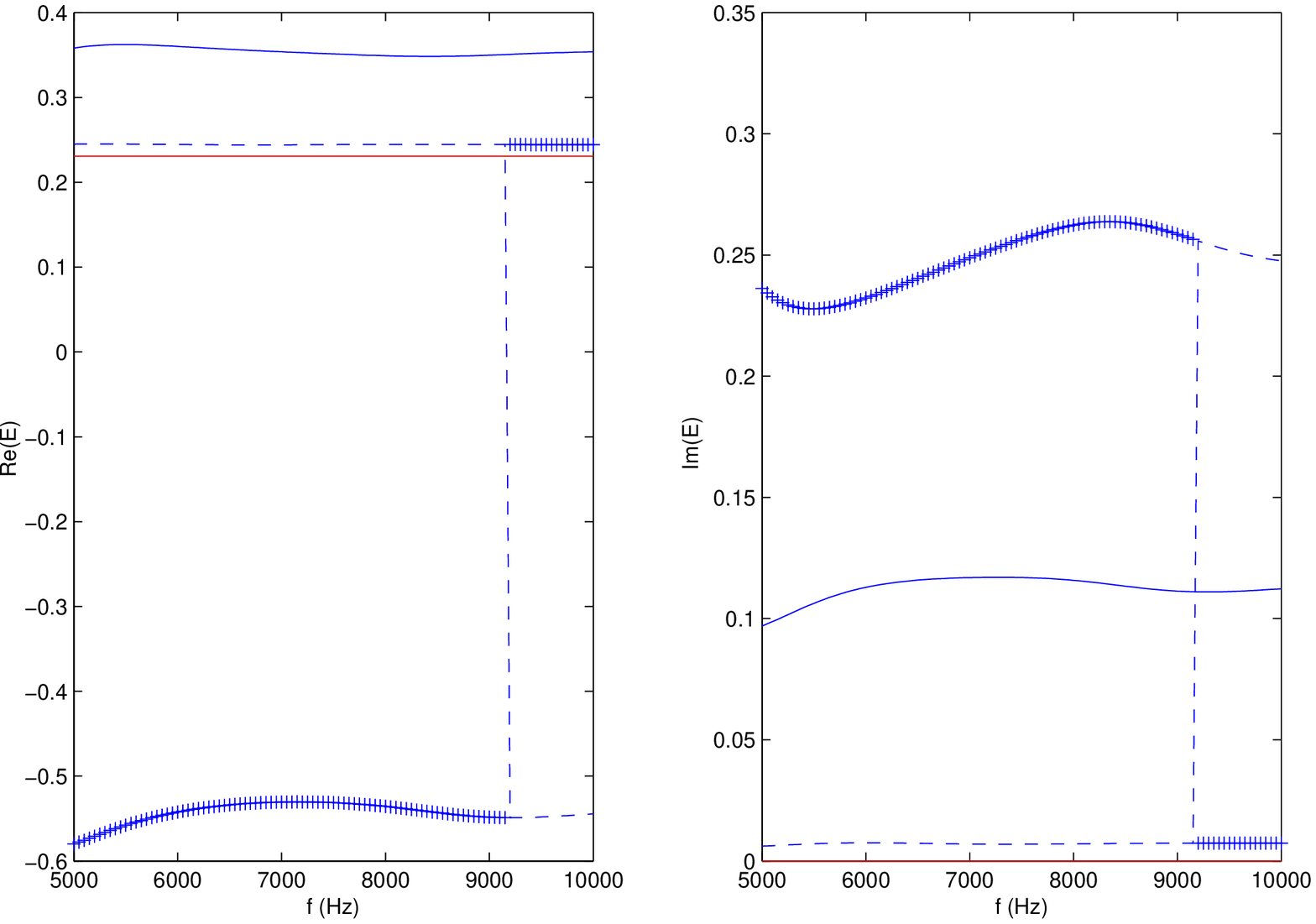}
\caption{Same as fig. \ref{fig20-01} except that $\boldsymbol{\Theta}^{i}=(0^{\circ}, 40^{\circ},3)$.}
\label{fig20-04}
\end{center}
\end{figure}
\begin{figure}[ptb]
\begin{center}
\includegraphics[width=0.65\textwidth]{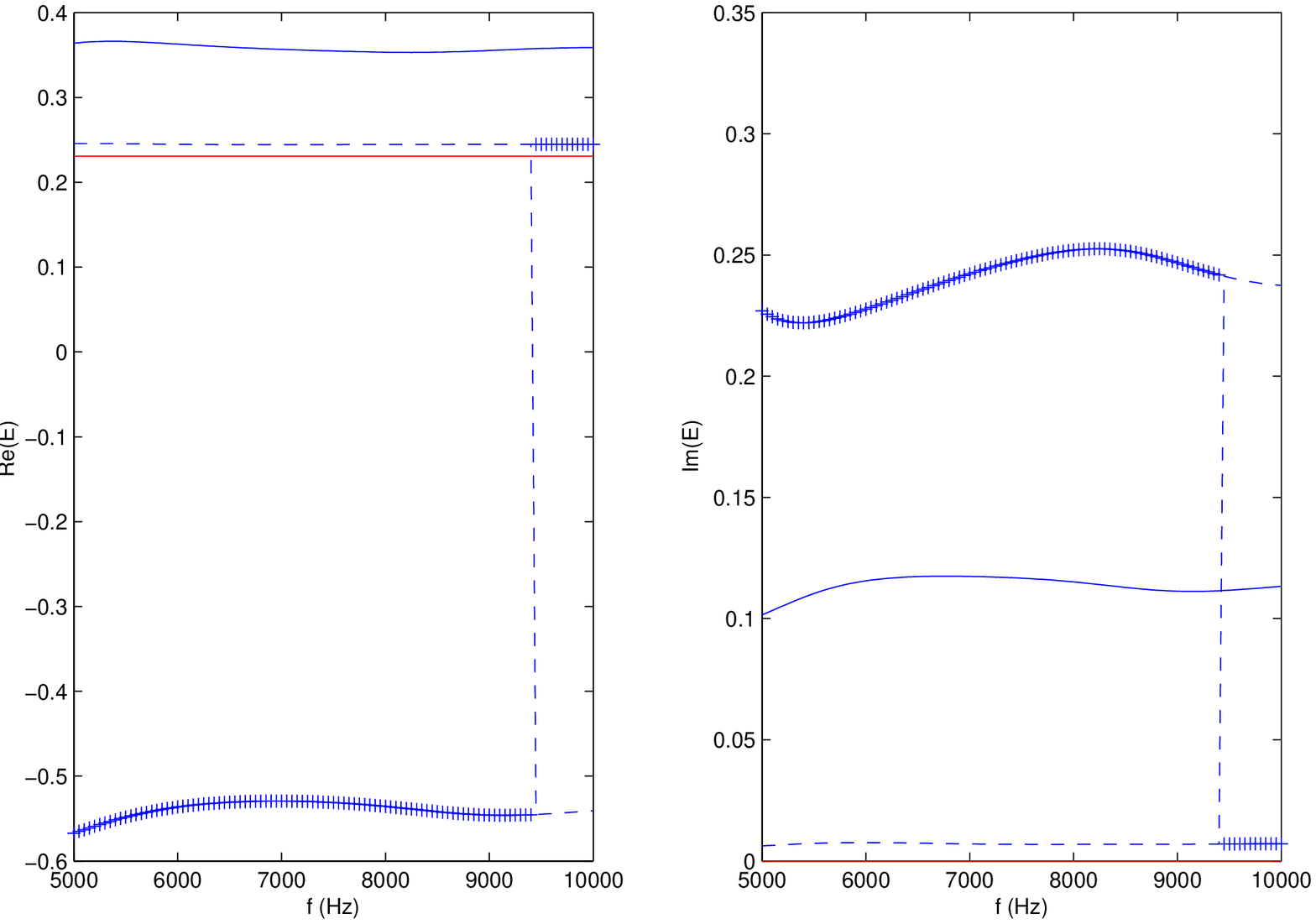}
\caption{Same as fig. \ref{fig20-01} except that $\boldsymbol{\Theta}^{i}=(0^{\circ}, 40^{\circ},5)$.}
\label{fig20-05}
\end{center}
\end{figure}
\begin{figure}[ptb]
\begin{center}
\includegraphics[width=0.65\textwidth]{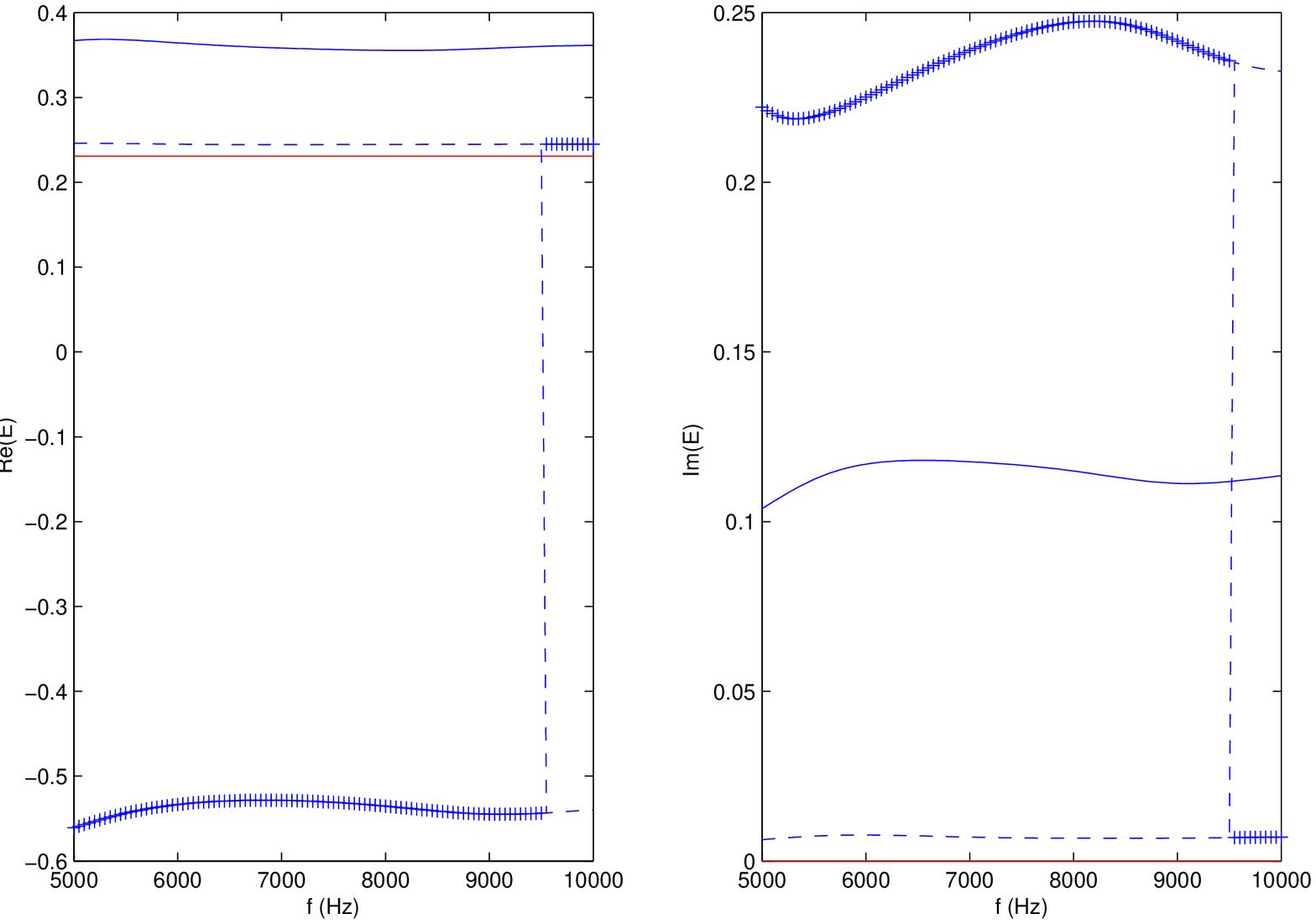}
\caption{Same as fig. \ref{fig20-01} except that $\boldsymbol{\Theta}^{i}=(0^{\circ}, 40^{\circ},9)$.}
\label{fig20-06}
\end{center}
\end{figure}
\begin{figure}[ptb]
\begin{center}
\includegraphics[width=0.65\textwidth]{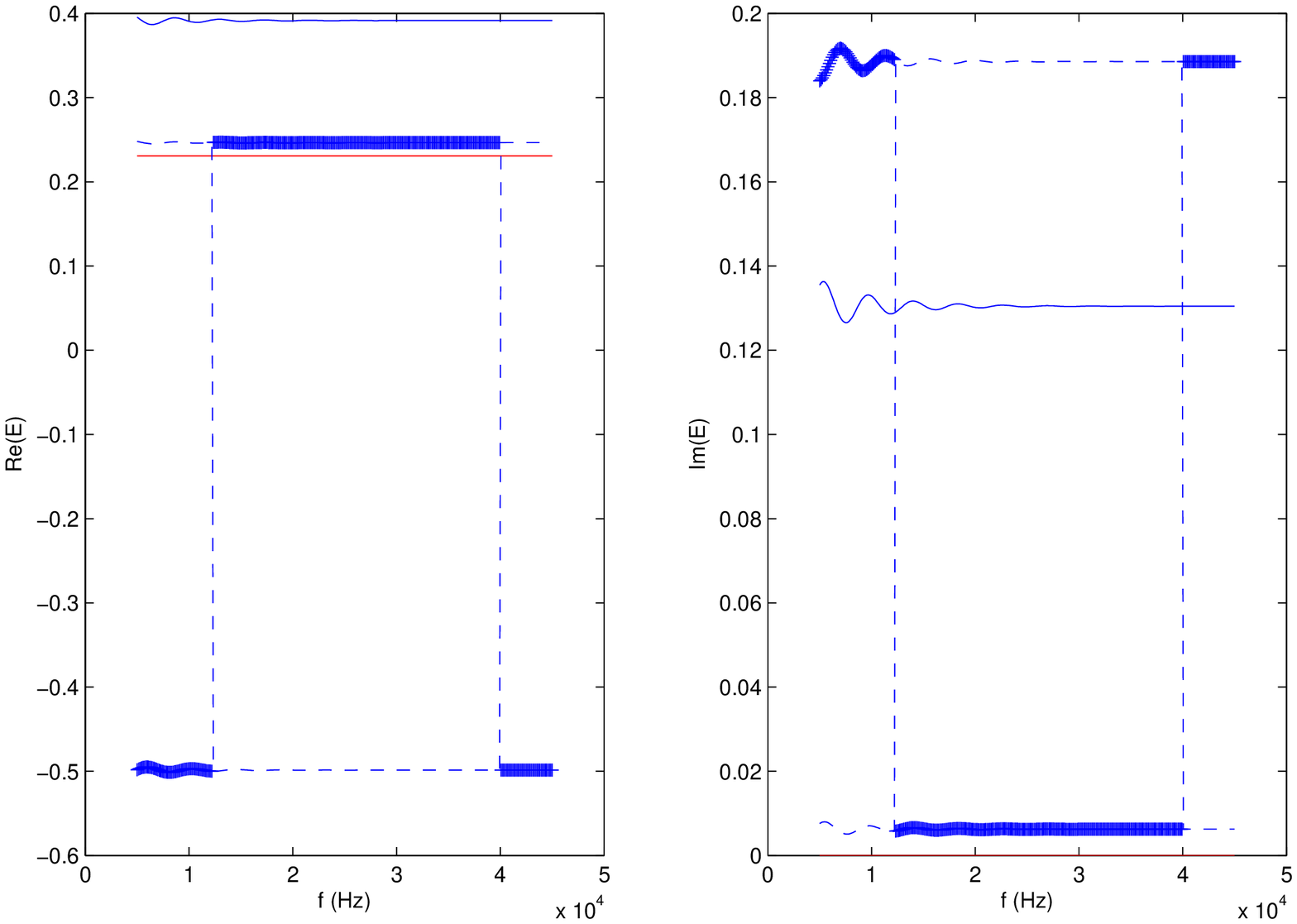}
\caption{Retrieved ($E$) mass density contrast compared to the actual ($\epsilon$) mass density contrast   as a function of, and for a different range of, the frequency $f$ .  The  left(right)-hand panels depict the real(imaginary) parts of  $\epsilon$ (red-----), $E^{(1)}$ (blue ------), $E^{(2)-}$ (blue - - - -), $E^{(2)+}$ (blue + + + +). Case $\rho^{[1]}=1300~Kgm^{-3}$, $c^{[1]}=1700-210i~ms^{-1}$, $h=0.2~m$, $\boldsymbol{\Theta}^{i}=(0^{\circ},0^{\circ},1)$.}
\label{fig21-01}
\end{center}
\end{figure}
\begin{figure}[ptb]
\begin{center}
\includegraphics[width=0.65\textwidth]{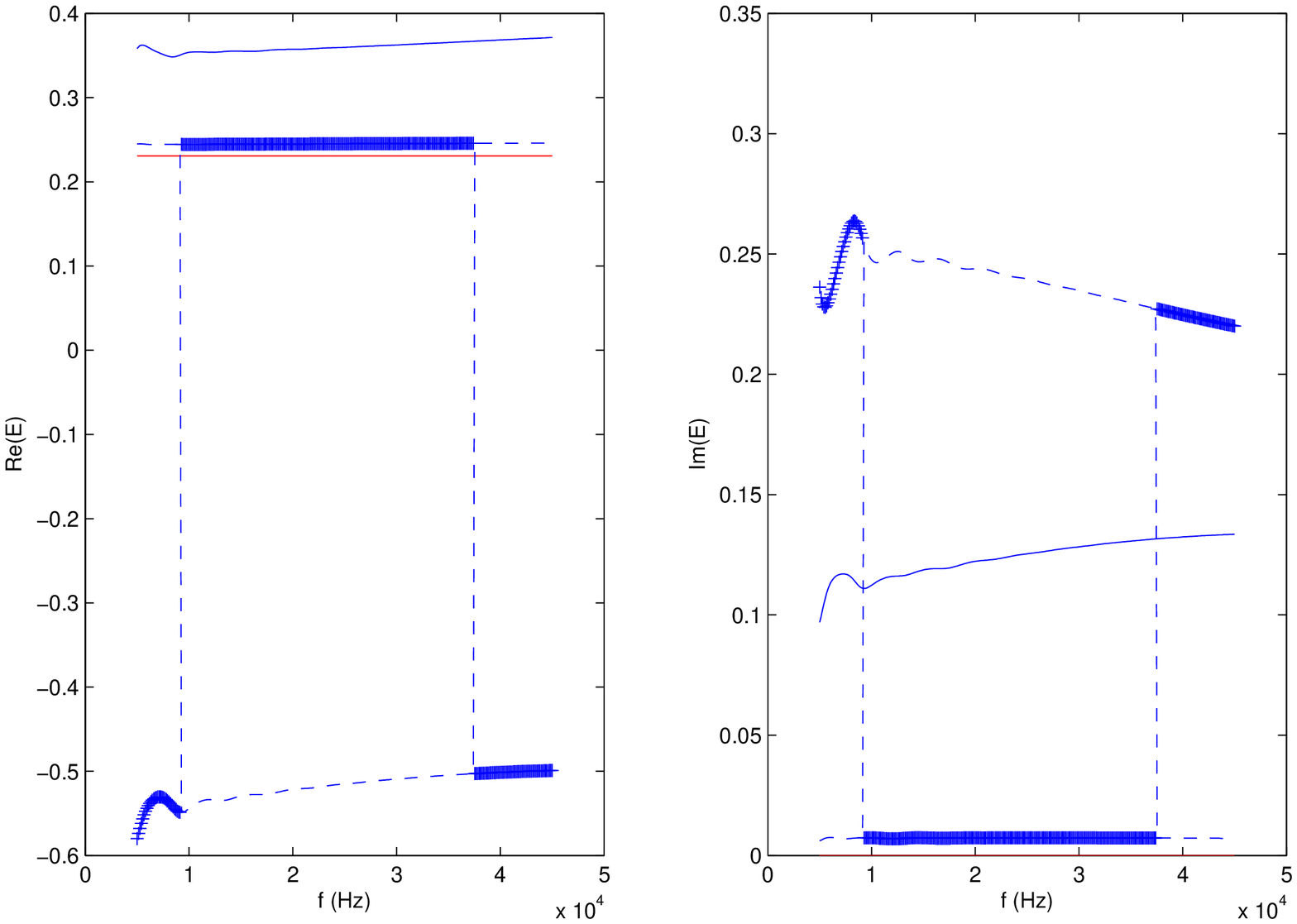}
\caption{Same as fig. \ref{fig21-01} except that $\boldsymbol{\Theta}^{i}=(0^{\circ}, 40^{\circ},3)$.}
\label{fig21-02}
\end{center}
\end{figure}
\clearpage
\begin{figure}[ptb]
\begin{center}
\includegraphics[width=0.65\textwidth]{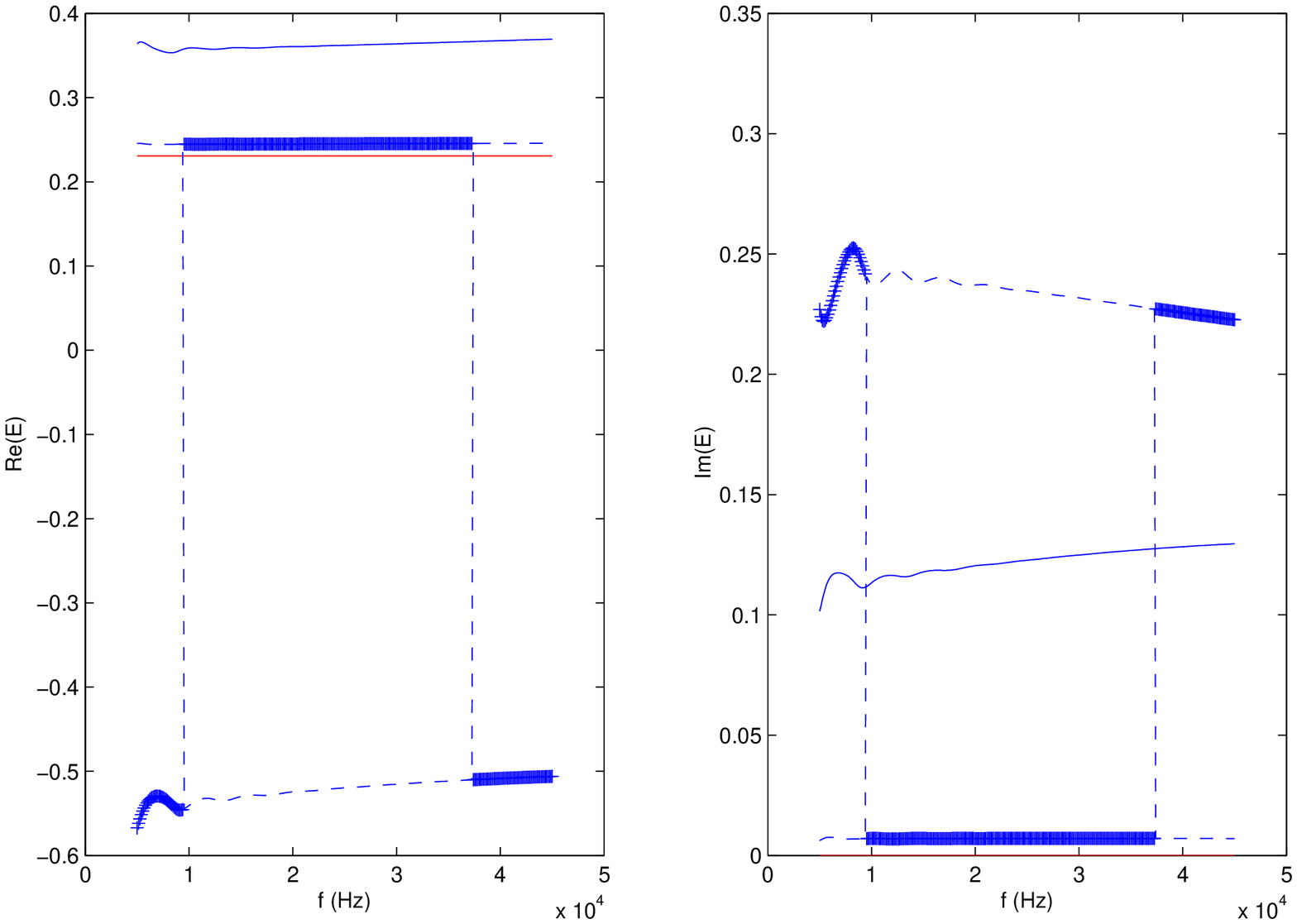}
\caption{Same as fig. \ref{fig21-01} except that $\boldsymbol{\Theta}^{i}=(0^{\circ}, 40^{\circ},5)$.}
\label{fig21-03}
\end{center}
\end{figure}
\begin{figure}[ptb]
\begin{center}
\includegraphics[width=0.65\textwidth]{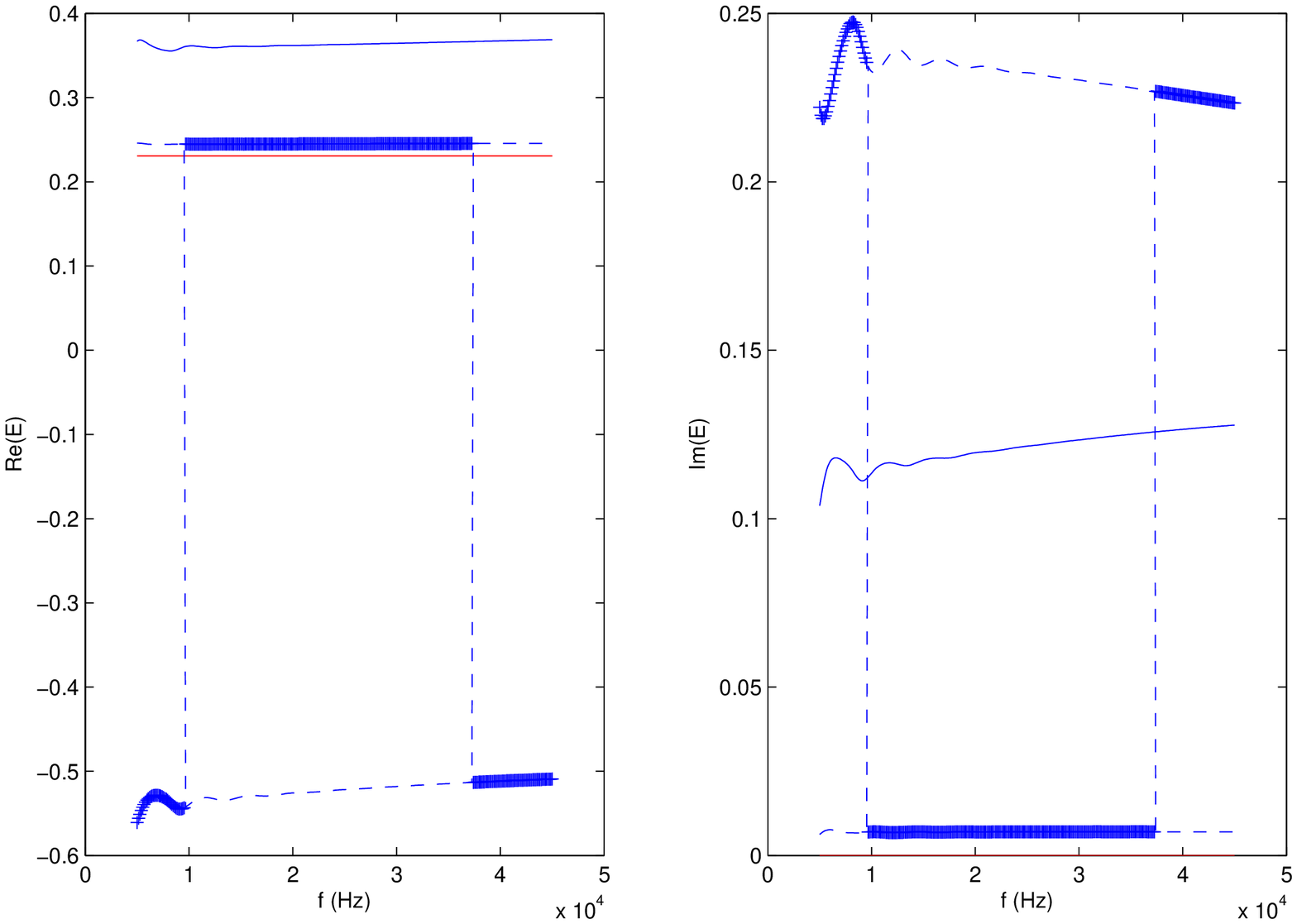}
\caption{Same as fig. \ref{fig21-01} except that  $\boldsymbol{\Theta}^{i}=(0^{\circ}, 40^{\circ},9)$.}
\label{fig21-04}
\end{center}
\end{figure}
\begin{figure}[ptb]
\begin{center}
\includegraphics[width=0.65\textwidth]{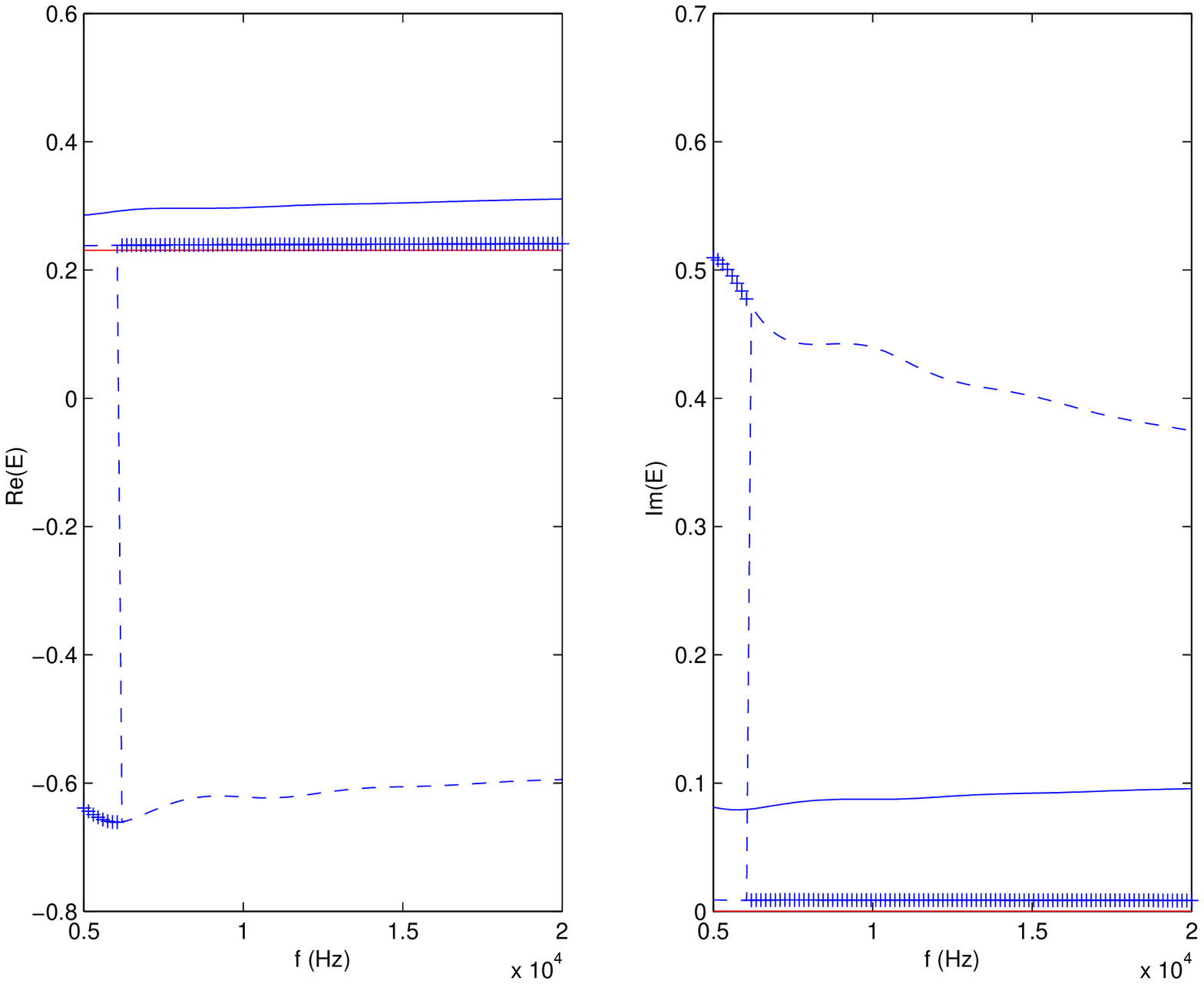}
\caption{Retrieved ($E$) mass density contrast compared to the actual ($\epsilon$) mass density contrast  as a function of, and for a different range of, the frequency $f$ .  The  left(right)-hand panels depict the real(imaginary) parts of  $\epsilon$ (red-----), $E^{(1)}$ (blue ------), $E^{(2)-}$ (blue - - - -), $E^{(2)+}$ (blue + + + +). Case $\rho^{[1]}=1300~Kgm^{-3}$, $c^{[1]}=1700-210i~ms^{-1}$, $h=0.2~m$, $\boldsymbol{\Theta}^{i}=(30^{\circ}, 60^{\circ},5)$.}
\label{fig22-01}
\end{center}
\end{figure}
\clearpage
\begin{figure}[ptb]
\begin{center}
\includegraphics[width=0.65\textwidth]{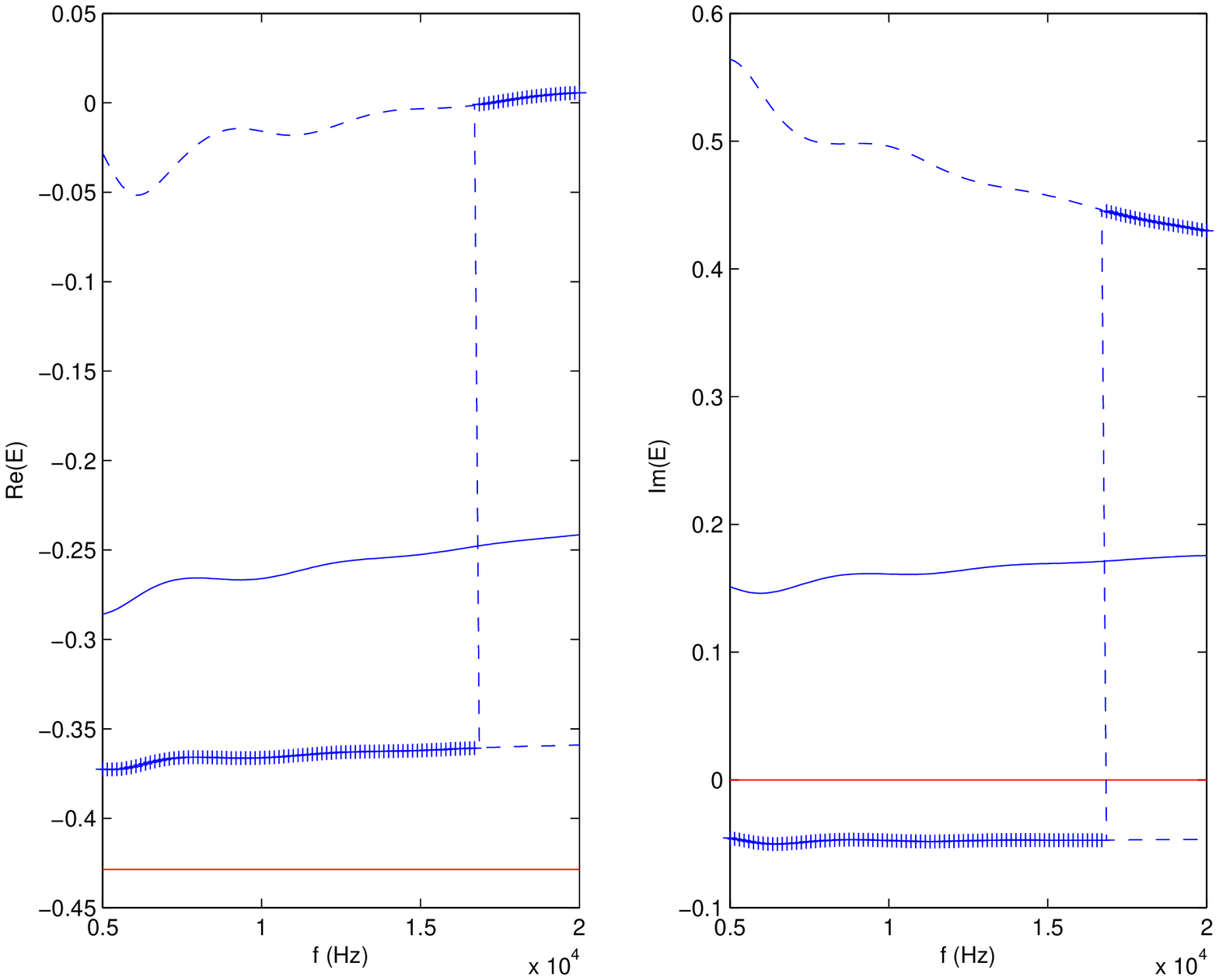}
\caption{Same as fig. \ref{fig22-01} except that $\rho^{[1]}=700~Kgm^{-3}$.}
\label{fig22-02}
\end{center}
\end{figure}
\begin{figure}[ptb]
\begin{center}
\includegraphics[width=0.65\textwidth]{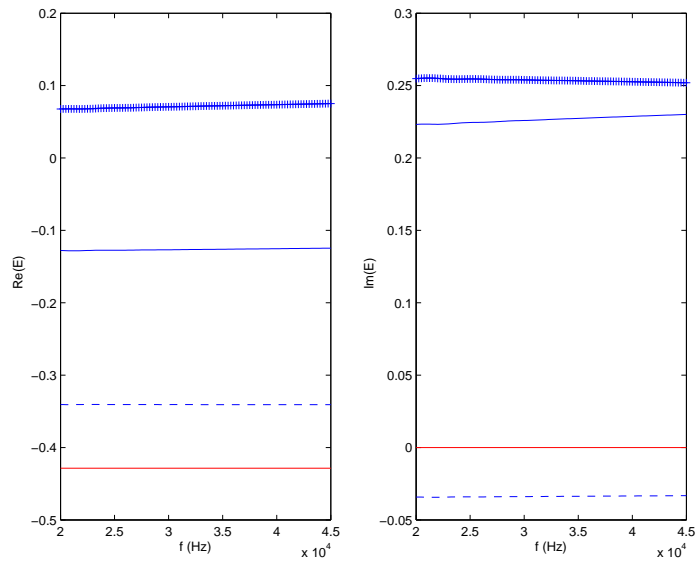}
\caption{Same as fig. \ref{fig22-02} except that $\boldsymbol{\Theta}^{i}=(0^{\circ}, 30^{\circ},5)$.}
\label{fig22-03}
\end{center}
\end{figure}
\clearpage
\newpage
What changes between the six figures \ref{fig20-01}-\ref{fig20-06} (relative to the narrow low-frequency range  $f=[5~KHz,10~KHz]$) and the four figures \ref{fig21-01}-\ref{fig21-04} (relative to the wider, including high-frequencies, range $f=[0~KHz,50~KHz]$) is $\boldsymbol{\Theta}^{i}$. The increase of the number of incident-angle realizations is observed to have the unexpected effect of decreasing the quality of the first-order mass-density retrievals. It is also observed that the second-order retrievals are divided into two widely-separated branches, one of which is rather close to the exact mass density $\epsilon$, with the $+$ and $-$ solutions suddenly shifting between these two branches as the frequency increases. These sudden shifts have been observed in other inverse problems and can be qualified as 'instability' which is known to be one of the characteristics (although usually associated with data, rather than trial model, error \cite{ta77,eg87}) of the 'solutions' of ill-posed mathematical problems. The  increase of the number $N$ of realizations seems not to affect the height of the branch closest to $\epsilon$, which suggests that this might be a useful indicator for distinguishing the 'good branch' from the 'bad' branch. Also the increase of $N$ seems, for small $N$, to shift to higher frequencies the moment of shift from one branch to the other, but for larger $N$, ceases to produce this possibly-useful effect. Besides this, the 'good branches' of the second-order retrievals are clearly-closer  than the corresponding first-order retrievals to the actual mass-density $\epsilon$, as one would expect, and clearly, the first-order retrievals are way off mark. Other than these remarks , the noticeable feature of figs. \ref{fig21-01}-\ref{fig21-04}, also discernible to some extent in figs. \ref{fig20-01}-\ref{fig20-06}, is the {\it dispersive nature} of the retrieved mass density contrast even though the actual mass density contrast was assumed to be non-dispersive (i.e., to not depend on $f$). This dispersive nature is seen to particularly affect: 1) the first-order retrieval and 2) the particular second-order retrieval (of the two possible second-order retrievals) which is the farthest from the actual mass density contrast. This suggests that the dispersion of the retrieval is induced by  trial model error and is all the larger the larger is the trial model error.

What changes between the two figures \ref{fig22-01}-\ref{fig22-02} is $\rho^{[1]}$. Other than the previous remarks which apply as well here as  for a different layer mass density (and therefore different $\epsilon$), one gets the impression that the effects of changing $\boldsymbol{\Theta}^{i}$ are not obviously-beneficial. These remarks also hold for figs. \ref{fig22-02}-\ref{fig22-03} which differ from each other by the choice (but not the number) of incident angles.
\subsubsection{First and second-order retrievals as a function of the actual density contrast $\epsilon$}
\begin{figure}[ht]
\begin{center}
\includegraphics[width=0.65\textwidth]{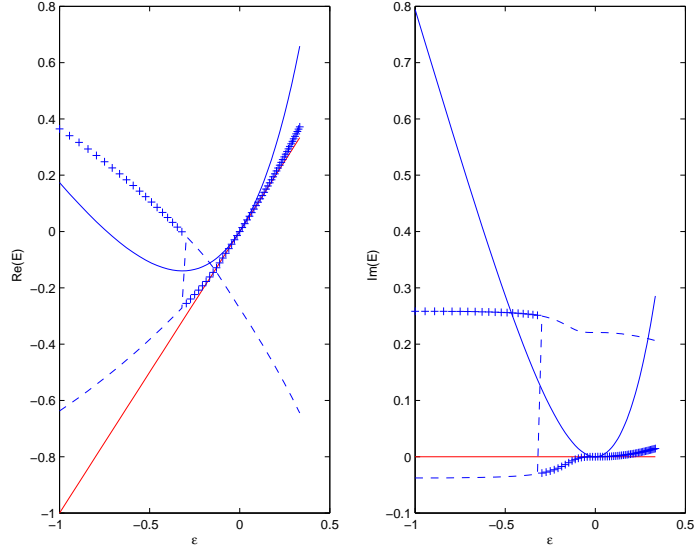}
\caption{Retrieved ($E$) mass density contrast compared to the actual ($\epsilon$) mass density contrast  as a function of the actual density contrast $\epsilon$.  The  left(right)-hand panels depict the real(imaginary) parts of  $\epsilon$ (red-----), $E^{(1)}$ (blue ------), $E^{(2)-}$ (blue - - - -), $E^{(2)+}$ (blue + + + +). Case $f=20000~Hz$, $c^{[1]}=1700-210i~ms^{-1}$, $h=0.2~m$, $\boldsymbol{\Theta}^{i}=(0^{\circ}, 30^{\circ},5)$.}
\label{fig24-01}
\end{center}
\end{figure}
\begin{figure}[ptb]
\begin{center}
\includegraphics[width=0.65\textwidth]{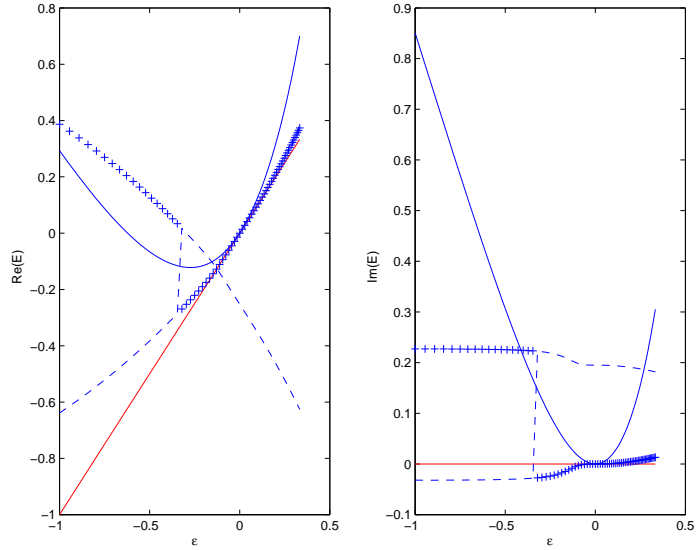}
\caption{Same as fig. \ref{fig24-01} except that  $\boldsymbol{\Theta}^{i}=(0^{\circ}, 0^{\circ},1)$.}
\label{fig24-02}
\end{center}
\end{figure}
\clearpage
\newpage
What changes between the two figures \ref{fig24-01} and \ref{fig24-02} is again $\boldsymbol{\Theta}^{i}$ which is seen to have a small effect on the first-order, but no visible effect on the second-order retrievals. Otherwise, the patterns are the same as previously: first-order retrieval rather far from the actual $\epsilon$ except for very small $\epsilon$, and double-branch second-order retrievals, the 'bad' one of which is really way off from $\epsilon$ except for very small $\epsilon$, and the 'good' one of which is quite close to $\epsilon$, especially in the interval $\epsilon\approx[-0.2,0.2]$, with the $+$ and $-$ solutions suddenly shifting between these two branches as $\epsilon$ increases. Due to this latter behavior there does not appear to exist a reliable way to spot the 'good' retrieval for a given $\epsilon$. This is, in fact, only possible when one disposes of rather narrow-range a priori information (such as: $E$ should lie in the interval $[E_{min}, E_{max}]$, with the difference of the two values being as small as possible) as to where the retrieved parameter should lie. However, if the trial model is rather crude, as in the case of the first-order model, the single retrieved solution might lie outside the range $[E_{min}, E_{max}]$ in which case one would be led to the conclusion that the retrieval (with this crude model) is impossible, this being an incorrect conclusion for a wider range  $[E_{min}, E_{max}]$, so that one should be cautious about the use of a priori information for the retrieval.
\subsubsection{First and second-order retrievals as a function of the layer thickness $h$}
\begin{figure}[ht]
\begin{center}
\includegraphics[width=0.65\textwidth]{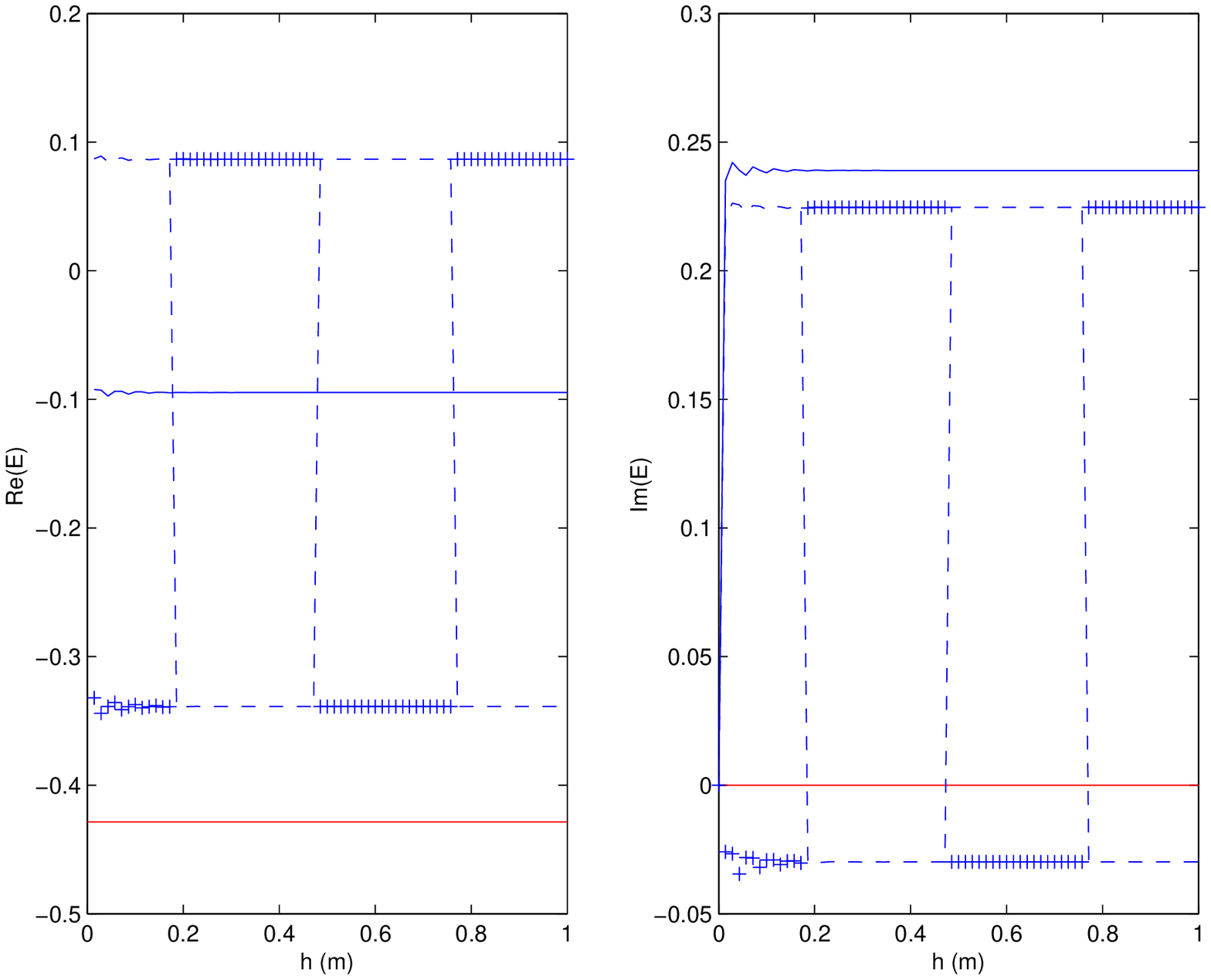}
\caption{Retrieved ($E$) mass density contrast compared to the actual ($\epsilon$) mass density contrast as a function of the layer thickness $h$.  The  left(right)-hand panels depict the real(imaginary) parts of  $\epsilon$ (red-----), $E^{(1)}$ (blue ------), $E^{(2)-}$ (blue - - - -), $E^{(2)+}$ (blue + + + +). Case $f=20000~Hz$, $\rho^{[1]}=700~Kgm^{-3}$, $c^{[1]}=1700-210i~ms^{-1}$, $\boldsymbol{\Theta}^{i}=(0^{\circ}, 0^{\circ},1)$.}
\label{fig26-01}
\end{center}
\end{figure}
\begin{figure}[ptb]
\begin{center}
\includegraphics[width=0.65\textwidth]{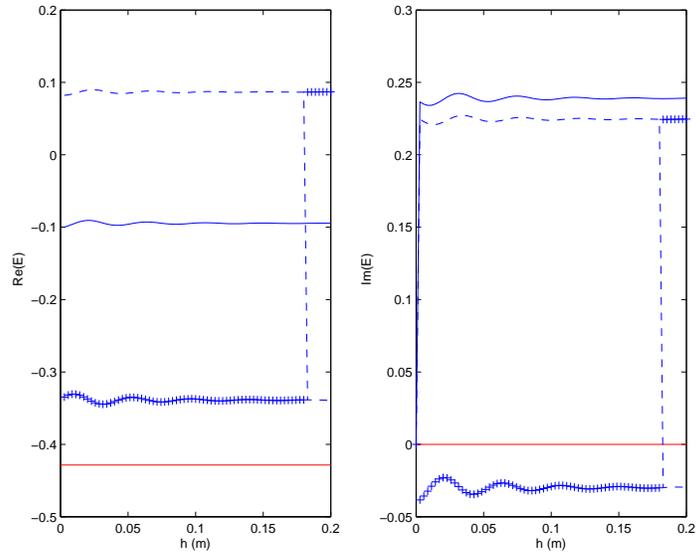}
\caption{Same as fig. \ref{fig26-01} for a different range of $h$.}
\label{fig26-02}
\end{center}
\end{figure}
\begin{figure}[ptb]
\begin{center}
\includegraphics[width=0.65\textwidth]{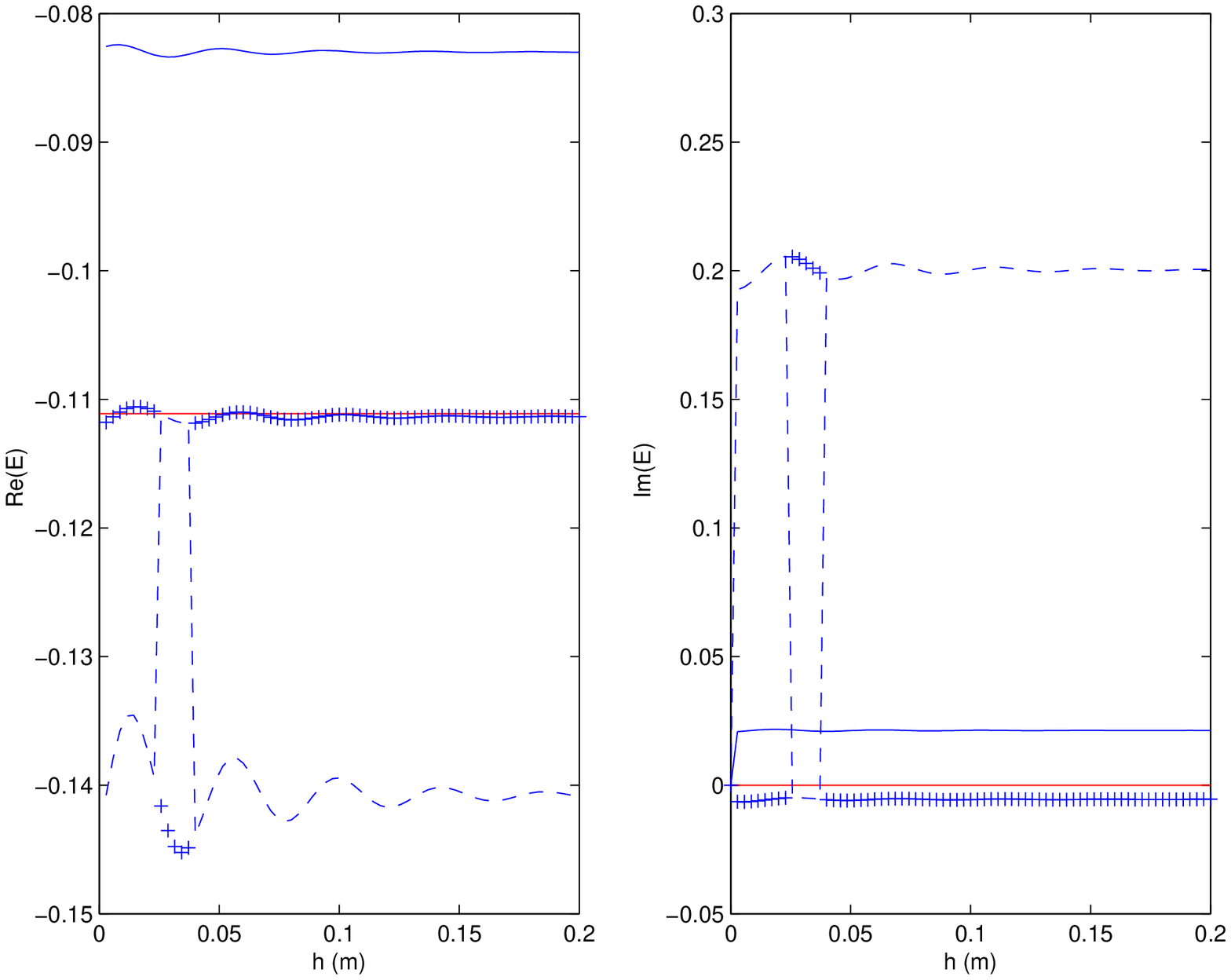}
\caption{Same as fig. \ref{fig26-02} except that  $\rho^{[1]}=900~Kgm^{-3}$.}
\label{fig26-03}
\end{center}
\end{figure}
\begin{figure}[ptb]
\begin{center}
\includegraphics[width=0.65\textwidth]{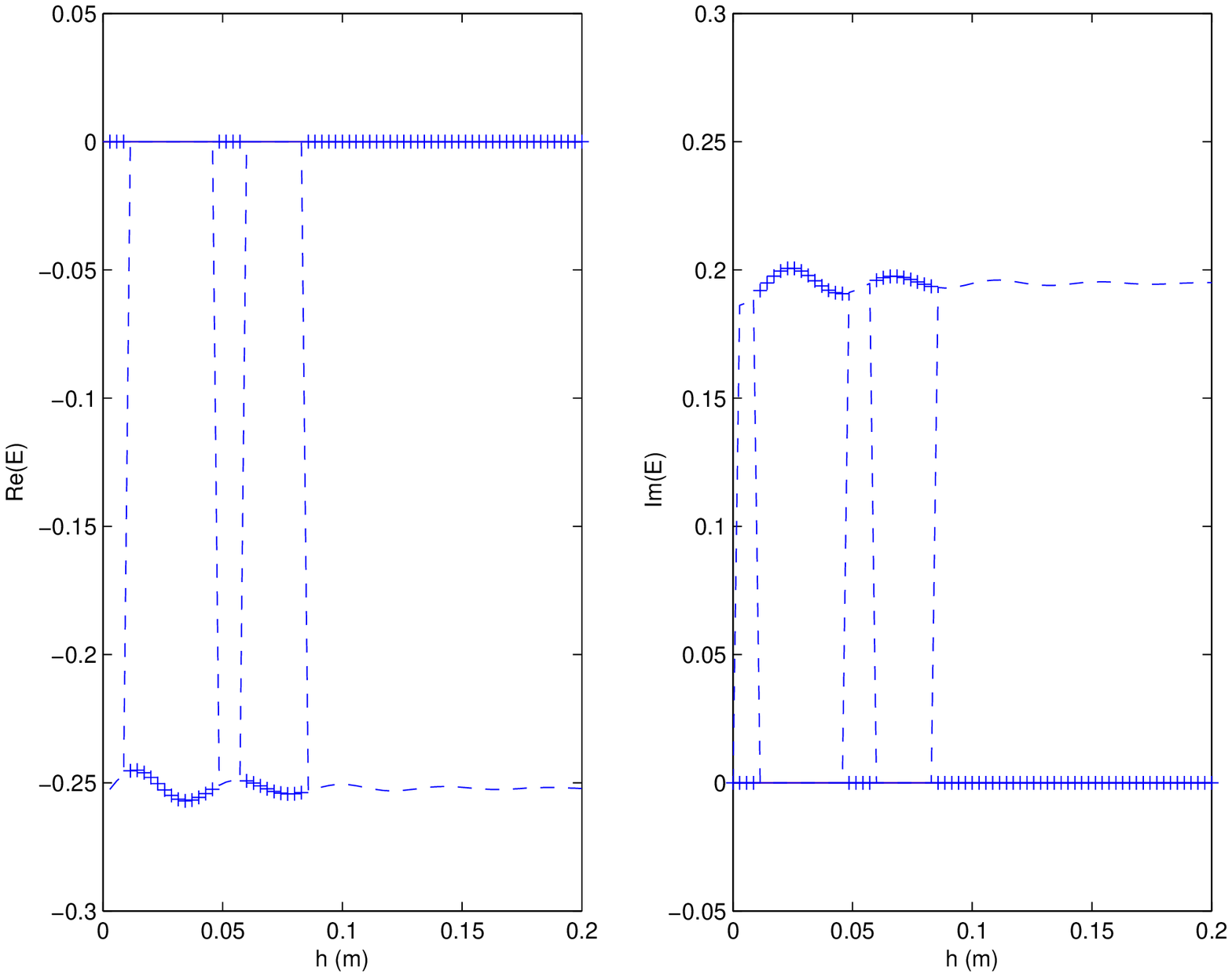}
\caption{Same as fig. \ref{fig26-02} except that  $\rho^{[1]}=1000~Kgm^{-3}$.}
\label{fig26-04}
\end{center}
\end{figure}
\begin{figure}[ptb]
\begin{center}
\includegraphics[width=0.65\textwidth]{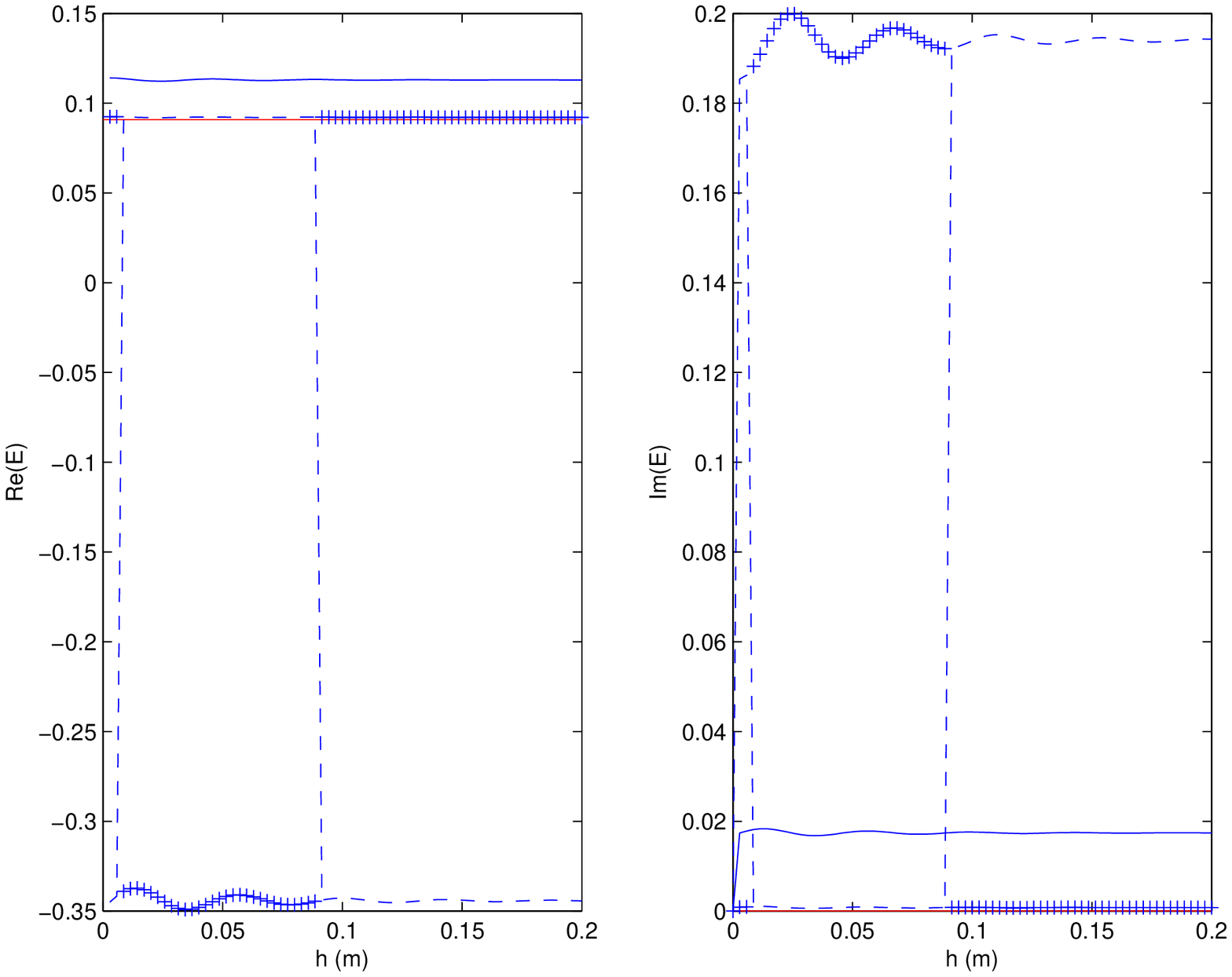}
\caption{Same as fig. \ref{fig26-02} except that  $\rho^{[1]}=1100~Kgm^{-3}$.}
\label{fig26-05}
\end{center}
\end{figure}
\begin{figure}[ptb]
\begin{center}
\includegraphics[width=0.65\textwidth]{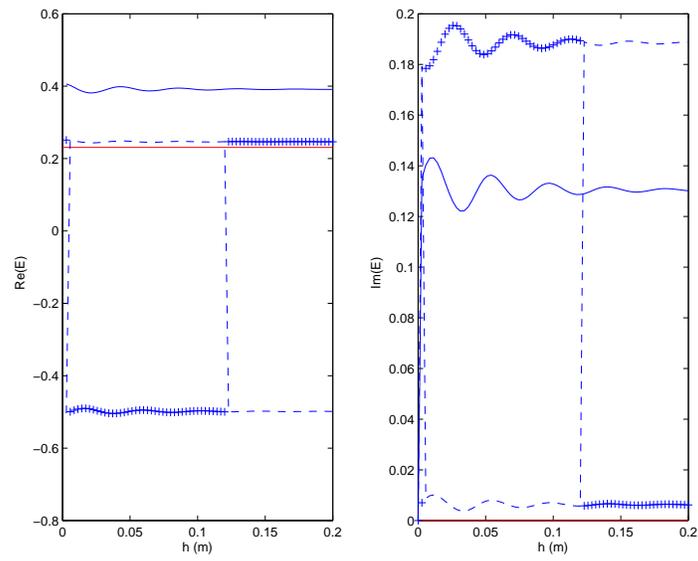}
\caption{Same as fig. \ref{fig26-02} except that  $\rho^{[1]}=1300~Kgm^{-3}$.}
\label{fig26-06}
\end{center}
\end{figure}
\clearpage
\newpage
What changes between fig. \ref{fig26-01} and fig. \ref{fig26-02} is the range of $h$, the latter figure constituting a sort of zoom to highlight the dispersive (with respect to $h$) behavior of the retrievals.

What changes between fig. \ref{fig26-02} and fig. \ref{fig26-06} is $\rho^{[1]}$, but this does not change the general pattern of behavior. Again, the dispersive (with respect to $h$) effects are quite apparent, more so for the first-order and second-order 'bad' branch retrievals than for the second-order 'good' branch retrievals, thus providing a possible manner of spotting the retrieval branch that is closest to the actual mass density contrast $\epsilon$.

In the retrievals of figs. \ref{fig26-01}-\ref{fig26-06} it should be noticed that we have renounced at considering more than one incident angle realization because the figures in the previous subsections showed that  taking more than one incident angle into account does not consistently improve the accuracy nor the uniqueness of the retrievals. For this reason, we shall not, in the next section, take more than one incident angle realization into account in our attempts to retrieve the real part of the layer wavespeed.
\subsection{Retrieval of the real part of the layer wavespeed}
Again, the simulated (via the DD-SOV model) field in the transmission half-space is
\begin{equation}\label{inv-210}
p(x,z)=a^{[2]}\exp[i(k_{y}^{i[0]}x-k_{y}^{i[0]}y)]
~.
\end{equation}
and once more, we assume the data is collected at $x=0,~y=-h$, so that the simulated pressure data is
\begin{equation}\label{inv-220}
p(0,-h)=a^{[2]}(c^{[1]'})\exp[ik_{y}^{i[0]}h]
~.
\end{equation}
wherein $c^{[1]'}=\Re\left(c^{[1]}\right)$. The trial fields, at the same location, for the three trial models, are,
due to the assumption of the identity of the priors, $K_{y}^{i[0]}=k_{y}^{i[0]}$, $H=h$:
\begin{equation}\label{inv-240}
P^{(l)}(0,-h)=A^{[2](l)}(C^{[1]'})\exp[ik_{y}^{i[0]}h]~;~l=0,1,2
~,
\end{equation}
wherein $C^{[1]'}=\Re\left(C^{[1]'}\right)$ is the trial parameter and
\begin{equation}\label{inv-245}
A^{[2](l)}=\sum_{j=0}^{l}A_{j}^{[2]}E^{j}~;~l=0,1,2
~,
\end{equation}
with the coefficients $A_{j}^{[2]}$ not depending on $E$. Now, the density contrast is a prior, and due to our assumption of the identity of the priors of the data simulation and trial models, $E^{(l)}=\epsilon~;~l=0,1,2$, so that the previous equation becomes
\begin{equation}\label{inv-247}
A^{[2](l)}=\sum_{j=0}^{l}A_{j}^{[2]}\epsilon^{j}~;~l=0,1,2
~,
\end{equation}
wherein the coefficients $A_{j}^{[2]}$ depend on the to-be-retrieved trial wavespeed $C^{[1](l)'}=\Re\left(C^{[1](l)}\right)$ in a complicated manner. For this reason, it is not obvious how to solve for $C^{[1](l)'}$ by the algebraic method previously employed for retrieving the density contrast.

However, the  cost functions are still a function of the difference of $P^{(l)}(0,-h)$ from $p(0,-h)$
\begin{equation}\label{inv-250}
\mathcal{K}^{(l)}(C^{[1](l)'})=\mathcal{F}\left(P^{(l)}(0,-h)-p(0,-h)\right)~;~l=0,1,2
~,
\end{equation}
the object being to find the trial layer wavespeed $C^{[1](l)'}$ that minimizes $\mathcal{K}^{(l)}$ for each $l$.
More often than not, in the inverse problem context, $\mathcal{F}$ is chosen to be $\mathcal{F}=\|(P^{(l)}(0,-h)-p(0,-h)\|^{2}$ and the optimal trial solution is the one for which the cost function
\begin{equation}\label{inv-260}
\mathcal{K}^{(l)}(C^{[1](l)'})=\|P^{(l)}(0,-h)-p(0,-h)\|^{2}
~,
\end{equation}
attains its absolute minimum.

Actually, searching for an absolute minimum is not as simple as it seems since the cost function usually has more than one minima and the number of these minima usually increases with the size of the search interval. Shrinking the search interval might be an appealing way to solve this problem, but, in fact, it is dangerous in that it increases the probability of excluding a priori a series of minima, one of which might be the absolute minimum one is looking for. In fact, there is no reasonable solution to this non-uniqueness problem unless one is sure at the outset that the to-be-retrieved parameter {\it must} lie between a given lower and upper bound.

A final remark: the proposed method for solving the wavespeed-retrieval problem is numerical and algorithmic in nature, in contrast to the mathematical, explicit, manner for solving the mass density contrast-retrieval problem; furthermore we have limited it to a single incident angle realization due to the inconclusive  previously-found results as to the possible amelioration of the retrievals afforded by multiple incident angle realizations.
\subsection{Numerical results for the retrieval of the real part of the layer wavespeed}
The aim of the computations was to compare numerically the actual real part of the wavespeed $\Re\left(c^{[1]}\right)$ in the layer to the retrievals thereof: (a) $\Re\left(C^{[1](0)}\right)$ obtained via the zeroth-order $\epsilon$ (i.e., constant-density) trial model, (b) $\Re\left(C^{[1](1)}\right)$ obtained via the first-order $\epsilon$ trial model, and (c) $\Re\left(C^{[1](2)}\right)$ obtained via the second-order $\epsilon$ trial model, as a function of the various parameters $\Re\left(c^{[1]}\right)$, $\epsilon$, $f$ and $\theta^{i}$,   the other parameters (recall that we are assuming identity of the priors of the trial models to those of the simulation model) being fixed at the following values: $\rho^{[0]}=1000~Kgm^{-3}$, $c^{[0]}=1500~ms^{-1}$, $h=0.2~m$, $a^{[0]}=1$, $\theta^{i}=0^{\circ}$,  $\Im\left(c^{[1](2)}\right)=-21~ms^{-1}$.

In the following figures of this section the lowest and highest ordinate represent the lower and upper bounds of the search interval within which the real part of the layer wavespeed is sought.
\subsubsection{Retrieved real part of the layer wavespeed $C^{[1]'}$ as a function of frequency $f$}
\begin{figure}[ht]
\begin{center}
\includegraphics[width=0.65\textwidth]{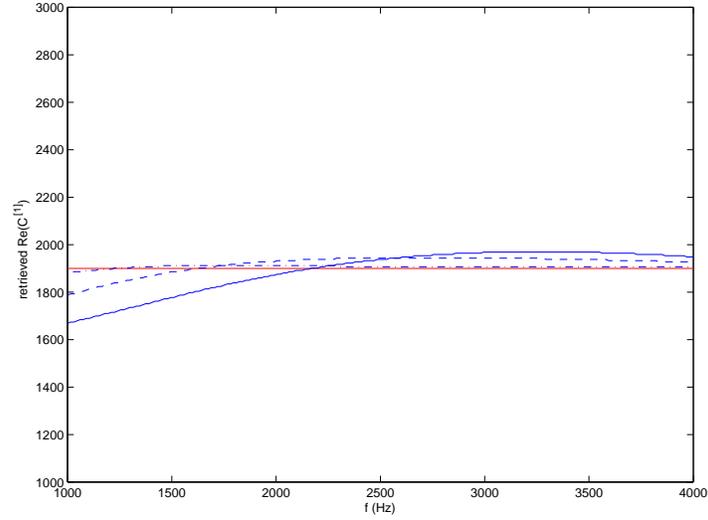}
\caption{Retrieved real part of the layer wavespeed ($C^{[1]'}$)  compared to the actual ($c^{[1]'}$) real part of the layer wavespeed  as a function of the frequency $f$.  The represented functions are: $c^{[1]'}$ (red-----), $C^{[1](0)'}$ (blue ------), $C^{[1](1)'}$ (blue - - - -), $C^{[1](2)'}$ (blue -.-.-.-). Case $\rho^{[1]}=1500~Kgm^{-3}$, $c^{[1]'}=1900~ms^{-1}$.}
\label{fig30-01}
\end{center}
\end{figure}
\begin{figure}[ptb]
\begin{center}
\includegraphics[width=0.65\textwidth]{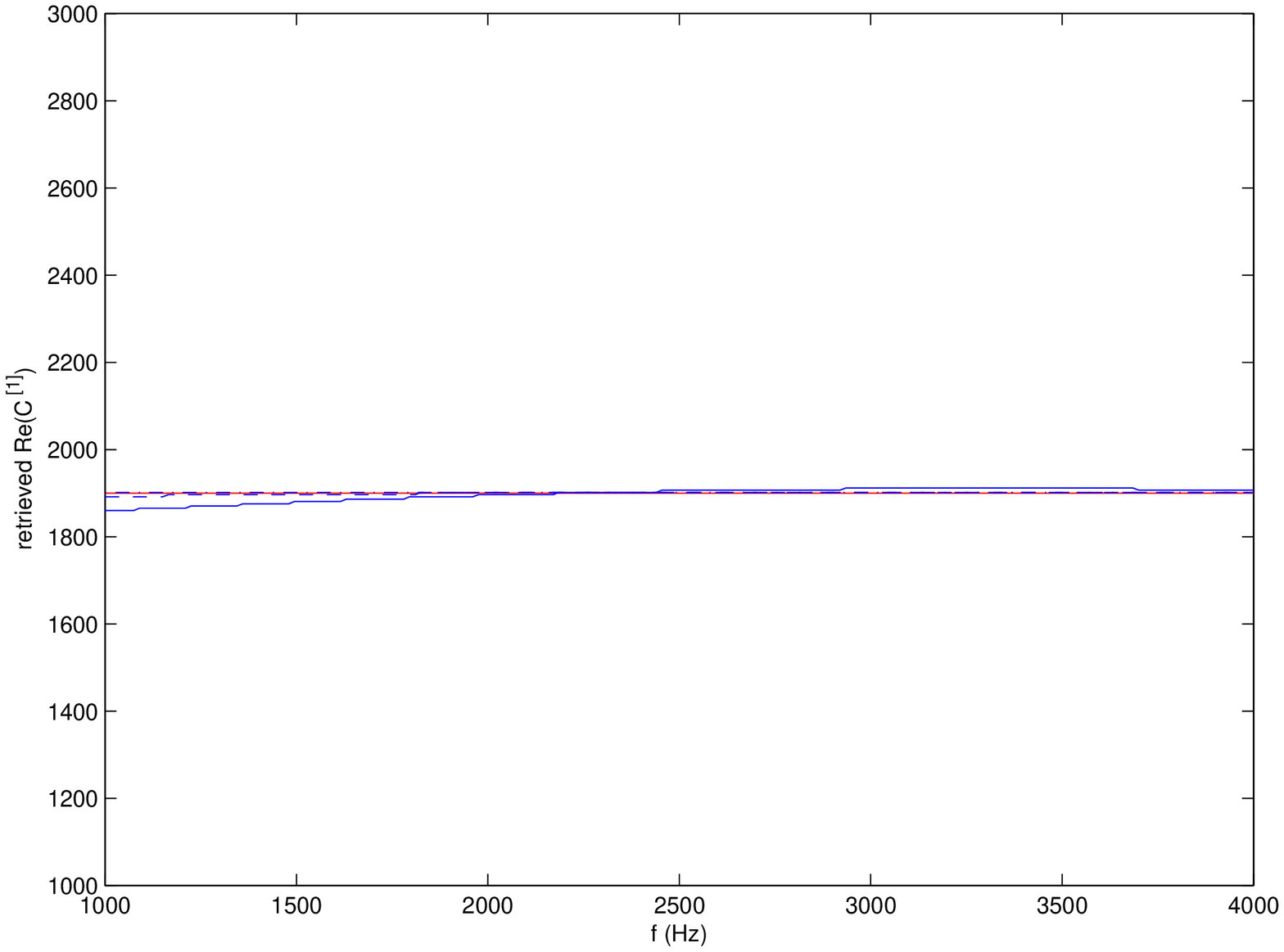}
\caption{Same as fig. \ref{fig30-01} except that $\rho^{[1]}=1100~Kgm^{-3}$.}
\label{fig30-02}
\end{center}
\end{figure}
\begin{figure}[ptb]
\begin{center}
\includegraphics[width=0.65\textwidth]{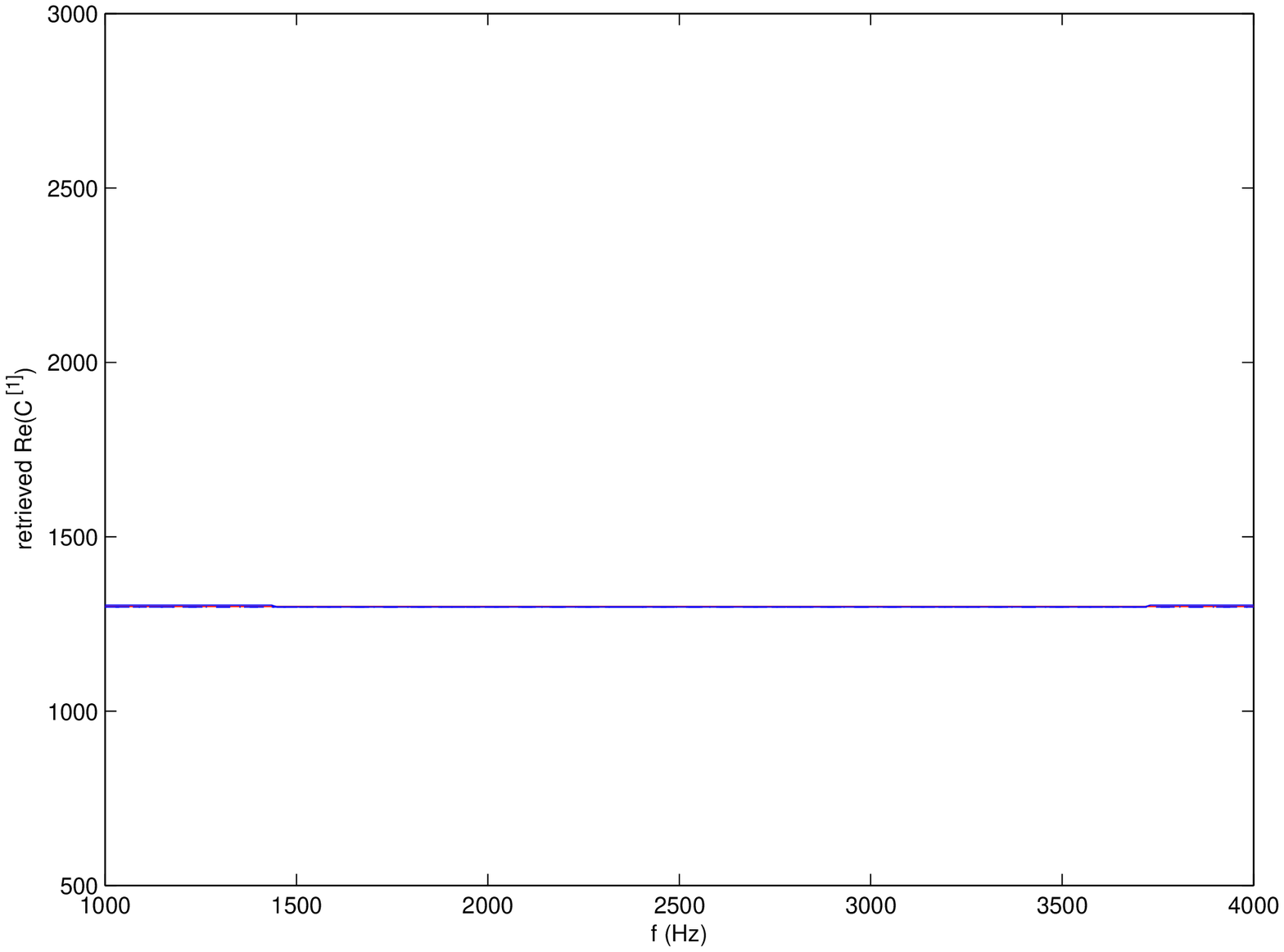}
\caption{Retrieved real part of the layer wavespeed ($C^{[1]'}$)  compared to the actual ($c^{[1]'}$) real part of the layer wavespeed  as a function of the frequency $f$.  The represented functions are: $c^{[1]'}$ (red-----), $C^{[1](0)'}$ (blue ------), $C^{[1](1)'}$ (blue - - - -), $C^{[1](2)'}$ (blue -.-.-.-). Case $\rho^{[1]}=1100~Kgm^{-3}$, $c^{[1]'}=1300~ms^{-1}$.}
\label{fig31-01}
\end{center}
\end{figure}
\begin{figure}[ptb]
\begin{center}
\includegraphics[width=0.65\textwidth]{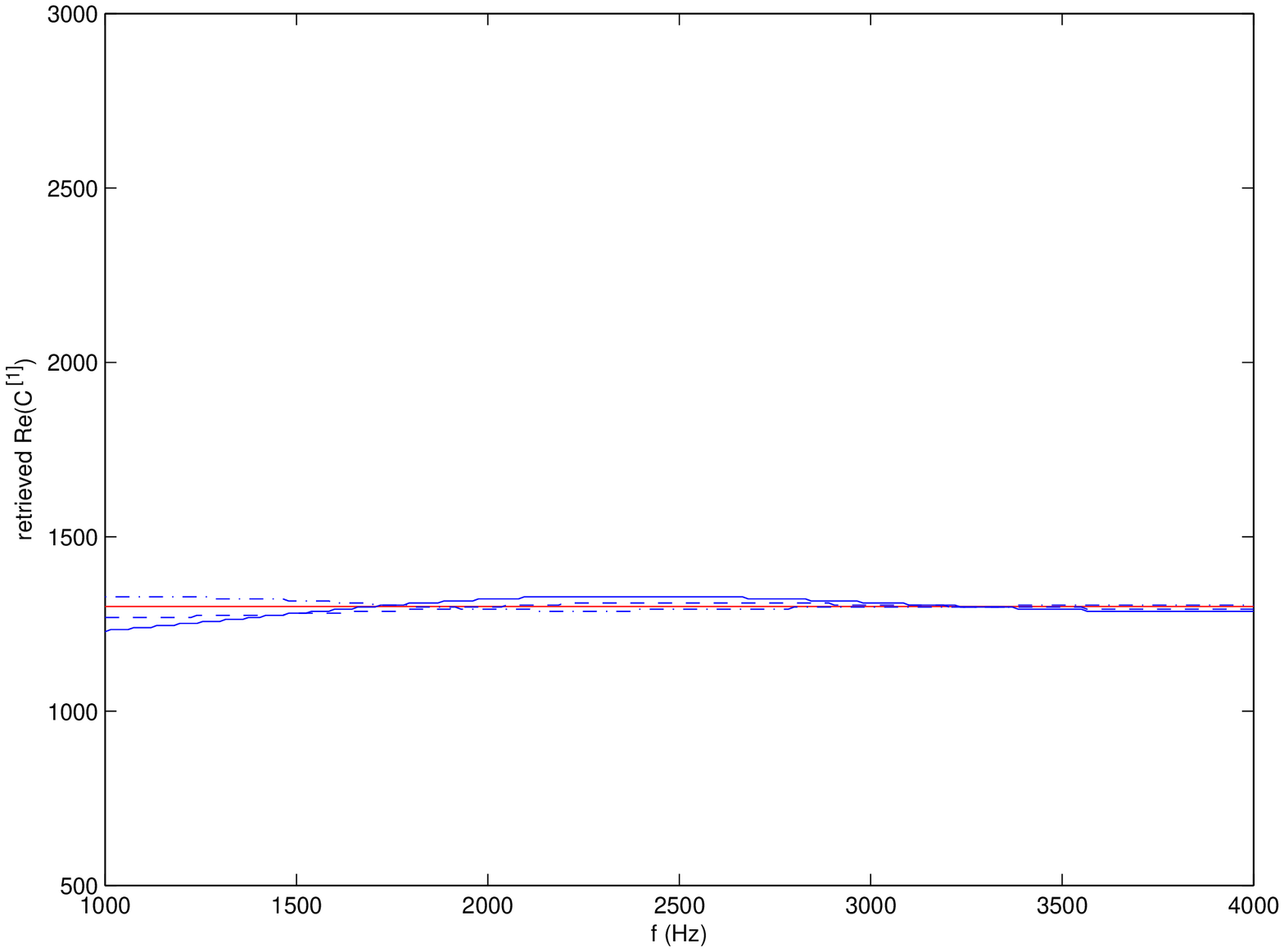}
\caption{Same as fig. \ref{fig31-01} except that $\rho^{[1]}=700~Kgm^{-3}$.}
\label{fig31-02}
\end{center}
\end{figure}
\begin{figure}[ptb]
\begin{center}
\includegraphics[width=0.65\textwidth]{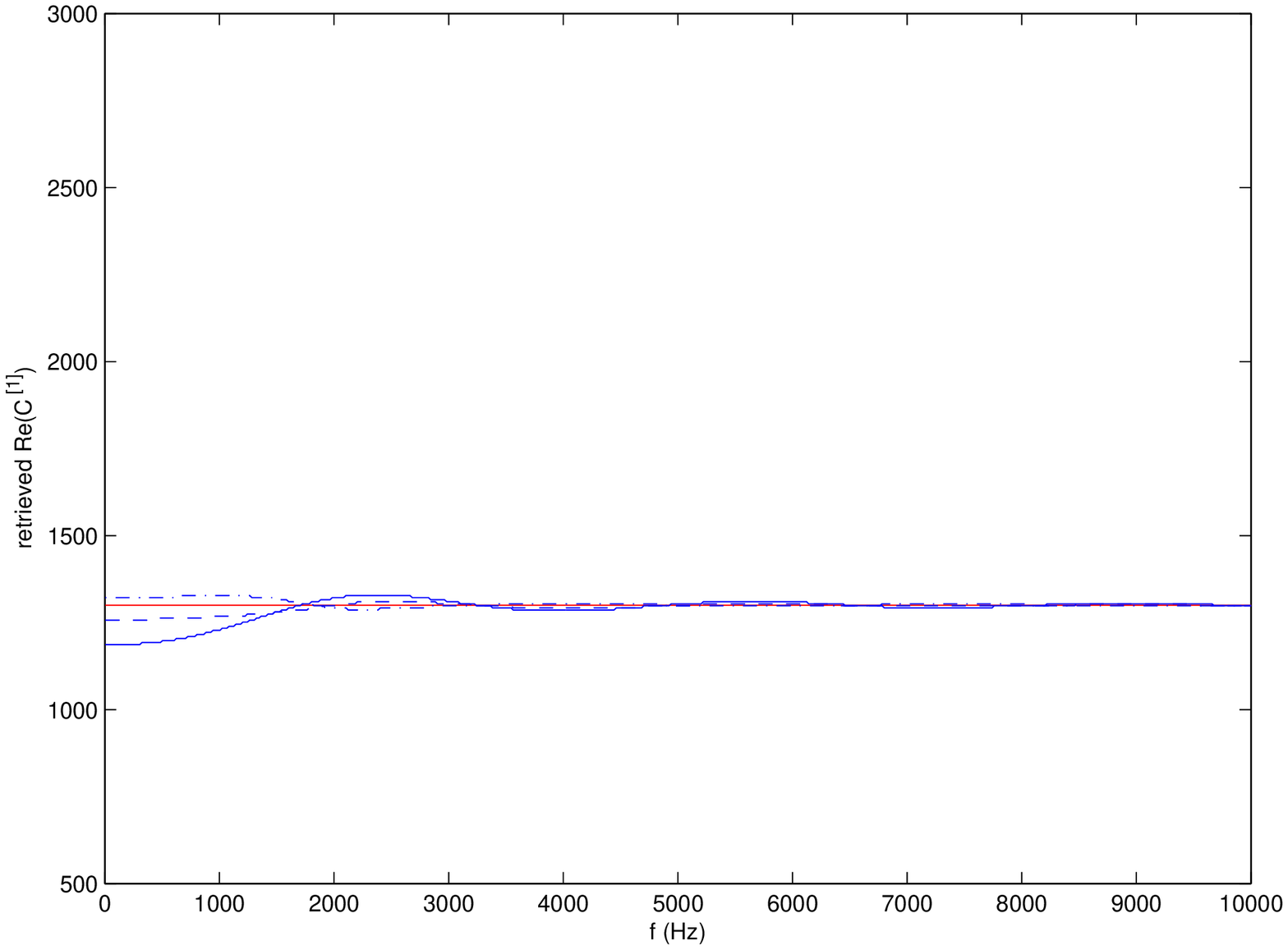}
\caption{Retrieved real part of the layer wavespeed ($C^{[1]'}$)  compared to the actual ($c^{[1]'}$) real part of the layer wavespeed  as a function of the frequency $f$.  The represented functions are: $c^{[1]'}$ (red-----), $C^{[1](0)'}$ (blue ------), $C^{[1](1)'}$ (blue - - - -), $C^{[1](2)'}$ (blue -.-.-.-). Case $\rho^{[1]}=700~Kgm^{-3}$, $c^{[1]'}=1300~ms^{-1}$.}
\label{fig32-01}
\end{center}
\end{figure}
\clearpage
\begin{figure}[ptb]
\begin{center}
\includegraphics[width=0.65\textwidth]{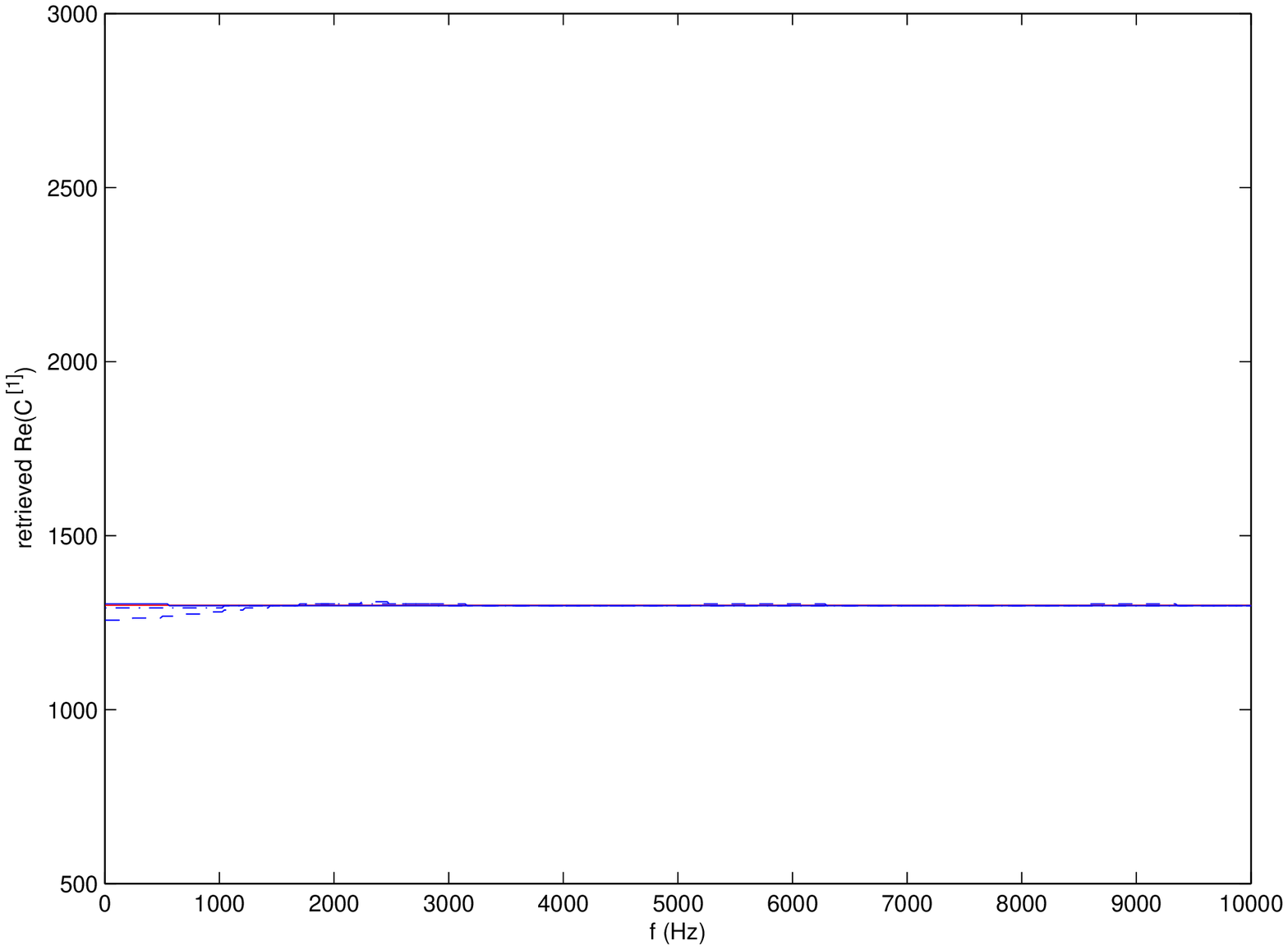}
\caption{Same as fig. \ref{fig32-01} except that $\rho^{[1]}=1300~Kgm^{-3}$,}
\label{fig32-02}
\end{center}
\end{figure}
\clearpage
\newpage
What changes between fig. \ref{fig30-01} and fig. \ref{fig30-02} is $\rho^{[1]}$. Throughout the (low) frequency range of the graphs, the zeroth order trial model is observed give inaccurate retrievals whereas  the first-order trial model is more accurate and the second-order model coincident with the actual $c^{[1]'}$. The closer is $\rho^{[1]}$ to $\rho^{[0]}$ (i.e., the closer is $\epsilon$ to zero), the more-accurate are all three retrievals.

What changes between fig. \ref{fig30-02} and fig. \ref{fig31-01} is $c^{[1]'}$. Owing to the small value of $\epsilon$, this change has little effect on the observed highly-accurate retrievals, particularly those obtained via the first and second order trial models.

What changes between fig. \ref{fig31-01} and fig. \ref{fig31-02} is $\rho^{[1]}$.
The same comments as for figs. \ref{fig30-01}-\ref{fig30-02} apply here except that the negative mass density contrast (case of fig. \ref{fig31-02}) introduces a more-pronounced dispersive behavior in the first and second-order retrievals.

What changes between fig. \ref{fig31-02} and fig. \ref{fig32-01} is the frequency range which in the second figure now extends from very small to high frequencies.
 The  dispersive behavior of the retrievals, largest for the zeroth and smallest for the second-order trial models means that this dispersion  is due to trial model error. Notice should be taken of the fact that all three retrievals tend asymptotically (i.e., for large frequencies) towards the same value $c^{[1]'}$ which means that retrieval inaccuracy, especially that resulting from the use of the zeroth-order trial model, is essentially a low-frequency problem.

What changes between fig. \ref{fig32-01} and fig. \ref{fig32-02} is $\rho^{[1]}$.
The latter figure applies to positive $\epsilon$ and the former to negative $\epsilon$ (both corresponding to the same $|\epsilon|$), and the comparison of the two indicates that the negative sign of $\epsilon$ is  an aggravating factor for the inaccuracy of the (especially the zeroth-order) retrievals.

Thus, to resume: the dispersive nature of the retrievals of the real part of the layer wavespeed  is due to (i.e., is induced by)  trial model error rather than by what is often thought to be of physical origin \cite{wi16c}, and these dispersive effects appear to be exacerbated by certain choices of the priors such as largely-negative $\epsilon$.
\subsubsection{Retrieved real part of the layer wavespeed $C^{[1]'}$ as a function of $c^{[1]}$}
\begin{figure}[ht]
\begin{center}
\includegraphics[width=0.65\textwidth]{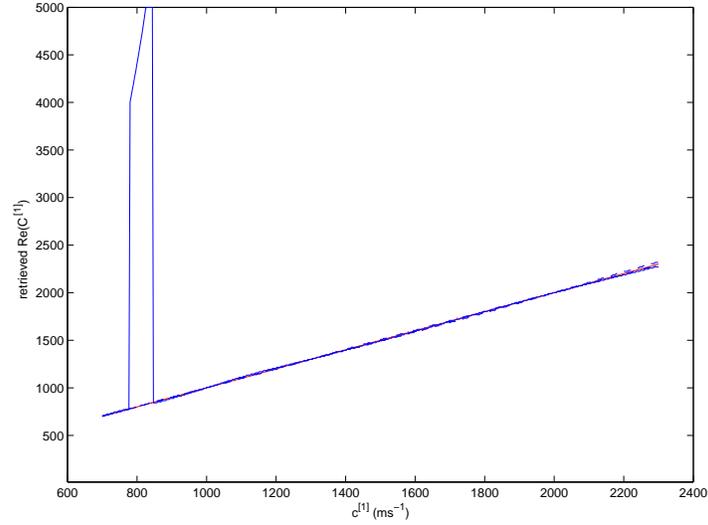}
\caption{Retrieved real part of the layer wavespeed ($C^{[1]'}$)  compared to the actual ($c^{[1]'}$) real part of the layer wavespeed  as a function of  the actual real part of the layer wavespeed $c^{[1]'}$.  The represented functions are: $c^{[1]'}$ (red-----), $C^{[1](0)'}$ (blue ------), $C^{[1](1)'}$ (blue - - - -), $C^{[1](2)'}$ (blue -.-.-.-). Case $f=5000~Hz$, $\rho^{[1]}=700~Kgm^{-3}$.}
\label{fig34-01}
\end{center}
\end{figure}
\begin{figure}[ptb]
\begin{center}
\includegraphics[width=0.65\textwidth]{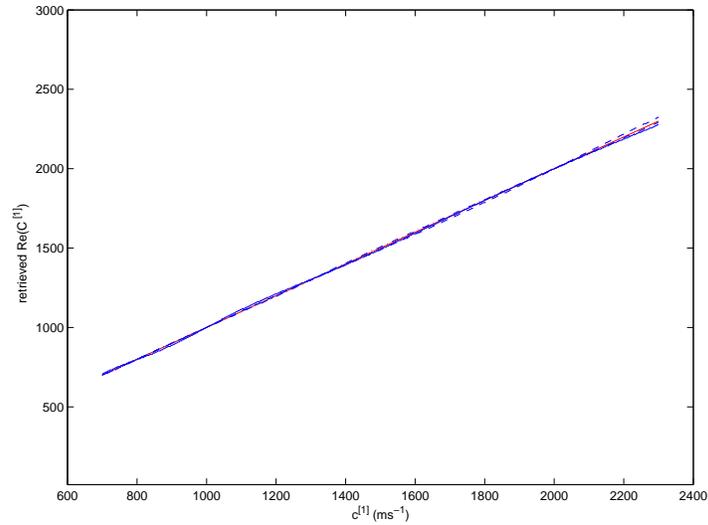}
\caption{Same as fig. \ref{fig34-01} except that search interval of $C^{[1]'}$ is narrowed.}
\label{fig34-02}
\end{center}
\end{figure}
\begin{figure}[ptb]
\begin{center}
\includegraphics[width=0.65\textwidth]{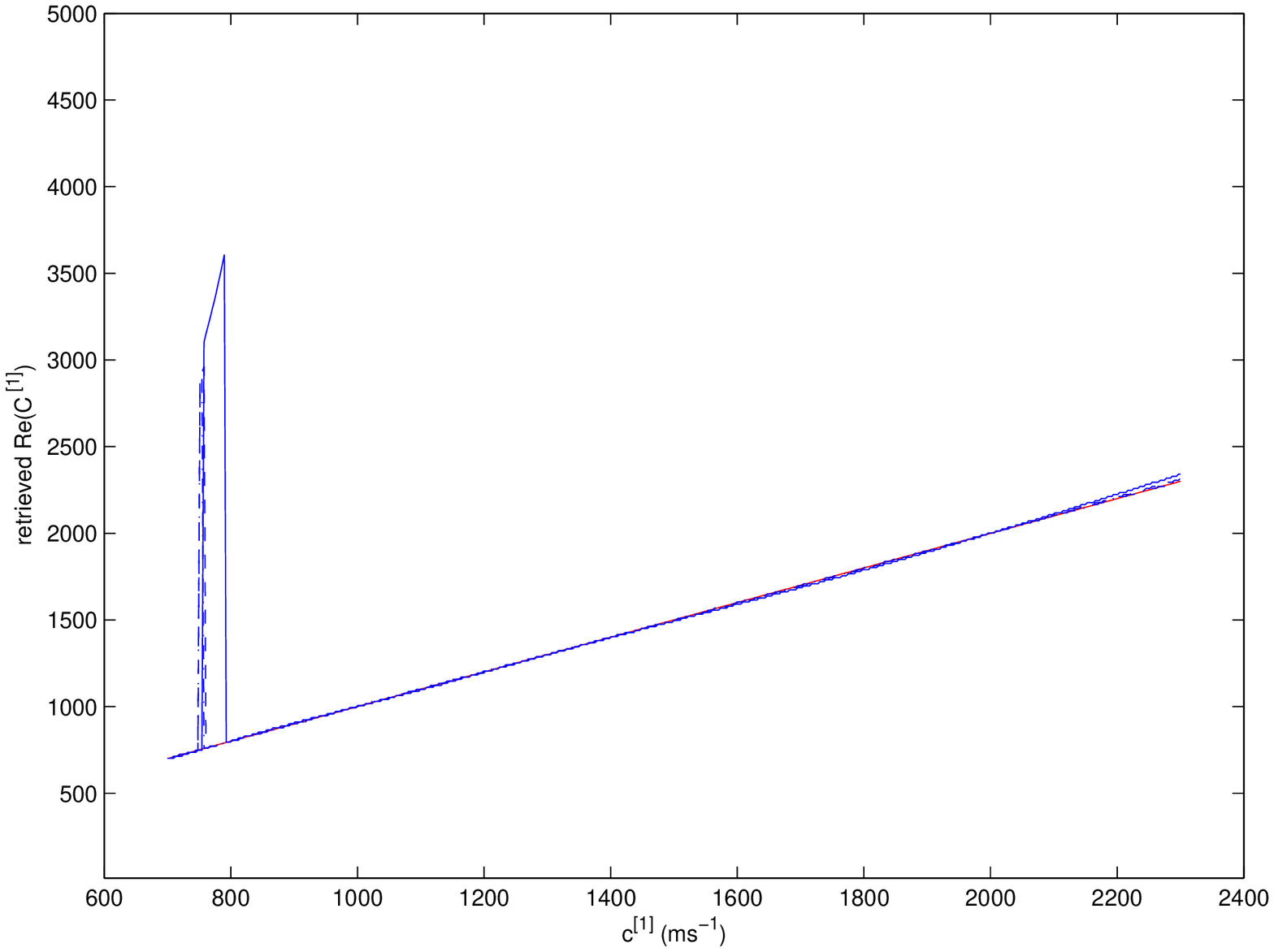}
\caption{Same as fig. \ref{fig34-01} for $\rho^{[1]}=1300~Kgm^{-3}$.}
\label{fig34-03}
\end{center}
\end{figure}
\begin{figure}[ptb]
\begin{center}
\includegraphics[width=0.65\textwidth]{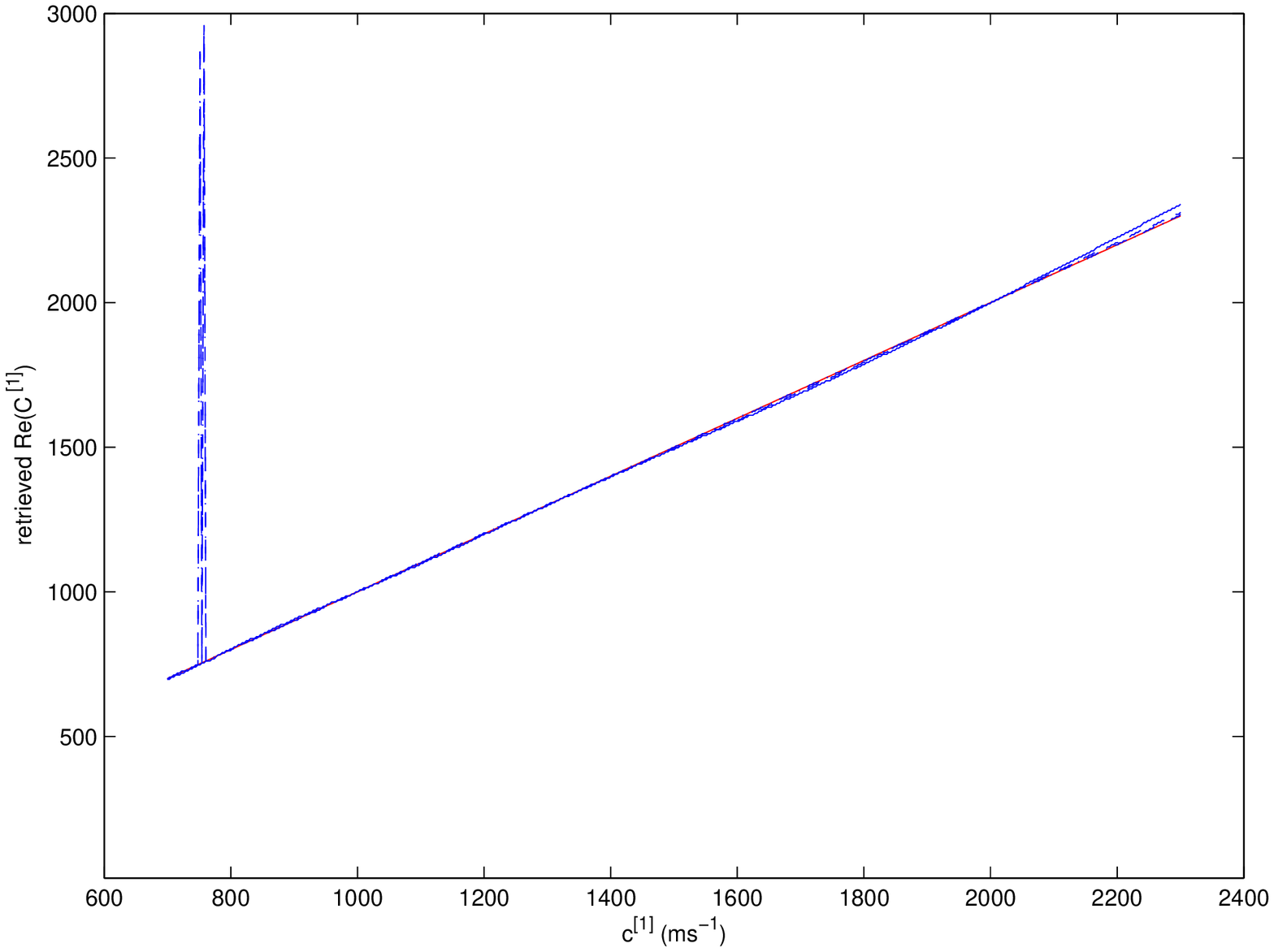}
\caption{Same as fig. \ref{fig34-03} for a narrower range of  the search interval.}
\label{fig34-04}
\end{center}
\end{figure}
\begin{figure}[ptb]
\begin{center}
\includegraphics[width=0.65\textwidth]{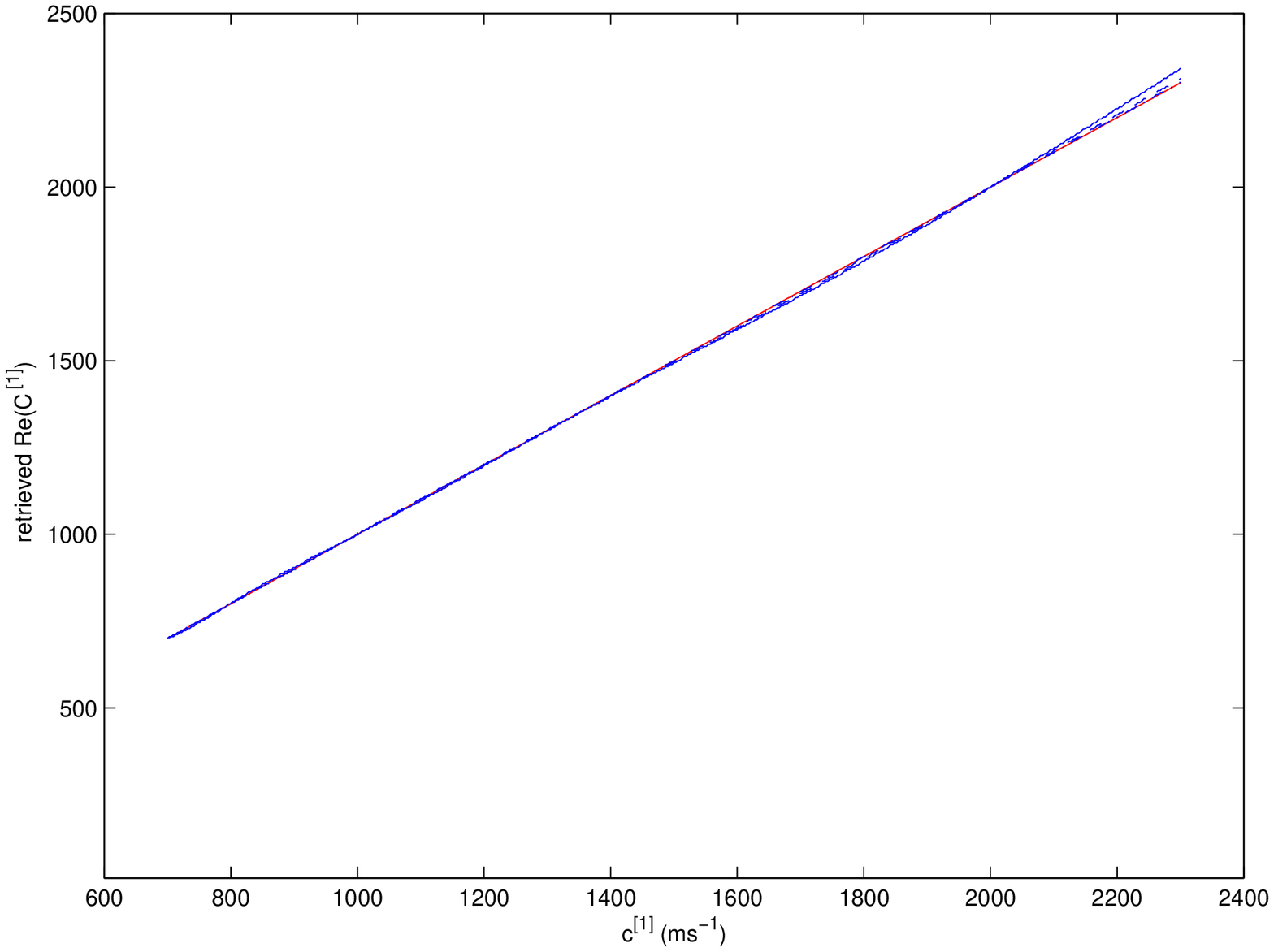}
\caption{Same as fig. \ref{fig34-04} for a narrower range of  the search interval.}
\label{fig34-05}
\end{center}
\end{figure}
\begin{figure}[ptb]
\begin{center}
\includegraphics[width=0.65\textwidth]{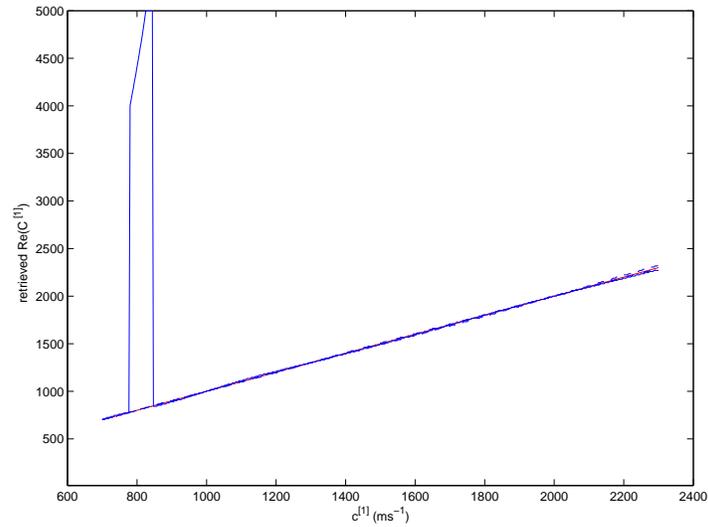}
\caption{Retrieved real part of the layer wavespeed ($C^{[1]'}$)  compared to the actual ($c^{[1]'}$) real part of the layer wavespeed   as a function of  the actual real part of the layer wavespeed $c^{[1]'}$.  The represented functions are: $c^{[1]'}$ (red-----), $C^{[1](0)'}$ (blue ------), $C^{[1](1)'}$ (blue - - - -), $C^{[1](2)'}$ (blue -.-.-.-).  Case $f=5000~Hz$, $\rho^{[1]}=700~Kgm^{-3}$. The search range is again very wide.}
\label{fig35-01}
\end{center}
\end{figure}
\begin{figure}[ptb]
\begin{center}
\includegraphics[width=0.65\textwidth]{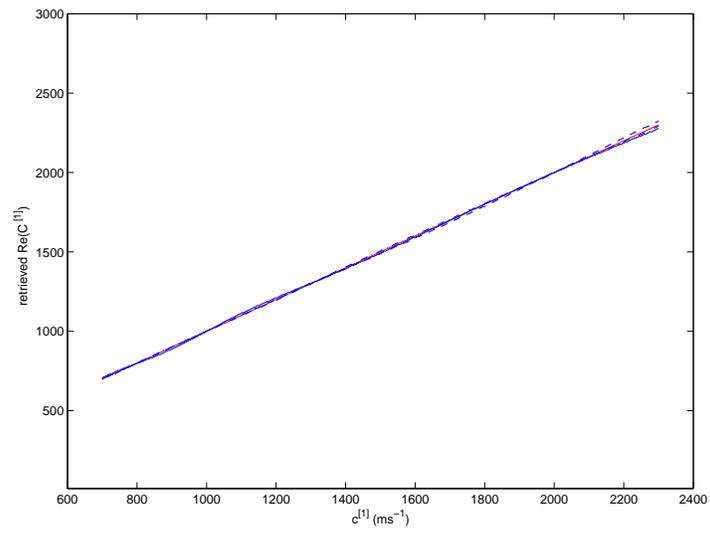}
\caption{Same as fig. \ref{fig35-01} except that the search range has been narrowed.}
\label{fig35-02}
\end{center}
\end{figure}
\clearpage
\newpage
What changes between fig. \ref{fig34-01} and fig. \ref{fig34-02} the search interval of $C^{[1]'}$ which is being narrowed.
 In fig. \ref{fig34-01} we observe  branch-hopping between two retrieval branches of the type formerly observed for the retrieval, via the second-order trial model, of the mass density contrast. But now the hopping occurs for all three trial models employed to obtain the retrievals. One of these branches appears to be 'bad'  and the second is 'good' insofar as it is linear (as is the actual red line). Narrowing the search interval  in fig.  \ref{fig34-02} results in the disappearance of the higher-lying branch, thus justifying its qualification as being 'bad'. In  fig.  \ref{fig34-02} we also observe near-coincidence of the three retrievals with the actual $c^{[1]'}$.

For a value of $\rho^{[1]}$ that is  different from its value in the former two figures, what changes between fig. \ref{fig34-03} and fig. \ref{fig34-05} is  again the search interval which is progressively narrowed. At first, it does not enable the elimination of the 'bad' branch, but finally succeeds in doing this in fig. \ref{fig34-05}  in which we again observe near-coincidence of the three retrievals with the actual $c^{[1]'}$.

For a value of $\rho^{[1]}$ that is  different from its value in the preceding three figures, what changes in figs. \ref{fig35-01}-\ref{fig35-02} is the search interval which is being narrowed. This narrowing produces the sought-for effect of eliminating the 'bad' retrieval branch, so that in fig. \ref{fig35-02} we again observe near-coincidence of the three retrievals with the actual $c^{[1]'}$.
\subsubsection{Retrieved real part of the layer wavespeed $C^{[1]'}$ as a function of $\epsilon$}
\begin{figure}[ht]
\begin{center}
\includegraphics[width=0.65\textwidth]{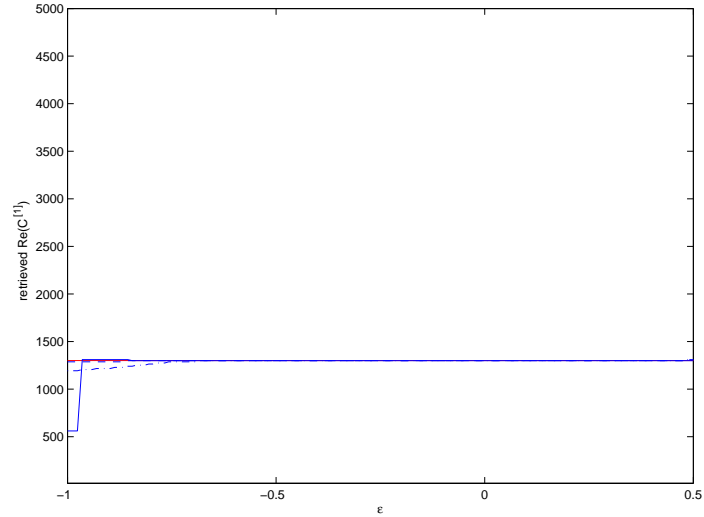}
\caption{Retrieved real part of the layer wavespeed ($C^{[1]'}$)  compared to the actual ($c^{[1]'}$) real part of the layer wavespeed  as a function of  the actual mass density contrast   $\epsilon$.  The represented functions are: $c^{[1]'}$ (red-----), $C^{[1](0)'}$ (blue ------), $C^{[1](1)'}$ (blue - - - -), $C^{[1](2)'}$ (blue -.-.-.-). Case $f=5000~Hz$, $c^{[1]}=1300~ms^{-1}$.}
\label{fig38-01}
\end{center}
\end{figure}
\begin{figure}[ht]
\begin{center}
\includegraphics[width=0.65\textwidth]{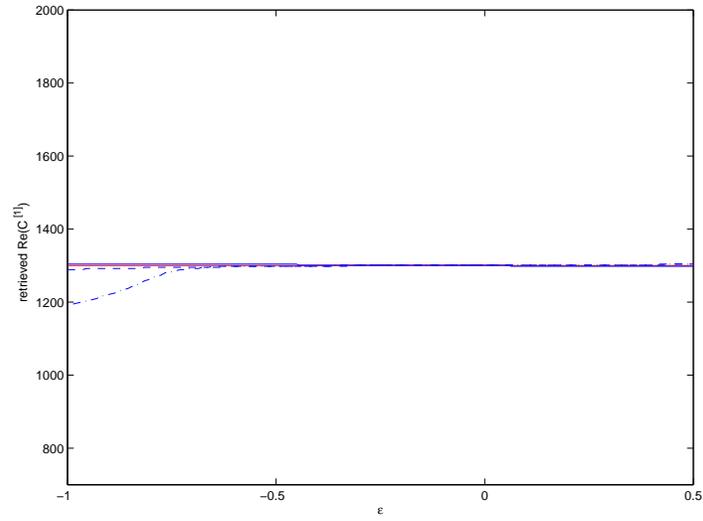}
\caption{Same as fig. \ref{fig38-01} for a smaller search interval.}
\label{fig38-02}
\end{center}
\end{figure}
\begin{figure}[ht]
\begin{center}
\includegraphics[width=0.65\textwidth]{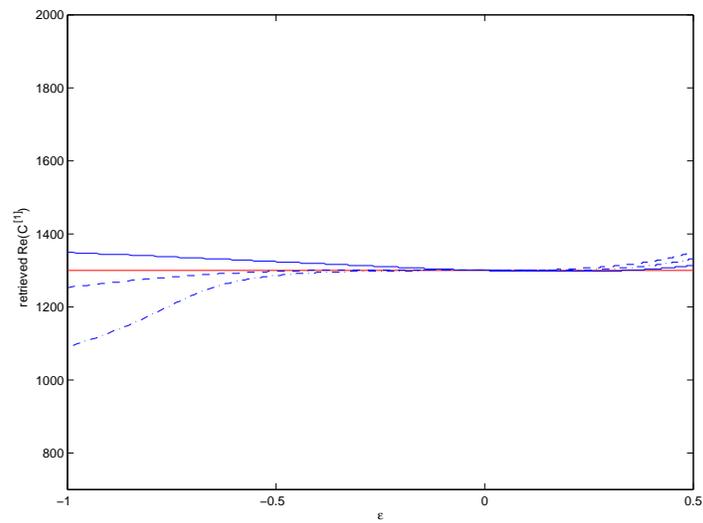}
\caption{Same as fig. \ref{fig38-02} for $f=2000~Hz$.}
\label{fig38-03}
\end{center}
\end{figure}
\begin{figure}[ptb]
\begin{center}
\includegraphics[width=0.65\textwidth]{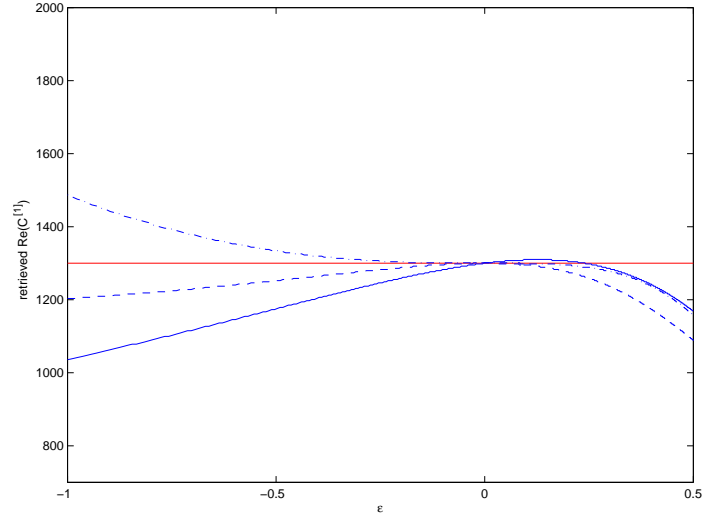}
\caption{Same as fig. \ref{fig38-02} for $f=500~Hz$.}
\label{fig38-04}
\end{center}
\end{figure}
\begin{figure}[ptb]
\begin{center}
\includegraphics[width=0.65\textwidth]{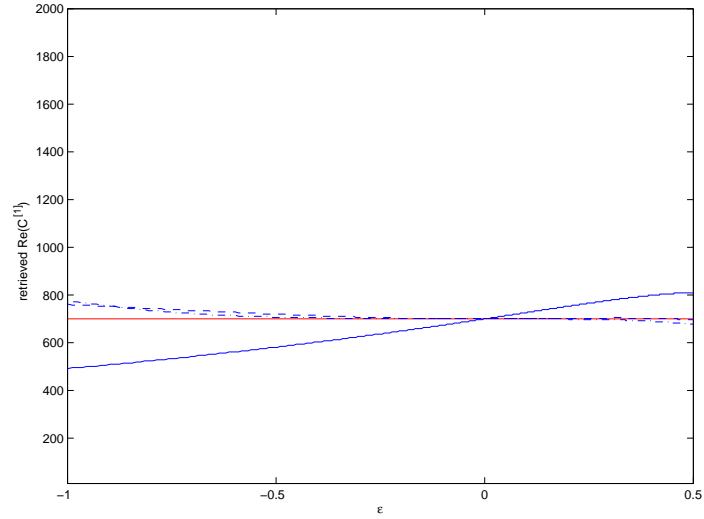}
\caption{Retrieved real part of the layer wavespeed ($C^{[1]'}$)  compared to the actual ($c^{[1]'}$) real part of the layer wavespeed  as a function of  the actual mass density contrast   $\epsilon$.  The represented functions are: $c^{[1]'}$ (red-----), $C^{[1](0)'}$ (blue ------), $C^{[1](1)'}$ (blue - - - -), $C^{[1](2)'}$ (blue -.-.-.-). Case $f=500~Hz$, $c^{[1]}=700~ms^{-1}$.}
\label{fig38-05}
\end{center}
\end{figure}
\clearpage
\newpage
What changes between fig. \ref{fig38-01} and fig. \ref{fig38-02} is the search interval which is being narrowed. In the first of these two figures we observe branch-hopping only for the zeroth-order retrieval, and in the second figure it is seen that the narrowing of the search interval enables the elimination of the 'bad' branch of the zeroth-order retrieval. In fig. \ref{fig35-02} we also observe near-concidence of the retrievals with the actual $c^{[1]'}$ over a wide range of $\epsilon$. In this figure we also observe that for very large negative $\epsilon$, the second-order retrieval can even be less-accurate than the zeroth and first-order retrievals.

What changes between figs. \ref{fig38-02}-\ref{fig38-04} is the frequency $f$. The reduction of frequency is seen to reduce the interval of $\epsilon$ in which the three retrievals are acceptable. These figures also show that the second-order retrieval is generally more accurate than the other two, albeit only for positive $\epsilon$.

What changes between fig. \ref{fig38-04} and fig. \ref{fig38-05} is $c^{[1]'}$ for the same low frequency $f=500~Hz$. Now, in conformity with expectations, the second-order retrieval is more accurate than the first-order retrieval which is more accurate than the zeroth-order retrieval, this being true over the whole range of $\epsilon$. In any case, the zeroth-order retrieval is significantly-inaccurate for  $|\epsilon|>0.2$.
\section{Conclusions}
This investigation dealt with the question of how small must the mass density contrast $\epsilon$ (which is unknown when it is the sought-for-parameter) be for the retrieval of $\epsilon$,  or of another constitutive parameter such as the wavespeed in the obstacle, to be reliable if at the outset the small(or zero)-$\epsilon$ assumption is incorporated in the trial model?

As written in the Introduction, the trial model is nothing but a (method of) solution of a forward problem. Since our inquiry had to do with a small-$\epsilon$ assumption, we first searched for a small-$\epsilon$, preferably mathematical, rather than numerical, solution of the forward-scattering problem. This could not be easily-done for an obstacle of arbitrary shape, so that the first idea was to treat the obstacle (as in the crude paradigm described in the Introduction for bioacoustic diagnosis) as being a flat-faced layer submitted to a plane wave. The mathematically-explicit solution to this forward-scattering  problem is easily-obtained, but the way $\epsilon$ intervenes in this solution is not easy to discern. For this reason, we chose to search for the solution via the domain integral formulation, whereby we found the same solution as previously, but in a form in which $\epsilon$ clearly emerges. Moreover, this new form of the solution lends itself to a perturbation analysis by which the solution, and thus the sought-for trial model (of similar form to what is found in \cite{wi02b,we13,ab14})  is expressed as a series of powers of $\epsilon$, the coefficients of which were found in explicit, algebraic form.

 What we called the zeroth-order trial model results from neglecting all except the first terms in the series, and this model (which does not involve $\epsilon$ and  therefore is as if $\epsilon$ were equal to zero) is nothing but the one resulting from the constant-density assumption. We showed that the inverse problem of the retrieval of $\epsilon$ is trivially-impossible using this zeroth-order trial model.

 Next, we called the first-order trial model the one that results from neglecting all except the first two terms in the series. The employment of this linearized (in terms of $\epsilon$) model enabled, via comparison with simulated data relative to the transmitted field, the retrieval problem to be cast as a linear algebraic equation which was easily-solved for the mass density contrast. The solution was thus found to exist and be unique, these being features of inverse problems that are often strived-for.

 Finally, we called the second-order trial model the one that results from neglecting all except the first three terms in the series. The employment of this model enabled, via comparison with simulated data relative to the transmitted field, the retrieval problem to be cast as a quadratic equation which was again easily and explicitly-solved for the mass density contrast. The solution was found to exist, but  not to be unique since the quadratic equation possesses two solutions, both of which are equally-plausible.

 Since there is no obvious relation between the solutions for the mass density contrast obtained via these three trial models, one can say that either we are not obtaining a solution, obtaining a single solution, or obtaining two solutions that all seem to differ from each other solely on account of the fact of employing different trial models to solve the inverse problem. This result is similar to  what happens when one or more of the priors (i.e., the parameters that are not retrieved, but assumed to be known) in the trial model are, in fact, different from the corresponding parameters in the model employed to simulate the scattered field data, or different from what is guessed to be these parameters in actually-measured data \cite{lm13,sw14}.

 We showed that the use of approximate trial models (i.e., those corresponding to the various orders of $\epsilon$) lead (this was shown numerically) to several remarkable features concerning the retrieved mass density contrast: \\
 1) complex retrievals of what is a real quantity,\\
 2) frequency dispersion of this parameter although the latter is assumed at the outset to be non-dispersive,\\
 3) branch-hopping for the second-order trial model solutions which means, for example, that the solution skips from one branch to another in apparently-haphazard fashion as the frequency (or another parameter) increases,\\
 4) first-order (i.e., based on a very small density contrast assumption) retrievals that are way off mark,\\
 5) one of the second-order solutions is  more affected by frequency dispersion, and is farther from the actual mass density contrast, than the other solution, which fact furnishes a possible means of spotting the 'right' retrieval, provided, of course, the to-be-retrieved parameter is known a priori to actually be non- or feebly-dispersive in the frequency range of interest.

 Not unsurprisingly, we found that the 'right' second-order retrievals were generally-closer than the first-order retrievals to the actual $\epsilon$ for a diversity of priors and frequencies. In fact, the 'right' second-order retrieval was found to be practically-coincident with the actual $\epsilon$ for $\epsilon\in[-0.2,0.2]$. Finally, we found that increasing the amount of angle of incidence realizations did not have a systematically-beneficial effect on the quality of the retrievals.

 The question of the effect of employing different-order $\epsilon$ models to retrieve the real part of the layer wavespeed $c^{[1]'}$ could not be treated in mathematical terms and thus was undertaken numerically. Consequently, we considered the retrieval of the single real parameter $c^{[1]'}$ to be  a non-linear optimization problem which we solved via a standard least-squares algorithm (for the minimization of the corresponding cost function). The retrievals obtained by this technique were found to have the following characteristics:\\
 (i) contrary to the case of the retrieval of $\epsilon$, the zeroth-order model enables the retrieval of the sought-for parameter,\\
 (ii)  the zeroth-order, and more so the two other trial models, enable quite-accurate retrievals of $c^{[1]'}$ except for very low frequencies and/or large $|\epsilon|$\\
 (iii) non-uniqueness (manifested by branch-hopping) was observed for all three types of retrievals (recall that it was only possible for the second-order retrievals of $\epsilon$) and was eliminated by narrowing  the search interval which is justified only if strong a priori information is available as to the actual value of $c^{[1]'}$,\\
 (iv) the retrievals of the real part of the layer wavespeed were found to be dispersive (with respect to frequency)  even though the actual value of this parameter did not depend on frequency, and this phenomenon was found to affect all three retrievals, although more so for the zeroth-order one, and to be particularly noticeable at low frequencies.

  It follows from the preceding remarks that the mass density contrast ($\epsilon$, which equals zero when the constant density assumption is made) is a constitutive parameter that plays an important role in the forward problem solution for the scattered field transmitted by the fluid-like layer solicited by a plane acoustic wave, and consequently this role is as, if not more, important in the inverse problem context in which one strives to retrieve  either the mass density contrast $\epsilon$ or the real part of the layer wavespeed $c^{[1]'}$ from transmitted field data (the same is probably true for the other constitutive parameters and perhaps for the characteristics of the solicitation). This means that the constant density assumption is unjustified, particularly in the inverse problem context, when  $\epsilon$ is poorly-known a priori and may therefore be large enough to cause large inaccuracies in the retrievals of the parameter(s) that is(are) sought-for via the constant density trial model.

  Another, more general, conclusion of this study (underlined in our previous investigation \cite{wi02b} for a much simpler scattering configuration), has to do with another aspect of the non-uniqueness of inverse problem 'solutions':  there exist as many of such 'solutions'    as the number of mathematical translations (here related to the number of terms in the $\epsilon$ series) of the trial models one might want to employ.

  Choosing the 'most accurate' trial model does not eliminate this paradox when real, measured, data is compared to the trial model, because it is not certain a priori that this data is as 'exact' (i.e., devoid of systematic and random errors) as the the trial model and this discrepancy can lead to a difficulty whose origin is described in the next paragraph. Moreover, the adopted trial model necessarily involves a number (which usually is larger the more exact is the trial model) of priors (i.e., the parameters which are not retrieved, but whose values are assumed to be known) and the uncertain knowledge one may have concerning these priors can induce the same type of non-uniqueness of the retrieval \cite{sw14} as the variety of mathematical expressions for the trial-model. As far as we know, no method,  including FWI (full-wave inversion \cite{we13}), are immune to these pathological aspects of inverse problem 'solving'.

   Finally, a nagging issue in the inverse problem context concerns how to obtain what one can qualify as the 'best' retrieval. If, for example, the sought-for parameter is the mass density contrast $\epsilon$, then the 'best' retrieval is obviously attained when the trial parameter $E$ is equal to $\epsilon$, in which case, whatever the definition of the cost function, the latter attains the value zero. In our study, we employed exact simulated data and compared it to approximate trial models with the objective of attaining a zero cost, the result of which generally led to a solution $E\ne\epsilon$ which cannot be qualified as 'best' even though it is associated with zero cost. The question is then: how to obtain $E=\epsilon$ and zero cost at the same time? The answer is that the simulated pressure field data must be identical to the  trial model prediction of the pressure field, this meaning, that we should have simulated the data with the same $\epsilon$ series as the one employed for the trial model (with, of course, the mass-density contrast being denoted by $\epsilon$ in the data series of $l+1$ terms, and by $E$ in the trial model series of $L+1$ terms), which would have led, after imposing the zero- cost condition, to $E=\epsilon$, whatever the number ($l+1=L+1>1$) of terms in both these series. From this one must conclude that the 'best' retrieval is obtained by tailoring the trial model to the data (this is called committing the 'inverse crime' \cite{wi04}), which means that if the data is 'bad', the trial model should be 'bad' and if the data is 'good', the trial model should be 'good'. Of course, tailoring of this sort is only possible when the data is simulated, which means, that when the data is real (as opposed to simulated), it is: 1) impossible to commit the 'inverse crime' and 2) impossible to obtain the 'best' retrieval of the sought-for parameter(s).
%


\begin{thebibliography}{199}
%
\bibitem{as68} Abramowitz M. and Segun I.A., {\em Handbook of Mathematical Functions}, Dover, New York (1968).
%
\bibitem{ab14} Ambrose D.M, Bona J.L. and Nicholls D.P., {\em On ill-posedness of truncated series models for water waves}, Proc. Roy. Soc. A, 470 doi.org/10.1098/rspa.2013.0849 (2014).
%
\bibitem{bh90} Barber P.W. and Hill S.C., {\em Light Scattering by Particles: Computational Methods}, World Scientific, Singapore (1990).
%
\bibitem{bt18} Benali L.A., Tribak A., Terhzaz J. and Mediavilla A., {\em Complex permittivity estimation for each layer in a bi-layer
dielectric material at Ku-band frequencies}, Prog. In Electromag. Res. M, 70, 109-116 (2018).
%
\bibitem{be10} Bergeot B., {\em Contribution \`a la caract\'erisation des mat\'eriaux poreux macroscopiquement inhomog\`enes sous l'approximation du squelette rigide}, Rapport de Stage, LAUM, Le Mans (2010).
%
\bibitem{be46} Bergman P.G., {\em The wave equation in a medium with a variable index of refraction}, J. Acoust. Soc. Am., 17(4), 329-333 (1946).
%
\bibitem{bg04}  Buchanan J.L., Gilbert R.P., Wirgin A. and Xu Y.,
{\it Marine Acoustics, Direct and Inverse Problems}, SIAM, Philadelphia (2004).
%
\bibitem{cc15} Cance P. and Capdeville Y.,  {\em Validity of the acoustic approximation for elastic
waves in heterogeneous media}, Geophys., 80(4) T161-T173 (2015).
%
\bibitem{cj15} Chen M.-Y., Jimura K., White C.N., Maddox W.T. and Poldrack R.A., {\em Multiple brain networks contribute to the acquisition of bias in perceptual decision-making}, Front. Neurosci.
     doi.org/10.3389/fnins2015.00063 (2015).
%
\bibitem{cg04}  Chen X.,  Grzegorczyk T.M.,  Wu B.-I.,  Pacheco Jr. J., and  Kong J.A., {\em Robust method to
retrieve the constitutive effective parameters of metamaterials}, Phys. Rev. E,  70, 016608 (2004).
%
\bibitem{cm17} Church G. and Marra V., {\em Manipulating and controlling sound: The development of acoustic metamaterials}, COMSOL News, Acoustics Edition (2017).
%
\bibitem{cc84} Coen S.,  Cheney M. and Weglein A.., {\em Velocity and density of a two-dimensional acoustic medium from point source surface data}, J. Math. Phys., 25(6), 1857-1861 (1984).
%
\bibitem{ch86} Cohen J.K.,  Hagin F.G.  and  Bleistein N., {\em Three-dimensional  Born  inversion  with  an  arbitrary  reference}, Geophys., 51(8), 1552-1558 (1986).
%
\bibitem{co19} Comsol X., {\em Acoustics Module Updates Version 5.2a}, https://www.comsol.de/release/5.2a/acoustics-module (2019).
%
\bibitem{dl00} Delamare S., Lefebvre J.-P. and Wirgin A., {\em Ghosts in the Born images of a layer probed by acoustic waves}, Ultrasonics, 37,  633-643 (2000).
%
\bibitem{dh07} Deouell L.Y.,  Heller A.S.,  Malach R., D'Esposito M. and  Knight R.T., {\em Cerebral responses to change in spatial location of unattended sounds}, Neuron, 55(6), 985-996 (2007).
%
\bibitem{dr08} De Ryck L., {\em Acoustical characterisation of macroscopically inhomogeneous porous materials}, Phd thesis,  Catholic Univ. Leuven, Leuven (2008).
%
\bibitem{dr07} De Ryck L., Groby J.-P., Leclaire P., Lauriks W., Wirgin A., Depollier C. and Fellah Z.E.A.,
{\em Acoustic wave propagation in a macroscopically inhomogenous porous medium saturated by a
fluid}, Appl. Phys. Lett., 90, 18901 (2007).
%
\bibitem{dl08} De Ryck L.,   Lauriks W., Leclaire P., Groby J.-P., Wirgin A. and  Depollier C., {\em Reconstruction of material properties profiles in one-dimensional macroscopically inhomogeneous rigid frame porous media in the frequency domain}, J. Acoust. Soc. Am., 124(3), 1591-1606 (2008).
%
\bibitem{dc96} Deshpande M.D., Cockrell C.R. and Reddy C.J., {\em Electromagnetic scattering analysis of arbitrarily shaped material cylinder by FEM-BEM method}, NASA Tech. Paper 3575, NASA, New York (1996).
%
\bibitem{de00} Docu A., Eremin Y. and Wriedt T., {\em Acoustic \& Electromagnetic Scattering Analysis Using Discrete Sources}, Academic, San Diego (2000).
%
\bibitem{dl85} Duch\^ene B., Lesselier D.  and Tabbara W., {\em Diffraction tomography approach to acoustical imaging and media characterization}, J. Opt. Soc. Am. A, 2(11), 1943-1953 (1985).
%
\bibitem{dt85} Duch\^ene B. and Tabbara W., {\em Tomographie ultrasonore par diffraction}, Rev. Phys. Appl., 20,  299-304 (1985).
%
\bibitem{el95} Ellis R.G.,  {\em Airborne electromagnetic 3D modelling and inversion}, Explor. Geophys., 26, 138-143 (1995).
%
\bibitem{et15} Elmajid H., Terhzaz J., Ammor H., Chalibi M. and Mediavilla A.,
{\em A new method to determine the complex permittivity and complex permeability of dielectric materials at X-band frequencies}, Int. J. Microwave Opt. Technol., 10(1), 34-39   (2015).
%
\bibitem{eg87} Engl H.W. and Groetsch C.W., Inverse and Ill-Posed Problems, Academic, Orlando (1987).
%
\bibitem{gv86} Gauthier O.,   Virieux J.  and    Tarantola A., {\em  Two-dimensional  nonlinear  inversion  of seismic  waveforms: Numerical  results}, Geophys.,  51(7), 1387-1403 (1986).
%
\bibitem{gv07}  G\'elis, Virieux J. and  Grandjean G.,  {\em Two-dimensional elastic full waveform inversion using Born and Rytov formulations in the frequency domain}, Geophys. J. Int.,  168, 605–633 (2007).
%
\bibitem{gd07} Glide C.,    Duric N. and   Littrup P., {\em Novel approach to evaluating breast density utilizing ultrasound tomography}, Med. Phys., 34(2), 744-753 (2007).
%
\bibitem{gr80} Greenberg M., {\em Focusing in on particle shape}, in Light Scattering By Irregularly Shaped Particles, Schuerman D.W.(Ed.), Plenum, New York,  7-24 (1980).
%
\bibitem{go10}  Groby J.-P.,  Ogam E., De Ryck L., Sebaa N. and Lauriks W., {\em Analytical method for the ultrasonic characterization of
homogeneous rigid porous materials from transmitted and reflected coefficients}, J. Acoust. Soc. Am., 127(2), 764-772 (2010).
%
\bibitem{go12} Groby  J.-P., Ogam E. , Wirgin A. and Xu Y., {\em Recovery of a
material parameter of a soft elastic layer}, Complex Var. Ellipt. Eqs. 57,  317-336 (2012).
%
\bibitem{gu00}  Guillermin R., {\em Caract\'erisation d'objets enfouis dans des s\'ediments marins par imagerie acoustique}, Phd thesis, Universit\'e Aix-Marseille II (2000).
%
\bibitem{gl01} Guillermin R., Lasaygues P., Sessarego J.P. and Wirgin A., {\em Inversion of synthetic and experimental acoustical scattering data for the comparison of two reconstruction methods employing the Born approximation}, Ultrasonics, 39, 121-131 (2001).
%
\bibitem{hh81} Hudson J.A. and Heritage J.R.  {\em The use of the Born approximation in seismic scattering problems}, Geophys. J.R. Astr. Soc., 66, 221-240 (1981).
%
\bibitem{hs11} Huthwaite P. and Simonetti F., {\em High-resolution imaging without iteration: A fast and robust
method for breast ultrasound tomography}, J. Acoust. Soc. Am., 130(3),  1721-1734 (2011).
%
\bibitem{ke16} 	Kerker M., {\em The Scattering of Light and Other Electromagnetic Radiation}, Elsevier, Amsterdam (2016).
%
\bibitem{kd91} Kormendi F. and Dietrich M., {\em Nonlinear waveform inversion of plane-wave seismograms in stratified
elastic media}, Geophys., 56(5), 664-674 (1991).
%
\bibitem{kr02a}  Kress R., {\em  Specific theoretical tools}, in Scattering, Pike R. and Sabatier P. (Eds.), Academic, San Diego,  37-51 (2002).
%
\bibitem{kr02b}  Kress R., {\em  Scattering by obstacles}, in Scattering, Pike R. and Sabatier P. (Eds.), Academic, San Diego,  52-73 (2002).
%
\bibitem{lv92} Lambar\'e  G., Virieux J., Madariaga R. and Jin S., {\em Iterative asymptotic inversion in the acoustic approximation},  Geophys., 57(9), 1138-1154 (1992).
%
\bibitem{ll07} Lasaygues P. and Le Marrec L., {\em  Ultrasonic reflection tomography vs. canonical body
approximation: Experimental assessment of an infinite elastic cylindrical tube}, Ultrason. Imag., 29,  182-194 (2007)
%
\bibitem{lo09} Lavarello R.J and Oelze M.L., {\em Density imaging using inverse scattering}, J. Acoust. Soc. Am. 125(2),  793-802 (2009).
%
\bibitem{lo18}  Lazri H.,  Ogam E.,  Amar B., Fellah Z.E.A., Sayoud N. and  Boumaiza Y.,
{\em Probing flexible thermoplastic thin films on a substrate using ultrasonic
waves to retrieve mechanical moduli and density: Inverse problem},   J. Phys.: Conf. Ser. 1017, 012004 (2018).
%
\bibitem{le85} Lefebvre J.-P., {\em La tomoographie d'imp\'edance acoustique}, Trait. Signal, 2(2), 103-110 (1985).
%
\bibitem{lm13} Lefeuve-Mesgouez G., Mesgouez A.,  Ogam E., Scotti T. and  Wirgin A., {\em Retrieval of the physical properties of an anelastic solid half space from seismic data}, J. Appl. Geophys., 88,   70-82 (2013).
%
\bibitem{ls50} Lippmann B.A. and Schwinger J., {Variational principles for scattering processes. I}, Phys. Rev., 79(3), 469-480 (1950).
%
\bibitem{mp17}  Maffucci A., Perrotta A.,  Rubinacci G., Tamburrino A. and  Ventre S. {\em Efficient numerical
evaluation of the electromagnetic scattering from arbitrarily-shaped objects by using the
Dirichlet-to-Neumann map} , Proc. of ICEAA 2017, Verona, Italy, 1616-1619, (2017).
%
\bibitem{mh99} Mast T.D., Hinkelman L.M., Metlay L.A., Orr M.J. and Waag R.C., {\em Simulation of ultrasonic pulse propagation, distortion, and attenuation in the human chest wall}, J. Acoust. Soc. Am., 106, 3665-3677 (1999).
%
\bibitem{mi18} Mindrinos L., {\em The electromagnetic scattering problem by a cylindrical doubly-connected domain at oblique incidence: the direct problem},  arXiv:1802.00225v1 [math.AP] (2018).
%
\bibitem{mt03}  Mishchenko M.I. and  Travis L.D., {\em Electromagnetic scattering by nonspherical particles}, in Exploring the Atmosphere by Remote Sensing Techniques, Guzzi R. (Ed.), Springer, Berlin, 77-127 (2003).
%
\bibitem{mc93}  Moghaddam M. and Chew W., {\em Variable density linear acoustic inverse
problem}, Ultrason. Imag.,  15(3), 255-266 (1993).
%
\bibitem{mv15} Mojabi P. and  Vetri J., {\em Ultrasound tomography for simultaneous reconstruction of acoustic density, attenuation, and compressibility profiles}, J. Acoust. Soc.Am., 137(4), 1813-1820 (2015).
%
\bibitem{mo08} M\"onk\"ol\"a S., {\em Spectral element method and controllability approach for time-harmonic wave propagation}, Phd thesis, Univ. Jyv\"askyl\"a, Jyv\"askyl\"a (2008).
%
\bibitem{mo87} Mora P., {\em Nonlinear two-dimensional elastic inversion of multioffset seismic data}, Geophys., 52(9), 1211-1228 (1987).
%
\bibitem{mf53} Morse P.M. and Feshbach H., {\em Methods of Theoretical Physics, vol. 1}, McGraw-Hill, New York (1953).
%
\bibitem{ns09} Nammour R. and Symes W., {\em Approximate constant density acoustic inverse scattering using dip-dependent scaling},  SEG Houston 2009 International Exposition and Annual Meeting, 2347-2351 (2009).
%
\bibitem{of11}  Ogam E. and Fellah Z.E.A., The direct and inverse problems of an air-saturated poroelastic
cylinder submitted to acoustic radiation, AIP Advances 1, 032174 (2011).
%
\bibitem{pa19} Pandey A. and Anand A., {\em Improved convergence of fast integral equation solvers for acoustic scattering by inhomogeneous penetrable media with discontinuous material interface}, J. Comput Phys., 376(1), 767-785  (2019).
%
\bibitem{pe80} Petit R., {\em Tutorial introduction}, in Electromagnetic Theory of Gratings, Petit R. (Ed.), Springer, Berlin (1980).
%
\bibitem{po12} Popov E. (Ed.), Gratings: Theory and  and Numeric Applications, PUP, Marseille (2012).
%
\bibitem{ro04} Ravaut C.,  Operto S., Improta L., Virieux J., Herrero A. and Dell'Aversana P., {\em
Multiscale imaging of complex structures from multifold wide-aperture seismic data by frequency-domain full-waveform tomography: application to a thrust belt}, Geophys. J. Int., 159, 1032-1056 (2004).
%
\bibitem{ra18}  Rayleigh L., The dispersal of light by a dielectric cylinder, Philos.Mag., 36, 365-376
(1918).
%
\bibitem{ra81} Raz S.,  {\em Direct reconstruction of velocity and density profiles from scattered field data},  Geophys., 46(6), 832-836 (1981).
%
\bibitem{ri65} Richmond J.H., {\em Scattering by a dielectric cylinder of arbitrary cross-section shape}, IEEE Trans. Anten. Prop., AP-13(3), 334-341 (1965).
%
\bibitem{rg86} Robinson B.S.  and Greenleaf J.F., {\em The  scattering  of ultrasound  by cylinders:  Implications for  diffraction  tomography}, J. Acoust. Soc. Am. 80(1), 40-49 (1986).
%
\bibitem{sw14} Scotti T. and Wirgin A., {\em Multiparameter identification of a lossy fluid-like object from its transient acoustic response}, Inv. Probs. Sci. Engrg., 22(8), 1228-1258  (2014).
%
\bibitem{sw04}  Scotti T. and Wirgin A., {\em Reconstruction of the three mechanical material constants
of a lossy fluid-like cylinder from low-frequency scattered acoustic fields}, C. R. M\'ecanique, 332, 717-724 (2004).
%
\bibitem{ss02} Smith D.R.,  Schultz S.,  Markos P., and  Soukoulis C.M., {\em Determination of effective permittivity and permeability of metamaterials from reflection and transmission coeffcients}, Phys. Rev. B,
65, 195104 (2002).
%
\bibitem{sv05}  Smith D.R.,  Vier D.C., Koschny T., and  Soukoulis C.M., {\em Electromagnetic parameter retrieval
from inhomogeneous metamaterials}, Phys. Rev. E,  71, 036617 (2005).
%
\bibitem{st92} Steinberg B.D., {\em A discussion of two wavefront aberration correction procedures}, Ultrason. Imag., 14, 387-397 (1992).
%
\bibitem{ta76} Tabbara W., {Etude de probl\`emes de diffraction inverse \'electromagn\'etique par analyse multifr\'equentielle des champs diffract\'es}, Phd thesis, Univ. Paris VI, Paris (1976).
%
\bibitem{ta84} Tarantola, A., {\em Inversion of seismic reflection data in the acoustic approximation}, Geophys., 49(8), 1259-1266 (1984).
%
\bibitem{ta86} Tarantola, A., {\em A strategy for nonlinear elastic inversion of seismic reflection data}, Geophys., 51(10), 1893-1903 (1986).
%
\bibitem{to99} Thierry P., Operto S. and  Lambar\'e G., {\em Fast 2-D ray+Born migration/inversion in complex media}, Geophys., . 64(1), 162-181 (1999).
%
\bibitem{th06} Thompson L.L., {\em A review of finite-element methods for time-harmonic acoustics}, J. Acoust. Soc. Am.,  119-131, 1315 (2006).
%
\bibitem{ta77} Tikhonov A.N. and Arsenin V.Y., Solutions of Ill-Posed Problems, Winston \& Sons, Washington (1977).
%
\bibitem{to10} Touhei T., {\em A volume integral equation method for the direct/inverse problem in elastic wave scattering phenomena},  in Wave Propagation in Materials for Modern Applications, Petrin A. (ed.), Intech, Croatia, 526 (2010).
%
\bibitem{ts99} Tsogka C., {\em Mod\'elisation math\'ematique et num\'erique
de la propagation des ondes \'elastiques tridimensionnelles
dans des milieux fissur\'es}, Phd thesis, Univ. Paris-Dauphine, Paris (1999).
%
\bibitem{ts18}  Tzarouchis D. and  Sihvola A., {\em Light scattering by a dielectric sphere: Perspectives on the Mie resonances}, Appl. Sci., 8, 184-195 (2018).
%
\bibitem{ut86} Umashankar K., Taflove A. and Rao S., {\em Electromagnetic scattering by  arbitrary shaped three-dimensional homogeneous lossy dielectric objects} , IEEE Trans. Anten. Prop., 34(6), 758-766 (1986).
%
\bibitem{ve12} Virieux J., Etienne V. and  Cruz-Atienza V.,  Brossier R.,  Chaljub E.,  Coutant O.,
 Garambois S.,  Mercerat D.,  Prieux V. and Operto S., {\em Modelling seismic wave propagation for
geophysical imaging}, Seismic Waves - Research and Analysis, Masaki
Kanao M. (Ed.), Ch.13, 253-304,  (2012).
%
\bibitem{we13} Weglein A.B., {\em A timely and necessary antidote to indirect methods and so-called
P-wave FWI}, The Leading Edge, Oct., 1192-1204 (2013).
%
\bibitem{wv86} Weglein A.B., Violette P.B. and Keho T.H., {\em Using multiparameter Born theory to obtain certain exact multilparameter inversion goals}, Geophys., 51(5),  1069-1074 (1986).
%
\bibitem{wi82} Wirgin A., {\em Iterative corrections to the Kirchhoff approximation in rough periodic surface sacattering}, 41(3), 154-158 (1982).
%
\bibitem{wi99}  Wirgin A., {\em Some quasi-analytic and numerical methods for acoustical imaging of complex media}, in Wavefield Inversion,  Wirgin A. (ed). Springer, Wien (1999).
%
\bibitem{wi02a}  Wirgin A., {\em  Acoustical imaging : classical and emerging methods for applications in macro-
physics},  Scattering, Pike R., Sabatier P. (Eds.), Academic, San Diego,  95-120 (2002).
%
\bibitem{wi02b} Wirgin A., {\em Ill-posedness and accuracy in connection with the
recovery of a single parameter from a single measurement}, Inverse Probs. Engrg.,
10(2), 105-115 (2002).
%
\bibitem{wi04} Wirgin A. {\em The inverse crime}, arXiv.org/abs/math-ph/0401050 (2004).
%
\bibitem{wi16a} Wirgin A., {\em  Retrieval of the equivalent acoustic constitutive parameters of an inhomogeneous fluid-like object by nonlinear full waveform inversion}, Ultrasonics, 65, 353-69 (2016).
%
\bibitem{wi16b} Wirgin A., {\em Retrieval of the frequency-dependent effective permeability and
permittivity of the inhomogeneous material in a layer}, Progr. in Electromag. Res.  70, 131-147 (2016).
%
\bibitem{wi16c} Wirgin A., {\em Apparent attenuation and dispersion arising in seismic body-wave velocity-retrieval}, Pure Appl. Geophys.,
173, 2473-2488 (2016).
%
\bibitem{wi18} Wirgin A., {\em Incorporation of macroscopic heterogeneity within a porous layer to enhance its acoustic absorptance}, arXiv:1810.02101,   physics.app-ph (2018).
%
\bibitem{wi19} Wirgin A., {\em On the constant constitutive parameter (e.g., mass
density) assumption in integral equation approaches to
(acoustic) wave scattering}, arXiv:1903.09573v1 (2019).
%
\bibitem{wb12} Wiskin J., Borup D.T., Johnson S.A., and Berggren M., {\em Non-linear inverse scattering: High resolution quantitative breast tissue tomography}, J. Acoust. Soc. Am. 131 (5), 3802-3813 (2012).
%
\bibitem{zb02}  Zhang X., Broschat S.L. and Flynn P.J., {\em A comparison of material classification techniques
for ultrasound inverse imaging}, J. Acoust. Soc. Am. 111 (1), Pt. 1, 457-467 (2002).
\end{thebibliography}
\end{document}